\documentclass[a4paper,11pt]{article}
\pdfoutput=1

\usepackage{jheppub}
\usepackage{graphicx}
\usepackage{amsmath}
\usepackage{xspace}

\newcommand{\eq}[1]{eq.~\eqref{eq:#1}}
\newcommand{\eqs}[2]{eqs.~\eqref{eq:#1} and \eqref{eq:#2}}
\renewcommand{\sec}[1]{section~\ref{sec:#1}}

\newcommand{\subsec}[1]{section~\ref{subsec:#1}}

\newcommand{\subsubsec}[1]{section~\ref{subsubsec:#1}}
\newcommand{\subsubsecs}[2]{sections~\ref{subsubsec:#1} and \ref{subsubsec:#2}}
\newcommand{\app}[1]{Appendix~\ref{app:#1}}
\newcommand{\fig}[1]{figure~\ref{fig:#1}}
\newcommand{\figs}[2]{figures~\ref{fig:#1} and \ref{fig:#2}}
\newcommand{\mycites}[1]{refs.~\cite{#1}}
\newcommand{\mycite}[1]{ref.~\cite{#1}}

\newcommand{\abs}[1]{\lvert#1\rvert}
\newcommand{\Abs}[1]{\bigl\lvert#1\bigr\rvert}
\newcommand{\ord}[1]{\mathcal{O}(#1)}

\newcommand{\df}{\mathrm{d}}
\newcommand{\img}{\mathrm{i}}

\newcommand{\tr}{\mathrm{tr}}

\newcommand{\la}{\lambda}

\newcommand{\vC}{\vec{C}}

\newcommand{\hq}{\hat{q}}
\newcommand{\hs}{\hat{s}}
\newcommand{\hS}{\widehat{S}}
\newcommand{\hT}{\widehat{T}}

\newcommand{\hga}{\widehat{\gamma}}

\newcommand{\cA}{\mathcal{A}}
\newcommand{\cI}{\mathcal{I}}
\newcommand{\cL}{\mathcal{L}}
\newcommand{\Tau}{\mathcal{T}}

\newcommand{\GeV}{\,\mathrm{GeV}}
\newcommand{\TeV}{\,\mathrm{TeV}}

\newcommand{\nn}{\nonumber}

\newcommand{\cut}{\mathrm{cut}}
\newcommand{\IR}{\mathrm{IR}}
\newcommand{\LO}{\mathrm{LO}}
\newcommand{\NLO}{\mathrm{NLO}}
\newcommand{\NNLO}{\mathrm{NNLO}}

\newcommand{\sing}{\mathrm{sing}}
\newcommand{\nons}{\mathrm{nons}}

\newcommand{\zero}{{(0)}}
\newcommand{\one}{{(1)}}
\newcommand{\two}{{(2)}}

\newcommand{\MSbar}{$\overline{\text{MS}}$\xspace}
\newcommand{\Ecm}{E_\mathrm{cm}}
\newcommand{\as}{\alpha_s}

\newcommand{\TauN}{\Tau_N}
\newcommand{\TauNcut}{{\Tau_N^\cut}}
\newcommand{\TauIR}{{\Tau_\delta}}
\newcommand{\delIR}{{\delta_\IR}}
\newcommand{\Tauoff}{{\Tau_\mathrm{off}}}

\newcommand{\C}{\mathcal{C}}
\newcommand{\fin}{\mathrm{fin}}
\newcommand{\meas}{X}
\newcommand{\rad}{\mathrm{rad}}

\newcommand{\geneva}{\textsc{Geneva}\xspace}
\newcommand{\vrap}{\textsc{Vrap}\xspace}
\newcommand{\hnnlo}{\textsc{HNNLO}\xspace}
\newcommand{\mcfm}{\textsc{MCFM}\xspace}

\newcommand{\conv}{\!\otimes\!}

\newcommand{\off}{\mathrm{off}}

\newcommand{\cusp}{\mathrm{cusp}}

\newcommand{\id}{\mathbf{1}}
\newcommand{\bC}{\mathbf{C}}
\newcommand{\bI}{\mathbf{I}}
\newcommand{\bL}{\mathbf{L}}
\newcommand{\bT}{\mathbf{T}}
\newcommand{\Tij}{\bT_i \!\cdot\! \bT_j}

\newcommand{\T}{\bar{T}}

\allowdisplaybreaks[3]

\title{N-jettiness Subtractions for NNLO QCD Calculations}

\author[a]{Jonathan R.~Gaunt,}
\author[a]{Maximilian Stahlhofen,}
\author[a]{Frank J.~Tackmann,}
\author[b,c]{and Jonathan R.~Walsh}

\emailAdd{jonathan.gaunt@desy.de}
\emailAdd{maximilian.stahlhofen@desy.de}
\emailAdd{frank.tackmann@desy.de}
\emailAdd{jwalsh@lbl.gov}

\affiliation[a]{Theory Group, Deutsches Elektronen-Synchrotron (DESY), Notkestra\ss e 85, D-22607 Hamburg, Germany}
\affiliation[b]{Lawrence Berkeley National Laboratory, University of California, Berkeley, CA 94720, USA}
\affiliation[c]{Berkeley Center for Theoretical Physics, University of California, Berkeley, CA 94720, USA}

\abstract{
We present a subtraction method utilizing the $N$-jettiness observable, $\Tau_N$,
to perform QCD calculations for arbitrary processes at next-to-next-to-leading order (NNLO).
Our method employs soft-collinear effective theory (SCET) to determine
the IR singular contributions of $N$-jet cross sections for $\TauN\to 0$,
and uses these to construct suitable $\TauN$-subtractions.
The construction is systematic and economic, due to being based on a physical observable.
The resulting NNLO calculation is fully differential and in a form directly
suitable for combining with resummation and parton showers.
We explain in detail the application to processes with an arbitrary number
of massless partons at lepton and hadron colliders together with the
required external inputs in the form of QCD amplitudes and lower-order calculations.
We provide explicit expressions for the $\TauN$-subtractions at NLO and NNLO. The required ingredients are
fully known at NLO, and at NNLO for processes with two external QCD partons.
The remaining NNLO ingredient for three or more external partons
can be obtained numerically with existing NNLO techniques.
As an example, we employ our results to obtain the NNLO rapidity spectrum for Drell-Yan and gluon-fusion Higgs production.
We discuss aspects of numerical accuracy and convergence and the practical implementation.
We also discuss and comment on possible extensions,
such as more-differential subtractions, necessary steps for going to N$^3$LO, and the treatment of massive quarks.
}


\begin{document}

\preprint{
\begin{flushright}
DESY 15-066\\
May 18, 2015
\end{flushright}
}

\maketitle

\section{Introduction}
\label{sec:intro}

The precise knowledge of QCD corrections is a key ingredient for interpreting the data from collider experiments.
In hadronic collisions, the inclusive QCD cross section for the production of a final state $X$ can, if the hard scale $Q$ associated with $X$ is large enough, be obtained in terms of a perturbatively calculable partonic cross section
convolved with parton distribution functions (PDFs).

Perturbative calculations performed using the leading order (LO) term in $\alpha_s$ typically suffer from large theoretical uncertainties due to missing higher-order perturbative corrections. Often, next-to-leading order (NLO) is the first order at which the normalization and in some cases the shape of cross sections can be considered reliable. As such, this level of accuracy has become standard for comparing with data from the LHC. For some processes the experimental uncertainties are becoming so small, or the perturbative uncertainties at NLO are still so large, that next-to-next-to-leading order (NNLO) computations are called for.

For many important benchmark processes, the required virtual amplitudes are known at NNLO. However, as is well known, the computation of the full cross sections beyond leading order is complicated by infrared (IR) divergences -- explicit divergences in virtual amplitudes, and divergences in the phase-space integration over the real-emission amplitudes in regions where particles become soft or collinear to other particles. These divergences only cancel after integrating the real-emission amplitudes over the phase space of unresolved particles and adding the result to the virtual loop amplitudes order by order.

To handle these divergences in practice one typically makes use of some subtraction method. That is, one subtracts terms from the real emission contributions that reproduce the IR soft and collinear behaviour of the real emissions, which then allows the phase-space integral of the full amplitude minus the subtraction terms to be performed numerically in $d=4$ dimensions, giving a finite result. The subtracted terms have to be sufficiently simple that they can be integrated over the phase space of emitted particles in $d=4-2\epsilon$ dimensions. They are then added back to the virtual contributions, where they cancel the explicit $1/\epsilon^n$ IR poles.

The goal of typical NLO subtraction schemes like FKS subtractions~\cite{Mangano:1991jk, Frixione:1995ms, Frixione:1997np} or CS subtractions~\cite{Catani:1996jh, Catani:1996vz, Catani:2002hc} is to construct subtraction terms that reproduce the correct IR-singular behaviour of the full real-emission amplitude point-by-point in phase space. Over the past decade enormous effort has been devoted to extend such local subtraction methods to NNLO using different approaches~\cite{Weinzierl:2003fx, Frixione:2004is, Somogyi:2005xz, Somogyi:2006da, Somogyi:2006db, Somogyi:2008fc, Aglietti:2008fe, Bolzoni:2009ye, Bolzoni:2010bt, DelDuca:2013kw, Somogyi:2013yk, DelDuca:2015zqa, GehrmannDeRidder:2004tv, GehrmannDeRidder:2005cm, Daleo:2006xa, Daleo:2009yj, Glover:2010im, Boughezal:2010mc, Gehrmann:2011wi, GehrmannDeRidder:2012ja, Currie:2013vh, Currie:2013dwa, Anastasiou:2003gr, Binoth:2004jv, Czakon:2010td, Czakon:2011ve, Boughezal:2011jf, Brucherseifer:2013iv, Boughezal:2013uia, Czakon:2013goa, Czakon:2014oma, Boughezal:2015dra}. This extension is very involved due to the many overlapping singularities at NNLO, which have to be isolated by appropriate phase-space parameterizations. At the same time, the subtractions have to remain simple enough that the $1/\epsilon^n$ IR poles can be extracted from the integrated subtractions.

The basic idea of our method, which we call $N$-jettiness subtractions, is to use a physical jet-resolution variable $\TauN$ to control the infrared behaviour of the cross section. The key point is that, if the (factorized) structure of the leading contribution to the $\TauN$-differential cross section in the IR limit $\TauN\to 0$ is known, the singular part can often be determined analytically and used to construct an IR subtraction term. A major advantage of using a physical observable is that the differential and integrated subtraction terms are then equivalent to the singular limits of a physical cross section, which can indeed be significantly easier to calculate than the full cross section. A well-known example of such a physical subtraction scheme is the $q_T$-subtraction method for color-singlet production in hadron collisions~\cite{Catani:2007vq}, which has been successfully applied to a variety of processes~\cite{Catani:2009sm, Catani:2011qz, Ferrera:2011bk, Ferrera:2014lca, Cascioli:2014yka, Gehrmann:2014fva, Grazzini:2013bna, Grazzini:2015nwa}. (It has also been suggested that this method can be applied to compute heavy-quark pair production at NNLO~\cite{Zhu:2012ts, Catani:2014qha}.) Our $N$-jettiness subtraction method generalizes this to arbitrary numbers of QCD partons in the initial and final state. It employs the $N$-jettiness global event shape~\cite{Stewart:2010tn} as the physical $N$-jet resolution variable. In this paper, we limit ourselves to massless quarks; the extension to massive quarks is in principle possible and commented on in \sec{conclusions}.

The key feature of $N$-jettiness is that it has very simple factorization properties in the singular limit. The factorization theorem for the $N$-jettiness cross section is known~\cite{Stewart:2009yx, Stewart:2010tn, Jouttenus:2011wh} from soft-collinear effective theory (SCET)~\cite{Bauer:2000ew, Bauer:2000yr, Bauer:2001ct, Bauer:2001yt, Bauer:2002nz, Beneke:2002ph}. It can be used to systematically compute the leading singular contributions (thus determining the subtraction terms) by performing standard fixed-order calculations of soft and collinear matrix elements in SCET. At NLO, all necessary ingredients have been known for some time, and by now, essentially all necessary NNLO ingredients are available. For processes with hadronic initial states a key ingredient that has become available recently are the two-loop quark and gluon beam functions~\cite{Gaunt:2014xga, Gaunt:2014cfa}.

The price one has to pay for using a single physical observable to describe the IR is that the subtraction does not act point-by-point in phase space, but only on a more global level after a certain amount of phase-space integration has been carried out. In essence, the large number of terms in a fully local subtraction method are projected onto a single, nonlocal subtraction term. In practice, this means that the numerical convergence may be slower than for the fully local case. However, this is compensated by the significant reduction in complexity of the subtractions. Furthermore, as we will discuss, it is possible to make the subtractions step-by-step more local by making the $N$-jettiness cross section more differential in additional variables. This is again possible by using SCET to factorize and calculate the singular contributions of more differential cross sections (see e.g.~\mycites{Bauer:2011uc, Jouttenus:2011wh, Jouttenus:2013hs, Procura:2014cba, Larkoski:2014tva, Larkoski:2015zka}).

There are several important benefits of using a physical observable as jet resolution variable, as already emphasized in \mycites{Alioli:2013hqa}. It allows one to directly reuse the existing NLO calculations for the corresponding $N+1$-jet cross sections, and the resulting NNLO calculation is automatically fully-differential in the Born phase space. Moreover, the calculation will be in a form which makes it directly suitable to be combined with higher-order resummation as well as parton showers by using the general methods developed in \mycites{Alioli:2012fc, Alioli:2013hqa}.

The idea of using $N$-jettiness as an $N$-jet resolution variable is not new. In fact, this is what largely motivated its invention in the first place.
It is already utilized in essentially the same context as here in the \geneva Monte-Carlo program~\cite{Alioli:2012fc}.
For color-singlet production, the $N$-jettiness subtraction method reduces to an analogue
of $q_T$ subtractions~\cite{Catani:2007vq} with an alternative physical resolution variable.
The differential version as a subtraction was used at NLO in \mycite{Gangal:2014qda}.

In its simplest form as a phase-space slicing, the $N$-jettiness subtraction method has been successfully applied already to calculate the top quark decay rate at NNLO~\cite{Gao:2012ja}.%
\footnote{A similar slicing method utilizing heavy-quark effective theory was also used in \mycite{Gao:2014nva, Gao:2014eea} to perform the fully-differential NNLO calculation for $e^+e^-\!\to t\bar t$.}
While this work was being finalized, this method was also suggested and applied to the NNLO calculations of $pp\to W/H+\text{jet}$ in Refs.~\cite{Boughezal:2015dva, Boughezal:2015aha}. These results clearly highlight the usefulness of the slicing method, even for complex $2\to2$ processes with three colored partons.

In this work we give a general description of how $N$-jet resolution variables, and specifically $N$-jettiness, can be used as subtraction terms to compute fixed-order cross sections.  In \sec{formalism}, we discuss how the IR singularities in QCD cross sections are encapsulated by an $N$-jet resolution variable.  We demonstrate that this naturally leads to subtraction terms for fixed-order calculations, and show how these can be used in phase-space slicing, as done in \mycites{Gao:2012ja, Gao:2014nva, Gao:2014eea, Boughezal:2015dva, Boughezal:2015aha}, and as differential subtractions, generalizing $q_T$-subtractions~ \cite{Catani:2007vq}.  In \sec{TauN}, we review the definition of $N$-jettiness and its general factorization theorem for $N$-jet production.  We show how the subtraction terms are defined in terms of functions in the factorization theorem. We explicitly construct the subtraction terms at NLO and NNLO for generic $N$-parton processes. We also discuss the extension to N$^3$LO and to more-differential subtractions.  In \sec{implementation}, we discuss how these subtractions may be implemented in parton-level Monte-Carlo programs. We also show results for Drell-Yan and gluon-fusion Higgs production at NNLO and use these as an example to discuss some of the numerical aspects.  We conclude in \sec{conclusions}.

\section{General Formalism}
\label{sec:formalism}

\subsection{Notation}

We denote the $N$-jet cross section that we want to compute by $\sigma(X)$. Here, $X$ collectively stands for all differential measurements and kinematic cuts applied at Born level. In particular, it contains the definitions of the $N$ identified signal jets in $\sigma(X)$ and all cuts required to stay away from any IR-singularities in the $N$-parton Born phase space.

The cross section at leading order (LO) in perturbation theory can then be written as
\begin{equation} \label{eq:sigmaLO}
\sigma^\LO(X) = \int\! \df\Phi_N\, B_N(\Phi_N) \, \meas(\Phi_N)
\,,\end{equation}
where the measurement function $\meas(\Phi_N)$ implements $X$ on an $N$-parton final state.
The Born contribution, $B_N(\Phi_N)$, is given by the square of the lowest-order amplitude, $\cA^\zero$, for the process we are interested in,%
\footnote{For a tree-level process, $\cA^\zero$ is given by the sum of the relevant tree-level diagrams. For a loop-induced process, like $gg\to H$, it is the sum of the relevant lowest-order IR-finite loop diagrams.}
\begin{equation}
B_N(\Phi_N)
= \sum_\mathrm{color} \Abs{\cA_N^\zero(\Phi_N)}^2
\qquad\text{or}\qquad
B_N(\Phi_N)
= f_a\,f_b\,\sum_\mathrm{color} \Abs{\cA_{ab\to N}^\zero(\Phi_N)}^2
\,,\end{equation}
where $\Phi_N$ denotes the complete dependence of the amplitude on the external state (including all dependence on momentum, spin, and partonic channel). For hadronic collisions, the PDFs $f_{a,b}$ are included in $B_N(\Phi_N)$ and $\Phi_N$ also includes the corresponding momentum fractions $x_{a,b}$. Correspondingly, the integral over $\df\Phi_N$ in \eq{sigmaLO} includes all phase-space integrals and sums over helicities and partonic channels. For simplicity, we also absorb into it flux, symmetry, and color and spin averaging factors. We use $N$ to denote the number of strongly-interacting partons in the final state. There can also be a number of additional nonstrongly interacting final states at Born level, which are included in $\Phi_N$ but we suppress for simplicity.

\subsection{Singular and nonsingular contributions}
 \label{subsec:SingAndNonsing}

Any $N$-jet cross section $\sigma(X)$ can also be measured differential in a generic $N$-jet resolution variable $\TauN$, which we write as $\df\sigma(X)/\df\TauN$.  Then $\sigma(X)$ may be written as
\begin{equation} \label{eq:dsigmadTau}
\sigma(X)
= \int_0\!\df\TauN\, \frac{\df\sigma(X)}{\df\TauN}
= \int_0^\TauNcut\!\df\TauN\, \frac{\df\sigma(X)}{\df\TauN}
+ \int_\TauNcut\!\df\TauN\, \frac{\df\sigma(X)}{\df\TauN}
\,,\end{equation}
dividing the more differential cross section into the region $0 \leq \TauN \leq \TauN^\cut$ and the region $\TauN \geq \TauN^\cut$.  For $\TauN$ to be an $N$-jet resolution variable it must satisfy the following conditions:
\begin{equation} \label{eq:TauNIRcondition}
\TauN(\Phi_N) = 0
\,,\qquad
\TauN(\Phi_{\geq N+1}) > 0
\,,\qquad
\TauN(\Phi_{\geq N+1} \to \Phi_N) \to 0
\,.\end{equation}
In words, $\TauN$ must be a physical IR-safe observable that resolves \emph{all additional} IR-divergent real emissions, such that the cross section $\df\sigma(X)/\df\TauN$ is physical and IR finite for any $\TauN > 0$, and the IR singular limit corresponds to $\TauN\to 0$.%
\footnote{For particular definitions of $\TauN$, there could also be regions of $\Phi_{\geq N+1}$ (far) away from any IR singularities where $\TauN$ is small or vanishing. Such regions do not pose a problem and are irrelevant for our discussion. The typical example for $\TauN\equiv q_T$ at NNLO are contributions from two hard real emissions that are back-to-back such that $q_T \to 0$. Another generic example are regions where two partons are collinear that cannot arise from a QCD singular splitting. Such cases can be avoided by defining $\TauN$ in a flavor-aware way.}
Hence, we have
\begin{equation} \label{eq:LO}
\frac{\df\sigma^\LO(X)}{\df\TauN} = \sigma^\LO(X)\,\delta(\TauN)
\,,\qquad
\frac{1}{\sigma^\LO(X)}\,
\frac{\df\sigma(X)}{\df\TauN} \bigg\rvert_{\TauN > 0} = \ord{\alpha_s}
\,.\end{equation}

We use the convention that $\TauN$ is normalized to be a dimension-one quantity, and for convenience we also define the dimensionless quantities
\begin{equation}
\tau = \frac{\TauN}{Q}
\,,\qquad
\tau^\cut = \frac{\TauNcut}{Q}
\,.\end{equation}
Here, $Q$ is a typical hard-interaction scale of the Born process (whose precise choice however is unimportant). For example, canonical choices would be $Q=\Ecm$ for $e^+e^-\to$ jets, $Q = \sqrt{q_{\ell\ell}^2}$ for Drell-Yan $pp\to V \to \ell \ell$, $Q = m_H$ for $gg\to H$, and $Q = p_T^\mathrm{jet}$ for $pp\to$ dijets.

We define the ``singular'' part of the $\TauN$ spectrum to contain all contributions that are singular in the $\TauN\to 0$ limit, i.e., all contributions which are either proportional to $\delta(\TauN)$ or that behave as $\ln^n(\tau)/\tau$ for $\tau\to 0$. It can be written as
\begin{align} \label{eq:dsigmasing}
\frac{\df\sigma^\sing(X)}{\df\tau}
&= \C_{-1}(X)\,\delta(\tau) + \sum_{n\geq 0} \C_n(X)\, \cL_n(\tau)
\,,\end{align}
where the $\cL_n(\tau)$ are the usual plus distributions. For a suitable test function $f(\tau)$:
\begin{align} \label{eq:cLn}
\cL_n(\tau) &= \biggl[\frac{\theta(\tau)\ln^n(\tau)}{\tau}\biggr]_+
\,,\nn \\
\int_{-\infty}^{\tau^\cut}\!\!\!\df\tau\, \cL_n(\tau)\,f(\tau)
&= \int_0^{\tau^\cut}\!\!\!\df\tau\, \frac{\ln^n(\tau)}{\tau}[f(\tau)-f(0)] + f(0)\,\frac{\ln^{n+1}(\tau^\cut)}{n+1}
\,.\end{align}
This logarithmic structure of the singular contributions directly follows from the IR singular structure of QCD amplitudes, the KLN theorem, and the fact that $\TauN$ is an IR-safe physical observable.  Since the infrared limit of the QCD amplitudes, and hence the IR singularities, depends only on the lower-order phase space, the singular coefficients $\C_n$ only depend on the underlying $\Phi_N$.  That is,
\begin{equation}
\C_n(X) = \int\!\df \Phi_N\, \C_n(\Phi_N)\, \meas(\Phi_N)
\,, \qquad
\frac{\df\sigma^\sing(X)}{\df\tau}
= \int\!\df \Phi_N\, \frac{\df\sigma^\sing(\Phi_N)}{\df\tau}\, \meas(\Phi_N)
\,.\end{equation}
We can therefore consider the singular distributions directly as a function of the full $\Phi_N$ and independently of the specific measurement $X$,
\begin{align} \label{eq:dsigmasingPhiN}
\frac{\df\sigma^\sing(\Phi_N)}{\df\tau}
&= \C_{-1}(\Phi_N)\,\delta(\tau) + \sum_{n\geq 0} \C_n(\Phi_N)\, \cL_n(\tau)
\nn \\
&= \sum_{m \geq 0} \biggl[\C_{-1}^{(m)}(\Phi_N)\,\delta(\tau) + \sum_{n = 0}^{2m-1} \C_n^{(m)}(\Phi_N) \cL_n(\tau) \biggr] \Bigl(\frac{\alpha_s}{4\pi} \Bigr)^m
\,.\end{align}
In the second line, we have expanded the singular coefficients in $\alpha_s$. At LO, the only nonzero coefficient is
\begin{equation}
\C_{-1}^\zero(\Phi_N) = B_N(\Phi_N)
\,,\end{equation}
so at LO the singular spectrum reproduces the LO cross section, consistent with \eq{LO},
\begin{equation} \label{eq:singLO}
\frac{\df\sigma^\sing_\LO}{\df\TauN} = \C_{-1}^\zero(X)\,\delta(\TauN) = \sigma^\LO(X)\,\delta(\TauN)
\,.\end{equation}
At NLO, the coefficients $\C_{-1,0,1}(\Phi_N)$ are nonzero, while at NNLO, the coefficients\linebreak[4]
$\C_{-1,0,1,2,3}(\Phi_N)$ contribute.

Writing the singular spectrum in terms of plus distributions as in \eqs{dsigmasing}{dsigmasingPhiN} precisely encodes the cancellation between real and virtual IR divergences. The $\C_{-1}$ coefficient contains the finite remnant of the virtual contributions after the real-virtual cancellation has taken place. By itself, it is not unique, but depends on the boundary conditions adopted in the definition of the plus distributions, which is encoded in the choice of $\tau$ (the choice of $Q$). Changing the boundary conditions is equivalent to rescaling the arguments of the plus distributions according to (see e.g. \mycite{Ligeti:2008ac})
\begin{equation} \label{eq:cLn_rescale}
\lambda\, \cL_n(\lambda \tau)
= \sum_{k = 0}^n \binom{n}{k} \ln^k\!\lambda\, \cL_{n-k}(\tau)
  + \frac{\ln^{n+1}\!\lambda}{n+1}\, \delta(\tau)
\,.\end{equation}
While this rescaling moves contributions between different $\C_n$, it does not change the overall $1/\TauN$ scaling, which implies that the sum of all terms in \eq{dsigmasing} is unique%
\footnote{It is unique in the sense that it has the minimal $\TauN$ dependence, only containing $\ln^n(\TauN)/\TauN$. One could in principle include some subleading $\TauN$ dependence in the coefficients, if this turns out to be useful or convenient. This would move some contributions between the singular contributions and the nonsingular remainder in \eq{nonsingular}.}
and in fact independent of the choice of $Q$.%
\footnote{The actual physical scales appearing together with $\TauN$ in the logarithms are set by the hard Born kinematics. The reason to think of $Q$ as a typical hard scale is that this provides the natural power suppression of the nonsingular terms.}
Once the singular spectrum is written in terms of distributions as in \eq{dsigmasing}, one can easily integrate it up to $\TauN \leq \TauNcut$ to obtain the singular cumulative distribution (or cumulant in short)
\begin{align} \label{eq:sigmasing}
\sigma^\sing(X, \TauNcut)
&\equiv \int_0^{\TauNcut}\!\!\df\TauN\,\frac{\df\sigma^\sing(X)}{\df\TauN}
= \C_{-1}(X) + \sum_{n\geq 0} \C_n(X)\, \frac{\ln^{n+1}(\tau^\cut)}{n+1}
\,.\end{align}

The ``nonsingular'' contributions are defined as the difference between total and singular contributions,
\begin{align} \label{eq:nonsingular}
\frac{\df\sigma^\nons(X)}{\df\TauN}
&= \frac{\df\sigma(X)}{\df\TauN} - \frac{\df\sigma^\sing(X)}{\df\TauN}
\,, \nn \\
\sigma^\nons(X, \TauNcut)
&= \int_0^{\TauNcut}\!\!\df\TauN\,\frac{\df\sigma^\nons(X)}{\df\TauN}
= \sigma(X, \TauNcut) - \sigma^\sing(X, \TauNcut)
\,.\end{align}
They start at $\ord{\alpha_s}$ relative to $\sigma^\LO(X)$ (which is part of $\df\sigma^\sing$).
By definition of the singular terms, the nonsingular spectrum contains at most integrable singularities for $\TauN\to 0$, the largest terms being $\df \sigma^\nons(X)/\df \TauN \sim \alpha_s^n\ln^{2n}(\tau)$. Equivalently, the nonsingular cumulant behaves for $\TauNcut\to 0$ as
\begin{equation}
\sigma^\nons(X, \TauNcut \to 0) \sim \tau^\cut\, \alpha_s^n \ln^{2n}(\tau^\cut) \to 0
\,.\end{equation}
Hence, also the underlying matrix-element contributions yielding the nonsingular terms can be safely integrated in the infrared.

\subsection{$\TauN$-subtractions}
 \label{subsec:TauNsubtract}

Up to this point, the decomposition of a cross section into singular and nonsingular terms is just notation and holds for any $\TauN$.  The key point of the $\TauN$-subtraction method is that if we have analytic control of the singular $\TauN$ dependence, we can turn the singular spectrum $\df\sigma^\sing(X)/\df\TauN$ and its integral $\sigma^\sing(X,\TauNcut)$ into subtractions, as discussed next.  This requires that for some $N$-jet resolution variable $\TauN$, the underlying coefficients $\C_n(\Phi_N)$ in \eq{dsigmasingPhiN} can be determined explicitly.%
\footnote{They do not necessarily have to be known fully analytically, and in general they will not be. All we really need is a sufficiently fast way to compute their numerical values for given $\Phi_N$ to in principle any desired accuracy.}
In particular, the ability to explicitly compute $\C_{-1}(\Phi_N)$ is precisely equivalent to being able to compute the integrated subtractions in a classical subtraction method. All these conditions are satisfied for $N$-jettiness, as we will discuss in \sec{TauN}.

\subsubsection{$\Tau_N$-slicing}

If the singular contributions for a given $\TauN$ are known, we can use $\TauNcut$ to divide the phase space into two regions: $\TauN < \TauNcut$ and $\TauN \geq \TauNcut$.  Taking $\TauNcut \to \TauIR = \delIR Q$, where $\delIR = \TauIR/Q$ is an (in-principle) arbitrarily small IR cutoff, the singular terms will numerically dominate the nonsingular for $\TauN < \TauNcut$.  In fact, since the nonsingular cumulant $\sigma^\nons(X, \TauIR)$ is of $\ord{\TauIR/Q} = \ord{\delIR}$, we can neglect it in this limit. Hence, we get
\begin{align} \label{eq:slicing}
\sigma(X)
&= \int_0^\TauIR \!\df\TauN\, \frac{\df\sigma(X)}{\df\TauN}
+ \int_\TauIR\!\df\TauN\, \frac{\df\sigma(X)}{\df\TauN}
\nn \\
&= \sigma^\sing(X, \TauIR) + \int_\TauIR\!\df\TauN\, \frac{\df\sigma(X)}{\df\TauN} + \ord{\delIR}
\,.\end{align}
This is precisely a phase-space slicing method, which we will call $\TauN$-slicing. Calculating $\sigma(X)$ to N$^n$LO in this way requires determining $\sigma^\sing(X, \TauIR)$ to N$^n$LO, which includes the N$^n$LO virtual contributions. Beyond that, since the $\TauN$ spectrum only starts at $\ord{\alpha_s}$ relative to $\sigma(X)$, the problem is reduced to the N$^{n-1}$LO calculation for the cross section $\df\sigma(X)/\df\TauN$ for $\TauN>\TauIR$.  Furthermore, if an N$^{n-1}$LO calculation is available, the slicing only needs to be performed for the pure N$^n$LO terms.

\subsubsection{Differential $\Tau_N$-subtractions}

It is instructive to rewrite the $\TauN$-slicing in \eq{slicing} in the form of a subtraction as follows,
\begin{align} \label{eq:slicingsubtraction}
\sigma(X)
&= \sigma^\sing(X, \Tauoff)
+ \biggl[\int_\TauIR\!\df\TauN\, \frac{\df\sigma(X)}{\df\TauN}\biggr]
- \biggl[\int_\TauIR^\Tauoff \!\df\TauN\, \frac{\df\sigma^\sing(X)}{\df\TauN}\biggr]
+ \ord{\delIR}
\,.\end{align}
This reorganization shows that the integral of the singular spectrum acts as a global subtraction for the integrated full spectrum, while the cumulant $\sigma^\sing(X,\Tauoff)$ is the corresponding contribution of the virtual terms (sitting at $\TauN = 0$) plus the integrated subtraction. The value of $\Tauoff$ is arbitrary and exactly cancels between the first and third terms. It determines the upper limit in $\TauN$ up to which the subtractions are used. The subtraction term in this case is maximally nonlocal, as it is applied after all phase-space integrations. Hence, one would naively expect the numerical cancellations to be maximally bad. This also shows that $\TauIR$ really is an IR cutoff below which only the singular (subtraction) terms are used, due to limited numerical precision.

Looking at \eq{slicingsubtraction}, we can also move the singular spectrum underneath the $\TauN$ integration,
\begin{align} \label{eq:TauNsubtraction}
\sigma(X)
&= \sigma^\sing(X, \Tauoff)
+ \int_\TauIR\!\df\TauN\, \biggl[\frac{\df\sigma(X)}{\df\TauN} - \frac{\df\sigma^\sing(X)}{\df\TauN}\, \theta(\TauN<\Tauoff) \biggr]
+ \ord{\delIR}
\nn \\ 
&=  \sigma^\sing(X, \Tauoff)
+ \int_\TauIR^\Tauoff\!\df\TauN\, \frac{\df\sigma^\nons(X)}{\df\TauN} + \int_\Tauoff \frac{\df\sigma(X)}{\df\TauN}
+ \ord{\delIR}
\,.\end{align}
which turns the singular spectrum into an actual subtraction which is local (point-by-point) in $\TauN$.  It is of course still nonlocal in the remaining real radiation phase space. To use \eq{TauNsubtraction}, one now has to explicitly calculate the singular differential spectrum. This requires essentially no additional effort, since the required singular coefficients are the same as in $\sigma^\sing(X,\TauNcut)$.

Writing it as in the second line of \eq{TauNsubtraction} shows explicitly that the numerical integral over $\TauN$ now only encounters an integrable singularity for $\TauN\to 0$ since the integrand is precisely the nonsingular contribution. This turns $\TauIR$ into a purely technical cutoff for the numerical integration, which is only necessary because the integrand is still given by the difference of two diverging integrands.  Finally, we note that the neglected contributions due to the numerical IR cutoff $\TauIR$ are precisely the same as in \eq{slicing} for the same value of $\TauIR$. The numerical error introduced by such a cutoff is discussed in the next section.

We stress that a technical IR cutoff analogous to $\delIR$ exists in any numerical fixed-order calculation using subtractions, since the QCD amplitudes (and their subtractions) become arbitrarily large in the IR. Below the cutoff, the full QCD amplitudes are always approximated by the subtraction terms, so that below the cutoff only the integral of the subtraction is used, while the nonsingular cross section below the cutoff is power suppressed by $\delIR$ and neglected.

Finally, note that separating the spectrum or cumulant into its singular and nonsingular parts, as we have done here, is in fact very well known and routinely used when performing the higher-order resummation for an IR-sensitive observable $\TauN$. In this context, the singular contributions are resummed to all orders in $\alpha_s$ and a given logarithmic order, while \eq{nonsingular} is used to determine the nonsingular contributions. At NNLO, this utilizes the result for $\df\sigma(X)/\df\TauN$ obtained from the NLO $N+1$-jet calculation and the NNLO singular contributions obtained from the NNLL$'$ resummation of $\TauN$. In \sec{TauN} we will employ the same techniques to compute directly the fixed-order singular contributions without resummation. This also makes it clear that if desired any NNLO calculation performed in this way can be straightforwardly improved with the corresponding higher-order resummation in $\TauN$.

\subsubsection{Estimating numerical accuracy}
\label{subsubsec:accuracy}

We can judge the numerical accuracy of the $\TauN$-slicing and differential $\TauN$-subtractions using some simple scaling arguments.  First, it is important to quantify the effect of the IR cutoff $\delIR$.  Using $N$-jettiness as an example, at N$^n$LO relative to the Born cross section, the most dominant singular terms in the spectrum and the cumulant are, for a given partonic channel,
\begin{align}
\frac{\df\sigma}{\df\tau}
&= \sigma^\LO \sum_{n\geq 1} \frac{2n}{n!} \Bigl( \frac{\as}{4\pi} \Bigr)^n \Bigl( -\sum_i C_i \Gamma_0 \Bigr)^n \cL_{2n-1} (\tau) + \dotsb
\,, \nn \\
\sigma(\TauN^\cut)
&= \sigma^\LO \sum_{n \geq 1} \frac{1}{n!} \Bigl( \frac{\as}{4\pi} \Bigr)^n \Bigl( - \sum_i C_i \Gamma_0 \Bigr)^n \ln^{2n} (\tau^\cut) + \dotsb
\,.\end{align}
Here, $\Gamma_0 = 4$ is the one-loop coefficient of the cusp anomalous dimension, $C_i = C_F$ for quarks and $C_i = C_A$ for gluons, and the ellipsis denote terms with fewer powers of logarithms at each order in $\alpha_s$.%
\footnote{In principle, subleading logarithmic terms can also be numerically important due to large numerical prefactors, especially for moderate $\TauIR$ values.  However, for small enough $\TauIR$ values, the leading logarithmic terms are a sufficient estimate.}
Correspondingly, the leading nonsingular term in the cumulant has the form
\begin{align} \label{eq:sigmanonsIR}
\sigma^\nons(\TauN^\cut)
&= \sigma^\LO \sum_{n \geq 1} \frac{1}{n!} \Bigl( \frac{\as}{4\pi} \Bigr)^n C_\nons^{(n)} \Bigl( - \sum_i C_i \Gamma_0 \Bigr)^n \tau^\cut \ln^{2n-1} (\tau^\cut) + \dotsb
\,.\end{align}
The coefficient $C_\nons^{(n)}$ is not known in general, but we take $C_\nons^{(n)} = 1$ here, which is the correct value for 2-jettiness in $e^+e^-$ (i.e. thrust).

We denote the missing nonsingular contribution due to approximating the full result by the singular contributions below $\TauN < \TauIR$ by $\Delta \sigma_\IR(\delIR)$ and expand it in $\alpha_s$ as
\begin{equation}
\sigma^\nons(\TauIR) \equiv \Delta \sigma_\IR(\delIR)
= \Delta \sigma_\IR^\one(\delIR)\, \frac{\alpha_s}{4\pi} +
\Delta \sigma_\IR^\two(\delIR) \Bigl(\frac{\alpha_s}{4\pi}\Bigr)^2 + \dotsb
\,.\end{equation}
The size of the dominant nonsingular terms in \eq{sigmanonsIR} at $\tau = \delIR$ is indicative of the size of $\Delta\sigma_\IR$.
For the production of a color singlet $X$ in the $pp \to X$ and $pp \to X+\text{jet}$ channels, the missing terms at NLO and NNLO scale as (plugging in the relevant color factors):
\begin{align} \label{eq:deltasigmaIRchannels}
q\bar{q}\to X:&&\!\!\!
\bigl\{ \Delta \sigma_\IR^\one(\delIR) \,,\, \Delta \sigma_\IR^\two(\delIR) \bigr\}
&\approx \sigma^\LO \bigl\{ -10.7\, \delIR \ln \delIR \,,\, 113.8\, \delIR \ln^3\! \delIR \bigr\}
, \nn \\
gg \to X:&&\!\!\!
\bigl\{ \Delta \sigma_\IR^\one(\delIR) \,,\, \Delta \sigma_\IR^\two(\delIR) \bigr\}
&\approx \sigma^\LO \bigl\{ -24\, \delIR \ln \delIR \,,\, 576\, \delIR \ln^3\! \delIR \bigr\}
, \nn \\
gq\to Xq,\, q\bar{q}\to Xg:&&\!\!\!
\bigl\{ \Delta \sigma_\IR^\one(\delIR) \,,\, \Delta \sigma_\IR^\two(\delIR) \bigr\}
&\approx \sigma^\LO \bigl\{ -22.7\, \delIR \ln \delIR \,,\, 513.8\, \delIR \ln^3\! \delIR \bigr\}
, \nn \\
gg\to Xg:&&\!\!\!
\bigl\{ \Delta \sigma_\IR^\one(\delIR) \,,\, \Delta \sigma_\IR^\two(\delIR) \bigr\}
&\approx \sigma^\LO \bigl\{ -36\, \delIR \ln \delIR \,,\, 1296\, \delIR \ln^3\! \delIR \bigr\}
.\end{align}

\begin{figure}
\centering
 \includegraphics[scale=0.6]{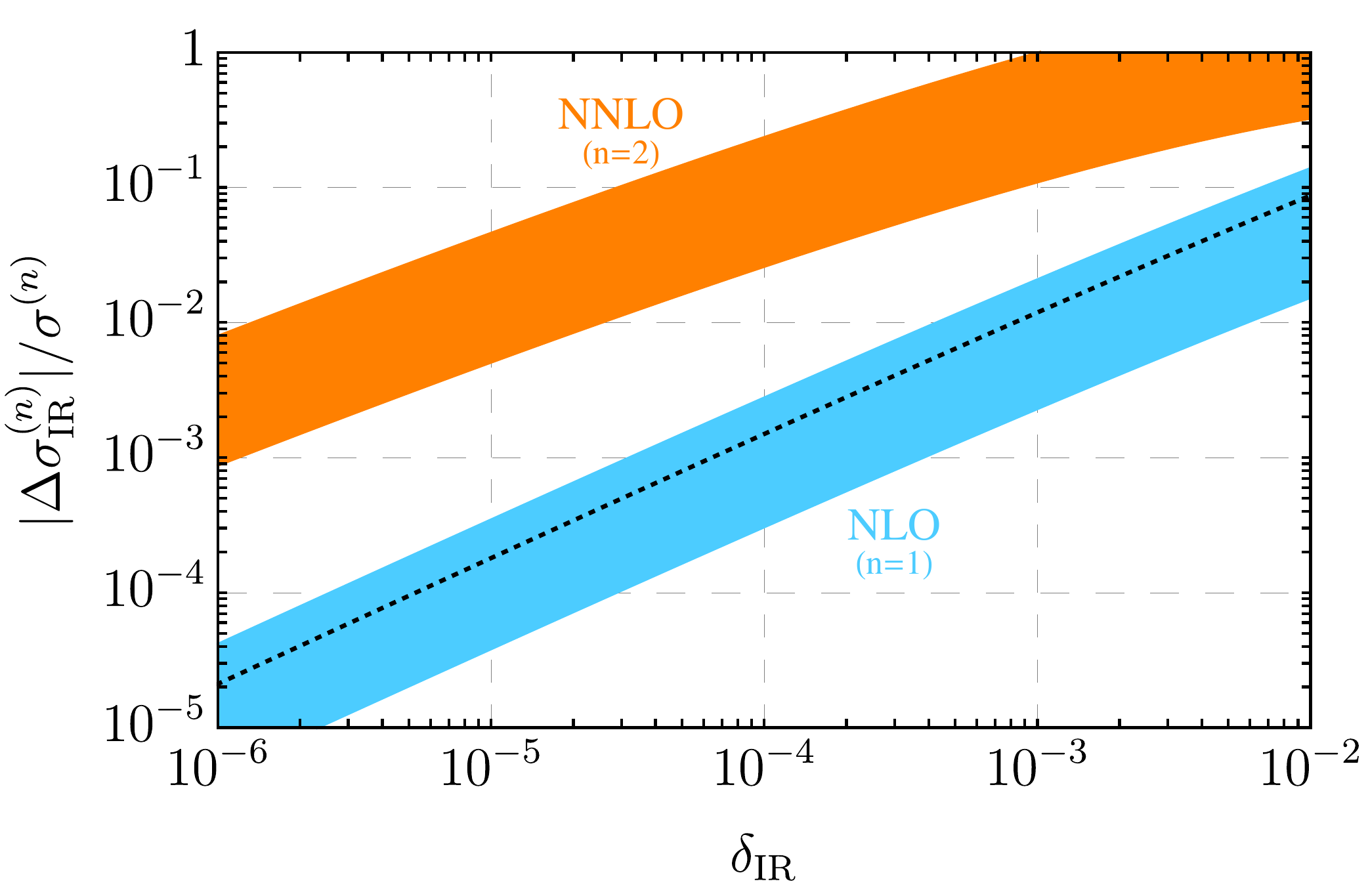}
\caption{Estimated size of the missing nonsingular terms below $\tau = \delIR$ as a fraction of the full correction at NLO (blue band) and NNLO (orange band), see \eq{deltasigmaIR}. The dashed line shows the known exact result for thrust.}
\label{fig:deltasigmaIR}
\end{figure}

To estimate the impact of these terms relative to the full NLO and NNLO contributions, we write the full result for the cross section as
\begin{equation}
\sigma = \sigma^\LO + \sigma^\one\, \frac{\as}{4\pi} + \sigma^\two \Bigl( \frac{\as}{4\pi} \Bigr)^2 + \dotsb
\,.\end{equation}
We assume that the $K$-factors at each order of perturbation theory for $q\bar{q} \to X$ and $q\bar{q} \to X g ,\, qg \to Xq$ processes are 10\%, so $\sigma^{(n)}/\sigma^\LO \approx 10^n$.  For $gg \to X$ and $gg \to Xg$ processes, we assume the $K$-factors are 30\%, so that $\sigma^{(n)}/\sigma^\LO \approx 30^n$ for these cases.  These factors roughly scale like the prefactors in \eq{deltasigmaIRchannels}. Hence, a rough estimate of the relative size of the missing terms at each order is given by
\begin{equation} \label{eq:deltasigmaIR}
\frac{\Delta \sigma_\IR^\one(\delIR)}{\sigma^\one} \approx a \, \delIR \ln \delIR
\,, \qquad
\frac{\Delta \sigma_\IR^\two(\delIR)}{\sigma^\two} \approx a \, \delIR \ln^3 \delIR
\,.\end{equation}
The dependence of these corrections on $\delIR$ is plotted in \fig{deltasigmaIR}, where we take $a$ between $1/3$ and $3$. The dashed line shows the known exact NLO result for thrust.
This implies that when working to NNLO, we need $\delIR \lesssim 10^{-3}-10^{-4}$ to have a reasonable $\lesssim\ord{10\%}$ determination of the $\alpha_s^2$ NNLO contribution to the cross section.  For typical applications with $Q\sim \ord{100\GeV}$ this implies that $\TauIR \lesssim 0.1-0.01 \GeV$.  To the extent that the NNLO terms are only a small part of the total cross section (as is the case for Drell-Yan, for example), a larger error on the NNLO terms might be tolerable. However, we stress that these estimates can only serve as an indication, and in practice one should carefully test the size of missing corrections, for example by studying the $\delIR$ dependence as discussed in \subsec{NNLOresults}.

An important comment concerns the fact that it is in principle possible and straightforward (though perhaps tedious in practice) to derive subleading factorization theorems for $N$-jettiness and other observables using SCET. These can then be used to systematically determine the next-to-singular $\ord{\tau}$ corrections and include them in the same way in the subtractions. This would substantially reduce the size of the missing nonsingular corrections by one power of $\delIR$.  A complete factorization theorem at subleading order for a single-jet process has been derived for semileptonic heavy quark decays in \mycite{Lee:2004ja}. For recent work in this direction for thrust in $e^+e^-$ see e.g.~\mycites{Freedman:2013vya, Freedman:2014uta, Kolodrubetztalk}.

A second important aspect concerns the required numerical precision in a practical implementation. For both $\TauN$-slicing and differential $\TauN$-subtractions, the full QCD and singular cross sections are probed in regions of phase space with $\TauN \gtrsim \TauIR$, where there are significant numerical enhancements due to the nearby IR singularity at $\TauN = 0$.  For $\delIR \sim 10^{-4}$, the cancellations between the full QCD and singular $\TauN$ distributions can easily reach the $\ord{10^4}$ level and only increase as $\delIR$ is lowered further. Getting a result at $\ord{10^{-k}}$ relative numerical precision in this case demands at least an $\ord{10^{-(k+4)}}$ relative numerical precision in the evaluation of the squared QCD amplitudes.

For $\TauN$-slicing, the numerical cancellations only happen after the $\TauN$ integration, which means that in the worst case the $\TauN$ integral itself may have to be carried out to the same high precision. In practice, this will strongly depend on the process and the chosen $\TauIR$, since the numerical cancellations actually happen between the two terms in \eq{slicing} rather than the the last two terms in \eq{slicingsubtraction}. In any case, using Monte-Carlo integration to determine the integral of the unsubtracted full result down to $\TauN \geq \TauIR$ very accurately requires very high statistics and good phase-space sampling. Since NLO codes are usually not designed for this purpose, this strongly limits how low $\TauIR$ can be taken.

For the $\TauN$-subtractions, the QCD amplitudes in the integrand still require the same high numerical precision at small $\TauN$ to obtain an accurate result for the nonsingular spectrum. However, since the cancellations now happen already at the integrand level, the $\TauN$ integration itself has to be carried out only to the nominal $\ord{10^{-k}}$ relative precision. Hence, the statistical requirements on the Monte-Carlo integration of the nonsingular spectrum in \eq{TauNsubtraction} are much more modest compared to the $\TauN$-slicing. This also means that $\TauIR$ can now be taken as low as the numerical precision in the integrand allows. The main nontrivial requirement now is that one must be able to sample phase-space for fixed $\TauN$, which we discuss further in \sec{implementation}.

\section{$N$-jettiness Subtractions}
\label{sec:TauN}

In this section, we now specify $\TauN$ to be $N$-jettiness and explicitly construct the $N$-jettiness subtractions.
We first discuss the Born kinematics and the definition of $N$-jettiness in \subsec{TauNdef}. In \subsec{TauNfact} we review the factorization theorem for the singular contributions in $\TauN$ and how the virtual QCD amplitudes enter into it. Then in \subsec{singlediff} we explicitly write out the $\TauN$ subtractions at NLO and NNLO. Finally, in \subsec{morediff} we discuss how the subtractions can be made more differential and thereby more local.

\subsection{Definition of $N$-jettiness}
\label{subsec:TauNdef}

\subsubsection{Born kinematics}

We always use the indices $a$ and $b$ to label the initial states, and $1,\ldots,N$ to label the final states. Unless otherwise specified, a generic index $i$ always runs over $a, b, 1, \ldots, N$. We denote the momenta of the QCD partons in the $\Phi_N$ Born phase space by $\{q_a, q_b; q_1, \ldots, q_N\}$ and the parton types (including their spin/helicity if needed) by $\{\kappa_a, \kappa_b; \kappa_1, \ldots, \kappa_N\}$. Thus, $\Phi_N$ corresponds to
\begin{align}
\Phi_N &\equiv \{(q_a, \kappa_a), (q_b, \kappa_b); (q_1, \kappa_1), \ldots, (q_N, \kappa_N); \Phi_L(q) \}
\,,\end{align}
where $\Phi_L(q)$ denotes the phase space for any additional nonhadronic particles in the final state, whose total
momentum is $q$. (For $ep$ or $ee$ collisions, one or both of the incoming momenta are considered part of $\Phi_L(q)$.)
We will mostly suppress the nonhadronic final state. For us, it is only relevant because it contributes to momentum conservation in $\Phi_N$, which reads
\begin{equation}
q_a^\mu + q_b^\mu = q_1^\mu + \dotsb + q_N^\mu + q^\mu
\,.\end{equation}
When there is no ambiguity, we will associate $\kappa_i \equiv i$ (e.g., we use $f_a \equiv f_{\kappa_a}$), and we use the collective label $\kappa$ to denote the whole partonic channel, i.e.,
\begin{equation}
\kappa \equiv \{\kappa_a, \kappa_b; \kappa_1, \ldots, \kappa_N\} \equiv \{a, b; 1, \ldots, N\}
\,.\end{equation}

We write the massless Born momenta $q_i$ as
\begin{equation}
q_i^\mu = E_i\, n_i^\mu
\,, \qquad
n_i^\mu = (1, \vec n_i)
\,, \qquad
|\vec n_i|=1
\,.\end{equation}
In particular, for the incoming momenta we have
\begin{equation} \label{eq:ISmom}
E_{a,b} = x_{a,b}\,\frac{\Ecm}{2}
\,, \qquad
n_a^\mu = (1, \hat z)
\,, \qquad
n_b^\mu = (1, - \hat z)
\,,\end{equation}
where $\Ecm$ is the total (hadronic) center-of-mass energy and $\hat z$ points along the beam axis.
The $x_{a,b}$ are the light-cone momentum fractions of the incoming partons, and momentum conservation implies
\begin{equation} \label{eq:xaxb}
x_a\Ecm = n_b \cdot (q_1 + \dotsb + q_N + q)
\,,\qquad
x_b\Ecm = n_a \cdot (q_1 + \dotsb + q_N + q)
\,.\end{equation}
The total invariant mass-squared $Q^2$ and rapidity $Y$ of the Born phase space are
\begin{align}
Q^2 &= x_a x_b \Ecm^2
\,, \qquad
Y = \frac{1}{2}\ln\frac{x_a}{x_b}
\,, \qquad
x_a\Ecm = Q\,e^Y
\,, \qquad
x_b\Ecm = Q\,e^{-Y}
\,.\end{align}

The complete $\df\Phi_N$ phase-space measure corresponds to
\begin{equation}
\int\!\df\Phi_N \equiv \frac{1}{2\Ecm^2} \int\!\frac{\df x_a}{x_a}\, \frac{\df x_b}{x_b}\, \int\! \df\Phi_N(q_a + q_b; q_1, \ldots, q_N, q)\,\frac{\df q^2}{2\pi}\, \df\Phi_L(q) \sum_\kappa s_\kappa
\,,\end{equation}
where $\df\Phi_N(...)$ on the right-hand side denotes the standard Lorentz-invariant $N$-particle phase space, the sum over $\kappa$ runs over all partonic channels, and $s_\kappa$ is the appropriate factor to take care of symmetry, flavor and spin averaging for each partonic channel.

\subsubsection{$N$-jettiness}

Given an $M$-particle phase space point with $M \geq N$, $N$-jettiness is defined as~\cite{Stewart:2010tn}
\begin{equation}\label{eq:TauN_def}
\Tau_N(\Phi_M) = \sum_{k=1}^M \min_i \Bigl\{ \frac{2 q_i\cdot p_k}{Q_i} \Bigr\}
\,,\end{equation}
where $i$ runs over $a, b, 1, \ldots, N$.
(Here we use a dimension-one definition of $\Tau_N$ following \mycites{Jouttenus:2011wh, Jouttenus:2013hs}.)
For $ep$ or $ee$ collisions, one or both of the incoming directions are absent.
The $Q_i$ are normalization factors, which are explained below.
The $p_k$ are the $M$ final-state parton momenta (so excluding the nonhadronic final state) of $\Phi_M$. The $q_i$ in \eq{TauN_def} are massless Born ``reference momenta'', and the corresponding directions $\vec n_i = \vec q_i/\abs{\vec q_i}$ are referred to as the $N$-jettiness axes. For later convenience we also define the normalized vectors
\begin{equation} \label{eq:qhat}
\hq_i = \frac{q_i}{Q_i}
\,.\end{equation}

The $q_i$ are obtained by projecting a given $\Phi_M$ onto a corresponding Born point $\hat\Phi_N(\Phi_M)$. For this purpose, any IR safe phase-space projection can be used. That is, in any IR singular limit where $\Phi_M\to \Phi_N$, the Born projection has to satisfy
\begin{equation} \label{eq:Bornprojection}
\hat\Phi_N(\Phi_M \to \Phi_N) \to \Phi_N
\,,\end{equation}
including the proper flavor assignments. In particular, for $M = N$, we simply have $\hat\Phi_N(\Phi_N) = \Phi_N$ and so
$q_i = p_i$, which implies $\TauN(\Phi_N) = 0$. For $M \geq N+1$, there is always at least one $p_k$ that cannot be exactly aligned with any of the $q_i$, which means that $\TauN(\Phi_M) > 0$. The minimization condition in \eq{TauN_def} ensures that for each $p_k$ the smallest distance to one of the $q_i$ enters the sum, which together with \eq{Bornprojection} implies that $\TauN(\Phi_M\to\Phi_N) \to 0$. Hence, $N$-jettiness satisfies all the criteria of an IR-safe $N$-jet resolution variable given in \eq{TauNIRcondition}.

Some examples of suitable Born projections are discussed in \subsubsec{Bornprojections} below.
Although the precise procedure to define the Born projection and the $q_i$ is part of the definition of $N$-jettiness, it is important that it does not actually affect the singular structure of the $\TauN$-differential cross section. Different choices only differ by power-suppressed effects, as explained in \mycite{Stewart:2010tn}, which means the precise choice only affects the nonsingular contributions. Hence, constructing the singular contributions and the subtraction terms does not actually require one to specify the Born projection, as they are constructed in the singular limit starting from a given $\Phi_N$.%
\footnote{In this regard, the $\TauN$-subtractions are FKS-like, namely they are intrinsically a function of the Born phase space $\Phi_N$ and an emission variable, which for us is $\TauN$, as opposed to starting from a given $\Phi_{\geq N+1}$ point.}
This fact provides considerable freedom in the practical implementation, which we will come back to in \sec{implementation}.

The singular structure of $\TauN$ is determined by the minimization condition in \eq{TauN_def} and the choice of the $Q_i$. The minimization effectively divides the $\Phi_M$ phase space into $N$ jet regions and up to $2$ beam regions, where each parton in $\Phi_M$ is associated (``clustered'') with the $q_i$ it is closest to, where the $Q_i$ determine the relative distance measure between the different $q_i$. We can then rewrite \eq{TauN_def} as follows,
\begin{align} \label{eq:TauNi_def}
\Tau_N &= \sum_i \Tau_N^i
\qquad\text{with}\qquad
\Tau_N^i = \sum_{k=1}^M \biggl[ \frac{2 q_i \cdot  p_k}{Q_i}
\prod_{j\neq i} \theta\Bigl(\frac{q_j \cdot p_k}{Q_j} - \frac{q_i \cdot p_k}{Q_i} \Bigr) \biggr]
\,,\end{align}
where the $\Tau_N^i$ are the contributions to $\Tau_N$ from the $i$th region.

The $Q_i$ can be chosen depending on the Born kinematics in $\Phi_N$ (subject to the constraint that the resulting distance measure remains IR safe). A variety of possible choices are discussed in detail in \mycites{Jouttenus:2011wh, Jouttenus:2013hs}. An ``invariant-mass'' measure is obtained by choosing common $Q_i = Q$. In this case, the sum of the invariant masses of all emissions in each region will be minimized. A class of ``geometric measures'' is obtained by choosing $Q_i$ proportional to $E_i$, which makes the value of $\TauN$ itself independent of the $E_i$, i.e.,
\begin{equation} \label{eq:geometric}
Q_i = 2\rho_i E_i
\qquad\Rightarrow\qquad
\hq_i = \rho_i\, \frac{n_i}{2}
\,,\qquad
\frac{2 q_i\cdot p_k}{Q_i} = \rho_i\, n_i\cdot p_k
\,,\end{equation}
where the $\rho_i$ are dimensionless numbers which determine the relative size of the different regions. In this case, the sum of the small light-cone momenta, $n_i\cdot p_k$, of all emissions relative to their associated $N$-jettiness axis are minimized. 

The singular structure of the cross section does explicitly depend on the distance measure. When discussing the singular contributions in the next section,  we will keep the $Q_i$ arbitrary, thus enabling various choices to be explored using our results. [As discussed in \mycite{Stewart:2010tn}, one can generalize $N$-jettiness further to use any IR-safe distance measure $d_i(p_k)$ in \eq{TauN_def}, which has been used for example in the application to jet substructure~\cite{Thaler:2010tr, Thaler:2011gf}.
For our purposes, the canonical form $d_i(p_k) = \hq_i \cdot p_k$ is suited well, because the simple linear dependence on $p_k$ simplifies the theoretical analysis and computations.]

\subsubsection{Example Born projections}
\label{subsubsec:Bornprojections}

To construct a generic Born projection, it suffices to use any IR-safe jet algorithm to cluster the $M$-parton final state into $N$ jets with momenta $P_i$. One can then define massless final-state $q_i^\mu = E_i n_i^\mu$ by taking ($i = 1, \ldots, N$)
\begin{equation}
\vec n_i = \frac{\vec{P_i}}{\abs{\vec P_i}}
\qquad\text{with}\qquad
E_i = P_i^0
\quad\text{or}\quad
E_i = \abs{\vec P_i}
\quad\text{or}\quad
2E_i = P_i^0 + \abs{\vec P_i}
\,,\end{equation}
where any of the choices for $E_i$ can be used.
To ensure that the total transverse momentum in the Born final state adds up to zero, one can then for example boost the hadronic system or recoil the leptonic final state in the transverse direction. Finally, the initial-state momenta $q_a$ and $q_b$, which always lie along the beam directions as in \eq{ISmom}, are determined by momentum conservation from \eq{xaxb}.

When using a geometric measure as in \eq{geometric}, the canonical way to determine the $N$-jettiness axes $\vec n_i$ is by an overall minimization of the total value of $\TauN$. Up to NNLO the relevant cases are $M=N+1$ and $M=N+2$, i.e., one and two extra emissions, in which case the overall minimization to find the $N$-jettiness axes is still fairly easy to work out explicitly.

Let us take $\rho_i = 1$ for simplicity and consider the case of hadron-hadron collisions, such that we have $N$ jet axes plus the two fixed beam axes $\vec n_{a,b} = \pm \hat z$. When $M=N+1$, it is easy to see that $N-1$ axes must be aligned with $N-1$ of the $p_k$ momenta. For the last axis, there are two possibilities, and the one which gives a smaller $\TauN$ is selected: Either it is aligned with one of the two remaining $p_k$ (this occurs if the last $p_k$ momentum lies close enough to one of the beam directions), or it lies along the direction of the sum of the two remaining $p_k$. The appropriate expression for $\Tau_N$ for $M=N+1$ is then:
\begin{equation}
 \Tau_N = \sum_{k=1}^{M} ( E_k - \abs{\vec{p}_k} )
 + \min\Bigl\{
  \min_{j \in 1..M} \bigl\{ \abs{\vec{p}_j} - \abs{p^z_{j}} \bigr\},
  \min_{jk\in 1..M} \bigl\{ \abs{\vec{p}_j} + \abs{\vec{p}_k} - \abs{\vec{p}_j+\vec{p}_k} \bigr\}
  \Bigr\} \,.
\end{equation}
The first term in the overall minimization corresponds to the first case above (extra emission clustered to the beam), whilst the second term corresponds to the second case (extra emission clustered to a jet).

When $M=N+2$ there are two extra emissions. Now, $N-2$ axes will always be aligned with $N-2$ of the $p_k$ momenta, and there are four possible cases how the remaining two axes can be chosen based on the remaining four $p_k$. The appropriate expression for $\Tau_N$ for $M=N+2$ is
\begin{align}
 \Tau_N = \sum_{j=1}^{M} ( E_j - \abs{\vec{p}_j} ) + \min  \Bigl\{ &
 \min_{jk\in 1..M} \bigl\{ \abs{\vec{p}_j} + \abs{\vec{p}_k} - \abs{p^z_{j}} - \abs{p^z_{k}} \bigr\},
 \\ \nn
 & \min_{jkl\in 1..M} \bigl\{ \abs{\vec{p}_j} + \abs{\vec{p}_k} + \abs{\vec{p}_l} - \abs{\vec{p}_j+\vec{p}_k} - \abs{p^z_{l}} \bigr\},
 \\ \nn
 & \min_{jkl\in 1..M} \bigl\{ \abs{\vec{p}_j} + \abs{\vec{p}_k} + \abs{\vec{p}_l} - \abs{\vec{p}_j+\vec{p}_k+\vec{p}_l} \bigr\},
 \\ \nn
 &\min_{jklm\in 1..M} \bigl\{ \abs{\vec{p}_j} + \abs{\vec{p}_k} + \abs{\vec{p}_l} + \abs{\vec{p}_m} - \abs{\vec{p}_j+\vec{p}_k} - \abs{\vec{p}_l+\vec{p}_m} \bigr\}
 \Bigr\} \,.
\end{align}
The first term in the overall minimization corresponds to both extra particles being clustered to a beam direction. The second term corresponds to one particle being clustered to a beam, and two particles being clustered together in a jet. The third term corresponds to clustering three particles together in a jet, and the final term corresponds to clustering two sets of two particles into two separate jets. In all cases the remaining jet directions are set by the remaining unclustered $p_k$ momenta.

\subsection{Factorization in the singular limit}
\label{subsec:TauNfact}

\subsubsection{Factorization theorem}

We start by writing the $N$-jettiness singular cross section differential in $\Phi_N$ and all individual $\TauN^i$ contributions,
\begin{align} \label{eq:dsigmasingfull}
\frac{\df\sigma^\sing(X)}{\df \TauN}
&= \int\!\df\Phi_N\, \frac{\df\sigma^\sing(\Phi_N)}{\df\TauN}\,X(\Phi_N)
\nn \\
\frac{\df\sigma^\sing(\Phi_N)}{\df\TauN}
&= \int\!\Bigl[\prod_i \df\TauN^i\Bigr] \frac{\df\sigma^\sing(\Phi_N)}{\df \TauN^a\, \df\TauN^b\dotsb\df\TauN^N}\,
\delta\Bigl(\TauN - \sum_i \TauN^i \Bigr)
\,.\end{align}
The factorization of the $N$-jettiness cross section in the singular limit [for the linear measures defined by \eq{TauN_def}] was derived in SCET in \mycites{Stewart:2009yx, Stewart:2010tn, Jouttenus:2011wh}. It takes the form
\begin{align} \label{eq:fact}
\frac{\df\sigma^\sing(\Phi_N)}{\df \Tau_N^a\, \df\Tau_N^b\dotsb\df\Tau_N^N}
&=
\int\!\df t_a\, B_a(t_a, x_a, \mu)
\int\!\df t_b\, B_b(t_b, x_b, \mu)
\biggl[\, \prod_{i=1}^N \int\!\df s_i\, J_i(s_i, \mu) \biggr]
\\\nn &\quad \times
\vC^\dagger(\Phi_N, \mu)\,
\hS_\kappa \biggl(\Tau_N^a - \frac{t_a}{Q_a}, \ldots, \Tau_N^N - \frac{s_N}{Q_N}, \{\hq_i \}, \mu\biggr)
\vC(\Phi_N, \mu)
\,.\end{align}
The first argument(s) of the beam, jet, and soft functions $B_i$, $J_i$, and $\hS_\kappa$ determine the contributions to the $\Tau_N^i$ from the respective collinear and soft sectors.
The beam function $B_a(t_a, x_a, \mu)$ contains all collinear emissions (virtual and real) from the incoming parton $a$, and depends on the parton's flavor $\kappa_a$ and light-cone momentum fraction $x_a$.
The jet function $J_i(s, \mu)$ contains all collinear emissions from the outgoing parton $i$, and depends on the parton's flavor $\kappa_i$.
The soft function $\hS_\kappa$ contains all soft emissions between all partons and depends on the directions $\hq_i$. It is a matrix acting in the color space of the partonic channel $\kappa$. More precisely, it acts in the color-conserving subspace of the full color space.
The hard Wilson coefficient $\vC(\Phi_N, \mu)$ is a vector in the same color space, and $\vC^\dagger(\Phi_N, \mu)$ is its conjugate (see below). It contains the QCD amplitudes for the $N$-parton process and depends on the full $N$-parton phase space $\Phi_N$.

All functions in the factorized cross sections have an explicit $\mu$ dependence (due to their nonzero anomalous dimensions). This $\mu$ dependence exactly cancels between the different functions at each order. The remaining internal $\mu$-dependence is the usual one due to the running of $\alpha_s(\mu)$ which cancels up to the order one is working at. In the general case, the $\mu$ dependence is used to resum the logarithms of $\TauN$ to all orders in $\alpha_s$ at a given order in logarithmic counting.
For our purposes, we require the strict fixed-order expansion in $\alpha_s(\mu)$ at NLO and NNLO.

We note in passing that starting at N$^4$LO, the partonic QCD cross section receives a contribution from noncancelling Glauber modes in graphs with the same structure as figure 5 in \mycite{Gaunt:2014ska}. Such contributions are not reproduced by \eq{fact}. However this is far beyond NNLO, which is the level we are concerned about here.

\subsubsection{QCD amplitudes and color space}

The hard coefficients $\vC(\Phi_N)$ contain the virtual $N$-parton amplitudes from QCD. They formally arise as the matching coefficients from QCD onto SCET. How this matching is performed in practice for generic processes using QCD helicity amplitudes is discussed extensively in \mycites{Stewart:2012yh, helicityops} (see also \mycites{Kelley:2010fn, Becher:2013vva, Moult:2014pja, Broggio:2014hoa}). We refer the reader there for details and only summarize the features relevant for our discussion here. The important point is that when working in pure dimensional regularization with \MSbar, the coefficients $\vC(\Phi_N)$ are given by the infrared-finite part, $\cA_\fin$, of the full $N$-parton QCD amplitude after UV renormalization.%
\footnote{The UV renormalization scheme must be the same for all functions appearing in the factorized cross section. The explicit results we give all use conventional dimensional regularization (CDR), which requires the QCD amplitudes to be renormalized in the CDR or 't Hooft-Veltman (HV) scheme.}
Hence, we have
\begin{equation}  \label{eq:matching_general}
C^{\alpha_a \dotsb \alpha_N}(\Phi_N)
= -\img \cA_{\fin}^{\alpha_a \dotsb \alpha_N}(\Phi_N)
\,,\end{equation}
where we have explicitly written out the color indices $\{\alpha_a, \ldots, \alpha_N\}$ of the external partons. (All remaining dependence on external helicities and momenta are contained in $\Phi_N$.)

The color indices $\{\alpha_i\}$ span the full color space for the partonic channel $\kappa$. We can now pick a complete basis of color structures $\T_k^{\alpha_a \dotsb \alpha_N}$, which span the color-conserving subspace. (For practical purposes, the basis can be overcomplete and does not have to be orthogonal.) For example, for $\kappa=gq\bar q$ the color-conserving subspace is still one-dimensional, since the only allowed color structure is $\T^{a \alpha \bar\beta} \equiv (T^a_{\alpha\bar\beta})$. For $\kappa = ggq\bar q$, one choice would be
\begin{equation} \label{eq:ggqqcol}
\T_k^{ ab \alpha\bar\beta}
= \Bigl(
   (T^a T^b)_{\alpha\bar\beta}\,,\, (T^b T^a)_{\alpha\bar\beta} \,,\, \tr[T^a T^b]\, \delta_{\alpha\bar\beta}
   \Bigr)
.\end{equation}
Given a basis $\T_k^{\alpha_a \dotsb \alpha_N}$, we write the hard coefficients in this basis as
\begin{equation} \label{eq:matching_LO}
C^{\alpha_a \dotsb \alpha_N}(\Phi_N)
= \sum_k \T_k^{\alpha_a \dotsb \alpha_N}\, C_k(\Phi_N)
\equiv \T^{\alpha_a \dotsb \alpha_N} \cdot \vC(\Phi_N)
\,.\end{equation}
This is in one-to-one correspondence to choosing a particular color decomposition for the $N$-parton amplitude, and so the coefficients $\vC$ are directly given by the IR-finite parts of the color-ordered (or color-stripped) amplitudes.
The precise form of the amplitude's color decomposition is irrelevant for our discussion and any convenient color basis can be used.

The conjugate $\vC^\dagger$ of the vector $\vC$ is defined by
\begin{align}
\vC^\dagger = \sum_{\alpha_a \dotsb\alpha_N} C^{*\alpha_a\dotsb\alpha_N} \T^{\alpha_a \dotsb\alpha_N}
= (\vC^*)^T \, \hT_\kappa
\,,\end{align}
where the superscript $T$ denotes the transpose and
\begin{equation}
\hT_\kappa = \sum_{\alpha_a \dotsb\alpha_N} (\T^{\alpha_a \dotsb\alpha_N})^\dagger \T^{\alpha_a \dotsb\alpha_N}
\,,\end{equation}
is the matrix of color sums for the basis chosen for the partonic channel $\kappa$. The typically used color bases are not orthonormal, in which case $\hT_\kappa$ is not equal to the identity operator $\id_\kappa$ and $\vC^\dagger$ is not just the naive complex conjugate transpose of $\vC$. We then have
\begin{equation}
\Abs{\vC(\Phi_N)}^2\equiv \vC^\dagger(\Phi_N)\, \vC(\Phi_N) =
\sum_\mathrm{color} \Abs{\cA_\fin(\Phi_N) }^2
\,.\end{equation}

\subsubsection{Leading order}

It is instructive to see how the LO cross section arises from \eq{fact}. At LO, we have
\begin{align}
J^\zero_i(s, \mu) &= \delta(s)
\,, \nn \\
B^\zero_a(t, x, \mu) &= \delta(t)\, f_a(x, \mu_F)
\,, \nn \\
\hS_\kappa^\zero(k_a, \ldots, k_N, \{\hs_{ij}\}, \mu) &= \id_\kappa \prod_i \delta(k_i)
\,,\end{align}
where the LO soft function is the identity operator in color space,
\begin{equation}
\id_\kappa  \equiv \delta^{\alpha_a \beta_a}\,\dotsb\,\delta^{\alpha_N \beta_N}
\,.\end{equation}
Plugging this back into \eq{fact} we get
\begin{align}
\frac{\df\sigma^\sing_\LO(\Phi_N)}{\df \Tau_N^a\, \df\Tau_N^b\dotsb\df\Tau_N^N}
&= f_a\, f_b\, \vC^{\dagger \zero}(\Phi_N)\,\id_\kappa \vC^{\zero}(\Phi_N) \prod_i \delta(\TauN^i)
\nn \\
&= \Bigl[ f_a\, f_b\sum_{\rm colors} \Abs{\cA^{\zero}(\Phi_N)}^2 \Bigr] \prod_i \delta(\TauN^i)
\equiv B_N(\Phi_N) \prod_i \delta(\TauN^i)
\,.\end{align}
Equation \eqref{eq:dsigmasingfull} then reproduces the LO cross section as in \eqs{LO}{singLO}.

\subsection{Single-differential subtractions}
\label{subsec:singlediff}

We now project onto the single-differential $N$-jettiness $\TauN$. Equations~\eqref{eq:dsigmasingfull} and \eqref{eq:fact} yield
\begin{align} \label{eq:factsing}
\frac{\df\sigma^\sing(\Phi_N)}{\df \Tau_N}
&=
\int\!\df t_a\, B_a(t_a, x_a, \mu)
\int\!\df t_b\, B_b(t_b, x_b, \mu)
\biggl[\, \prod_{i=1}^N \int\!\df s_i\, J_i(s_i, \mu) \biggr]
\\\nn &\quad \times
\vC^\dagger(\Phi_N, \mu)\,
\hS_\kappa \biggl(\Tau_N - \frac{t_a}{Q_a} - \frac{t_b}{Q_b} - \sum_{i=1}^N \frac{s_i}{Q_i}, \{\hq_i \}, \mu\biggr)
\vC(\Phi_N, \mu)
\,,\end{align}
where the single-differential soft function is the projection of the multi-differential one appearing in \eq{fact}, see \eq{softsinglediff}. We expand this singular contribution to the $N$-jettiness cross section as [cf. \eq{dsigmasingPhiN}]
\begin{align} \label{eq:dsigmasingPhiN2}
\frac{\df\sigma^\sing(\Phi_N)}{\df\TauN}
&= \C_{-1}(\Phi_N, \xi)\,\delta(\TauN) + \sum_{n \geq 0} \C_n(\Phi_N, \xi)\, \frac{1}{\xi} \cL_n\Bigl(\frac{\TauN}{\xi}\Bigr)
\\\nn
&= \sum_{m \geq 0} \biggl[\C_{-1}^{(m)}(\Phi_N, \xi, \mu)\,\delta(\TauN) + \sum_{n = 0}^{2m-1} \C_n^{(m)}(\Phi_N, \xi, \mu)\, \frac{1}{\xi} \cL_n\Bigl(\frac{\TauN}{\xi}\Bigr) \biggr] \Bigl(\frac{\alpha_s(\mu)}{4\pi} \Bigr)^m
\,.\end{align}
The $\cL_n(\tau)$ are the usual plus distributions defined in \eq{cLn}.
Here we explicitly denote the dependence of the subtraction coefficients $\C_n^{(m)}(\Phi_N, \xi, \mu)$ on the renormalization scale $\mu$. The individual coefficients also depend on the arbitrary dimension-one parameter $\xi$, which drops out exactly in the sum of all coefficients at each order in $\alpha_s$. (In \subsec{SingAndNonsing} we used $\xi\equiv Q$.) Finally, the coefficients also depend on the $N$-jettiness measures $Q_i$, which we suppress for simplicity.

To determine the subtraction coefficients, we simply expand all the functions in the factorization theorem \eq{factsing} in terms of $\alpha_s(\mu)$,
\begin{align}
\label{eq:JBSCexp}
J_i(s, \mu) &= \delta(s) + \sum_{m\ge 1} J_i^{(m)}(s, \mu) \Bigl(\frac{\alpha_s(\mu)}{4\pi} \Bigr)^m
\,, \nn \\
B_a(t, x, \mu) &= \delta(t)\, f_a(x, \mu_F) +  \sum_{m\ge 1} B_a^{(m)}(t, x, \mu, \mu_F) \Bigl(\frac{\alpha_s(\mu)}{4\pi} \Bigr)^m
\,, \nn \\
\hS_\kappa(k, \{\hs_{ij}\}, \mu)
&= \id_\kappa \, \delta(k) +  \sum_{m\ge 1} \hS_\kappa^{(m)}(k, \{\hs_{ij}\}, \mu) \Bigl(\frac{\alpha_s(\mu)}{4\pi} \Bigr)^m
\,, \nn \\
\vC(\Phi_N, \mu)
&= \vC^\zero(\Phi_N, \mu)  +  \sum_{m\ge 1} \vC^{(m)}(\Phi_N, \mu) \Bigl(\frac{\alpha_s(\mu)}{4\pi} \Bigr)^m
\,,\end{align}
plug these back, and collect all contributions to each order in $\alpha_s$ and each power in $\ln\TauN$.
Explicit results for the jet, beam, and soft functions through $\ord{\alpha_s^2}$ are given in \app{Ingredients}.

\subsubsection{NLO subtractions}
\label{subsubsec:NLOsub}

At NLO, the differential subtractions require the subtraction coefficients $\C_0^\one$ and $\C_1^\one$, which are the coefficients of the $1/\TauN$ and $(\ln\TauN)/\TauN$ contributions. They are given by (with $n = 0, 1$),
\begin{align} \label{eq:CdiffNLO}
\C_{n}^\one(\Phi_N, \xi, \mu)
&= \Abs{\vC^\zero(\Phi_N, \mu)}^2 \biggl[
 f_a(x_a, \mu_F)\, f_b(x_b, \mu_F) \sum_{i=1}^N\! J_{i,n}^\one\!\Bigl(\frac{Q_i \xi}{\mu^2}\Bigr)
\nn \\ & \quad
+ B_{a,n}^\one \Bigl(x_a,\mu,\mu_F, \frac{Q_a \xi}{\mu^2}\Bigl)\, f_b(x_b, \mu_F)
   + f_a(x_a, \mu_F)\, B_{b,n}^\one \Bigl(x_b,\mu,\mu_F,\frac{Q_b \xi}{\mu^2}\Bigl)
\biggr]
\nn \\ & \quad
   + f_a(x_a, \mu_F)\, f_b(x_b, \mu_F)\,
   \vC^{\dagger\zero}(\Phi_N, \mu)\, \hS_{\kappa,n}^\one\Bigl(\!\{\hq_i\}, \frac{\xi}{\mu}\Bigr)\, \vC^{\zero}(\Phi_N, \mu)
\,.\end{align}
The jet function contributions in the first line effectively correspond to collinear final-state subtractions, while the beam function contributions in the second line effectively correspond to collinear initial-state subtractions. The soft function contribution in the last line effectively corresponds to a soft subtraction.

As explained in \sec{formalism}, the coefficient $\C_{-1}$ determines the integrated subtractions plus the virtual contributions,
which becomes obvious when choosing $\xi = \Tauoff$. At NLO, we have
\begin{align} \label{eq:CintNLO}
\C_{-1}^\one(\Phi_N, \xi, \mu)
&= f_a(x_a, \mu_F)\, f_b(x_b, \mu_F)\, \bigl(\vC^{\dagger\one}\vC^{\zero} + \vC^{\dagger\zero}\vC^{\one}\bigr)(\Phi_N, \mu)
\nn \\ & \quad
+ \Abs{\vC^\zero(\Phi_N, \mu)}^2 \biggl[
f_a(x_a, \mu_F)\, f_b(x_b, \mu_F)\, \sum_{i=1}^N J_{i,-1}^\one\Bigl(\frac{Q_i \xi}{\mu^2}\Bigr) 
\nn \\ & \quad
+ B_{a,-1}^\one \Bigl(x_a,\mu,\mu_F, \frac{Q_a \xi}{\mu^2}\Bigl) f_b(x_b, \mu_F) + f_a(x_a, \mu_F) B_{b,-1}^\one  \Bigl(x_b,\mu,\mu_F,\frac{Q_b \xi}{\mu^2}\Bigl) \biggr]
\nn \\ & \quad
+ f_a(x_a, \mu_F)\, f_b(x_b, \mu_F)\,
\vC^{\dagger\zero}(\Phi_N, \mu)\, \hS_{\kappa,-1}^\one\Bigl(\!\{\hq_i \}, \frac{\xi}{\mu}\Bigr)\, \vC^{\zero}(\Phi_N, \mu)
\,.\end{align}
The first line contains the IR-finite virtual one-loop amplitudes in $\vC^\one(\Phi_N)$. The remaining lines effectively correspond to the integrated collinear and soft subtractions.
The NLO beam, jet, and soft function coefficients, $B^\one_{a,n}(x,\mu,\mu_F,\la)$, $J^\one_{i,n}(\la)$, and $\hS^\one_{\kappa,n}(\{\hq_i \},\la)$ appearing in \eqs{CdiffNLO}{CintNLO} are all known and are collected in \app{Ingredients}. The PDF factorization scale ($\mu_F$) dependence only enters via the beam functions and the PDFs.

One can see that the structure of the subtraction terms has a close resemblance with FKS subtractions. The important difference is that here one does not have to divide up phase space in order to individually isolate all possible IR singular regions. Instead, all the singular regions are projected onto the single variable $\TauN$. An analogous phase-space division for the soft emissions now happens in the calculation of the $N$-jettiness soft function. It is also important to note that there are no overlaps (i.e. double counting) between the soft and collinear subtraction terms. In principle, such overlaps can exist and must be removed, which in SCET corresponds to removing so-called zero-bin contributions~\cite{Manohar:2006nz}. A nice feature of $N$-jettiness is that all such overlap contributions automatically vanish in pure dimensional regularization at all orders in perturbation theory.

\subsubsection{NNLO subtractions}
\label{subsubsec:NNLOsub}

For simplicity of the presentation, we define the abbreviations
\begin{align}
{\rm J}^{(m)}_{i,n} \equiv J_{i,n}^{(m)}\Bigl(\frac{Q_i \xi}{\mu^2}\Bigr)
\,, \qquad
{\rm B}^{(m)}_{i,n} &\equiv B^{(m)}_{i,n}  \Bigl(x_i,\mu,\mu_F, \frac{Q_i \xi}{\mu^2}\Bigl)
\,, \qquad
{\rm f}_i \equiv  f_i(x_i, \mu_F)
\,,\nn\\
\widehat {\rm S}_{n}^{(m)} \equiv \hS_{\kappa,n}^{(m)}\Bigl(\!\{\hq_i\},\frac{\xi}{\mu}\Bigr)
\,, \qquad
\vec {\rm C}^{(m)} &\equiv \vC^{(m)}(\Phi_N, \mu)
\,,\end{align}
where we use roman letters (B, J, S, C, f) to avoid any confusion with some of the coefficients listed in \app{Ingredients}.
The NNLO coefficients ${\rm J}^\two_{i,n}$ and ${\rm B}^\two_{i,n}$ as well as the soft function coefficients $\widehat {\rm S}_{n\ge0}^\two$ are all known analytically, see \app{Ingredients}.

The $N$-jettiness soft function describes how the soft radiation is split into the different $N$-jettiness regions. Obtaining the two-loop soft constant $\hS_{\kappa,-1}^\two$ (the coefficient of the $\delta(k)$, see \eq{softLexp}) is the remaining principal challenge. It is known analytically for processes with two external partons, see \eqs{softtwoloopconstqbarq}{softtwoloopconstgg}, but it is currently unknown for generic $N$-jet processes. It can however be determined numerically by extending the NLO calculation in \mycite{Jouttenus:2011wh} using existing NNLO results. A procedure to do so has been outlined recently in \mycite{Boughezal:2015eha}, where numerical results for $1$-jettiness in $pp$ collisions were presented.

We conveniently denote the genuine $m$-loop contributions from jet, beam, and soft functions to the $\C_{n}^{(m)}$ as
\begin{align} \label{eq:genuineX}
X^{(m)}_n &\equiv \Abs{\vec {\rm C}^\zero}^2 \, \biggl( {\rm f}_a\, {\rm f}_b \sum_{i=1}^N {\rm J}^{(m)}_{i,n}
\,+\, {\rm B}^{(m)}_{a,n} \, {\rm f}_b
+ {\rm f}_a \, {\rm B}^{(m)}_{b,n} \biggr)
\,+\,  {\rm f}_a\, {\rm f}_b\, \vec {\rm C}^{\dagger\zero} \, \widehat {\rm S}_{n}^{(m)}\, \vec {\rm C}^\zero\,.
\end{align}
Using this notation, we write the two-loop cross terms related to real-virtual contributions and involving the one-loop virtual amplitudes in $\vec{\rm C}^\one$ as
\begin{align}
X^{(1+1)}_n &\equiv \bigl( \vec {\rm C}^{\dagger\zero} \vec {\rm C}^\one + \vec {\rm C}^{\dagger\one} \vec {\rm C}^\zero \bigr) \,
\biggl( {\rm f}_a\, {\rm f}_b \sum_{i=1}^N {\rm J}^\one_{i,n}
\,+\, {\rm B}^\one_{a,n} \, {\rm f}_b + {\rm f}_a \, {\rm B}^\one_{b,n} \biggr)
\nn\\ &\quad
+ {\rm f}_a\, {\rm f}_b\,  \bigl( \vec {\rm C}^{\dagger\zero} \, \widehat {\rm S}_{n}^\one\, \vec {\rm C}^\one
\,+\, \vec {\rm C}^{\dagger\one} \, \widehat {\rm S}_{n}^\one\, \vec {\rm C}^\zero \bigr)
\,.\end{align}
Finally, the cross terms with two one-loop coefficients of jet, beam, or soft functions from the associated $\cL_n \otimes \cL_m$ convolution are denoted as
\begin{align}
X^{(1+1)}_{n, m} &\equiv \Abs{\vec {\rm C}^\zero}^2 \, \biggl(
{\rm f}_a\, {\rm f}_b \sum_{i < j=1}^N {\rm J}^\one_{i,n}{\rm J}^\one_{j,m} 
\,+\, {\rm B}^\one_{a,n}\, {\rm f}_b \sum_{i=1}^N {\rm J}^\one_{i,m} 
\,+\, {\rm f}_a \, {\rm B}^\one_{b,n} \sum_{i=1}^N {\rm J}^\one_{i,m}
\,+\, {\rm B}^\one_{a,n} \, {\rm B}^\one_{b,m} \biggr)
\nn\\ &\quad
+\, {\rm f}_a\, {\rm f}_b \sum_{i=1}^N {\rm J}^\one_{i,n} \, \vec {\rm C}^{\dagger\zero} \, \widehat {\rm S}_{m}^\one\, \vec {\rm C}^\zero
\,+\, {\rm B}^\one_{a,n}\, {\rm f}_b \, \vec {\rm C}^{\dagger\zero} \, \widehat {\rm S}_{m}^\one\, \vec {\rm C}^\zero
\,+\, {\rm f}_a \, {\rm B}^\one_{b,n}\, \vec {\rm C}^{\dagger\zero} \, \widehat {\rm S}_{m}^\one\, \vec {\rm C}^\zero 
\,.\end{align}
With these definitions, the NNLO subtraction coefficients read
\begin{align}
\label{eq:L3Coefftwo}
\C_{3}^\two(\Phi_N, \xi, \mu)
&= X^\two_3 + X^{(1+1)}_{1,1}
\,, \\[1 ex]
\C_{2}^\two(\Phi_N, \xi, \mu)
&= X^\two_2 + \frac32 \bigl(X^{(1+1)}_{0,1} + X^{(1+1)}_{1,0}\bigr)
\,, \\[1 ex]
\C_{1}^\two(\Phi_N, \xi, \mu)
&= X^\two_1 + X^{(1+1)}_1 + 2\, X^{(1+1)}_{0,0} - \frac{\pi^2}{3} X^{(1+1)}_{1,1}
\,, \\[1 ex]
\C_{0}^\two(\Phi_N, \xi, \mu)
&= X^\two_0 + X^{(1+1)}_0 - \frac{\pi^2}{6} \bigl(X^{(1+1)}_{0,1} + X^{(1+1)}_{1,0}\bigr) +2 \zeta_3 X^{(1+1)}_{1,1}
\,, \\[1 ex]
\label{eq:deltaCoefftwo}
\C_{-1}^\two(\Phi_N, \xi, \mu)
&= {\rm f}_a\, {\rm f}_b\, \bigl( \vec {\rm C}^{\dagger\zero} \vec {\rm C}^\two + \vec {\rm C}^{\dagger\one} \vec {\rm C}^\one + \vec {\rm C}^{\dagger\two} \vec {\rm C}^\zero \bigr)
\\\nn  &\quad
+ X^\two_{-1} + X^{(1+1)}_{-1} - \frac{\pi^2}{6} X^{(1+1)}_{0,0} + \zeta_3 \bigl(X^{(1+1)}_{0,1} + X^{(1+1)}_{1,0}\bigr) - \frac{\pi^4}{360} X^{(1+1)}_{1,1}
\,.\end{align}
The $\delta(\Tau_N)$ coefficient $\C_{-1}^\two$ again corresponds to the integrated NNLO subtraction piece and contains the full IR-finite $\ord{\alpha_s^2}$ virtual $N$-parton amplitudes in $\vec{\rm C}^\two$.

\begin{table}[t]
\begin{center}
\begin{tabular}{c|ccc}
\hline\hline
$k$         & $V_k^{00}$ & $V_k^{01}=V_k^{10}$ & $V_k^{11}$ \\ \hline
$-1$ & $-\pi^2/6$            & $\zeta_3$             & $-\pi^4/360$  \\
$0$    & $0$                   & $-\pi^2/6$            & $2\,\zeta_3$  \\
$1$    & $2$                   & $0$                   & $-\pi^2/3$  \\
$2$    & $0$                   & $3/2$                 & $0$  \\
$3$    & $0$                   & $0$                   & $1$  \\ \hline\hline
\end{tabular}
\end{center}
\caption{Coefficients $V_k^{mn}$ for the convolution $\cL_m \otimes \cL_n$ according to \eq{ExpLnLm}.}
\label{tab:Lconvolutions}
\end{table}

The constants multiplying the $X^{(1+1)}_{n,m}$ in equations~\eqref{eq:L3Coefftwo}-\eqref{eq:deltaCoefftwo} are the coefficients $V_k^{mn}$ arising in the convolution $\cL_m \otimes \cL_n$,
\begin{align} \label{eq:ExpLnLm}
(\cL_m \otimes \cL_n)(\tau)
 &\equiv \int\!\df \tau' \, \cL_m(\tau - \tau')\, \cL_n(\tau')
= V_{-1}^{mn}\, \delta(\tau) + \sum_{k=0}^{m+n+1} V_k^{mn}\, \cL_k(\tau)
\,.\end{align}
They are given in table~\ref{tab:Lconvolutions} for $m,n\leq 1$. Their expression for general $m,n$ can be found in Appendix B of \mycite{Ligeti:2008ac}.

\subsubsection{Toward N$^3$LO subtractions}
\label{subsubsec:N3LOsub}

Using the notation introduced in the previous subsection it is straightforward to also write down the N$^3$LO $N$-jettiness subtraction terms. Besides the genuine three-loop terms $X_n^{(3)}$ according to \eq{genuineX}, we now have ``two-loop times one-loop'' cross terms $X^{(1+2)}_n=X^{(2+1)}_n$ and $X^{(1+2)}_{n, m}=X^{(2+1)}_{n, m}$
as well as the ``(one-loop)$^3$'' cross terms $X^{(1+1+1)}_n$, $X^{(1+1+1)}_{n,m}$, and $X^{(1+1+1)}_{n,m,l}$, where the latter is associated with the convolution $\cL_n \otimes \cL_m \otimes \cL_l$.

The N$^3$LO subtraction coefficients then schematically take the form
\begin{align}
\label{eq:deltaCoeffthree}
\C_{5}^{(3)}(\Phi_N, \xi, \mu)
&= X^{(3)}_5 \,+\, \text{cross terms}\,,  \nn \\
&\;\;\vdots \\
\C_{0}^{(3)}(\Phi_N, \xi, \mu)
&= X^{(3)}_0 \,+\, \text{cross terms}\,, \nn \\[1 ex]
\C_{-1}^{(3)}(\Phi_N, \xi, \mu)
&= X^{(3)}_{-1} +  {\rm f}_a\, {\rm f}_b\, \bigl( \vec {\rm C}^{\dagger\zero} \vec {\rm C}^{(3)} + \vec {\rm C}^{\dagger\one} \vec {\rm C}^\two + \vec {\rm C}^{\dagger\two} \vec {\rm C}^\one + \vec {\rm C}^{\dagger(3)} \vec {\rm C}^\zero \bigr) \,+\, \text{cross terms}
\,.\nn
\end{align}
The cross terms in \eq{deltaCoeffthree} are a linear combination of the above listed $X$'s, whose numerical coefficients can be easily worked out by evaluating the relevant convolutions among the $\cL_{n\le3}$ distributions in analogy to the NNLO case.

For processes with only two colored external partons, so $e^+ e^- \!\to\! q\bar q$, DIS, or $pp\to$ color singlet, analytic expressions for all $X^{(3)}_{n\ge0}$ are in fact available. This is because the three-loop anomalous dimensions of jet and beam functions, the PDFs, and the hard function are known~\cite{Vogt:2004mw, Moch:2004pa, Moch:2005id, Moch:2005tm, Becher:2006mr, Becher:2009th, Stewart:2010qs, Berger:2010xi}, which also fixes the three-loop soft anomalous dimension. This means the complete set of $\ord{\alpha_s^3}$ logarithmic ($\cL_n$) terms of the renormalized jet, beam, and soft functions are determined by their RGE. The only coefficient that is not fully known is $\C_{-1}^{(3)}$, associated with the integrated N$^3$LO subtractions plus virtual corrections. The ${\rm C}^{(3)}$ are known from the IR-finite parts of the three-loop quark and gluon form factors~\cite{Baikov:2009bg, Gehrmann:2010ue}. As the cross terms only involve known lower-order contributions, the only missing piece in \eq{deltaCoeffthree} then is $X^{(3)}_{-1}$, for which one has to compute the three-loop $\mu$-independent constants of the jet, beam, and soft functions.

\subsection{Constructing more-differential subtractions}
\label{subsec:morediff}

As mentioned already, the $\TauN$-subtractions we have defined thus far are nonlocal, in the sense that all the singular regions are projected onto the single variable $\TauN$ and the subtraction acts only after the corresponding phase-space integrations.
It is conceivable that in order to improve the numerical stability or convergence of the NNLO calculation one might wish to use a more local subtraction -- indeed, many of the available NNLO subtraction schemes utilize highly local subtraction terms.

In our approach, it is straightforward, at least conceptually, to progressively increase the locality of the subtractions. All one needs to do is split $\Tau_N$ up into further IR-safe observables that cover the phase space and which are sensitive to emissions in different regions, and/or introduce further observables that resolve the nature of emissions, e.g. allowing one to discriminate between double-real and single-real(+virtual) emissions in a given region. The subtraction is then given by the singular cross section differential in all of these observables. In practice, this requires the relevant factorization theorem for this more-differential cross section.

Let us demonstrate how this works for a simple example. The factorization theorem in \eq{fact} is already differential in the individual $N$-jettiness contributions $\TauN^i$. For simplicity, we take $N = 0$ and consider the $X + 0j$ NNLO cross section.
In this case, $0$-jettiness (aka beam thrust) effectively splits the event into two hemispheres (beam regions) $a$ and $b$, whose $N$-jettiness axes are defined by the beam directions. The total $0$-jettiness is given by $\Tau_0 = \Tau_0^a + \Tau_0^b$, where $\Tau_0^a$ and $\Tau_0^b$ are the contributions from the two hemispheres [cf. \eq{TauNi_def} and its discussion].

Following the procedure in \subsec{singlediff} we can use the total $\Tau_0$ to construct a subtraction. However, instead of taking the sum, we can also consider $\Tau_a \equiv \Tau_0^a$ and $\Tau_b \equiv \Tau_0^b$ separately, and perform the subtraction differential in both of these observables. Each of them is then sensitive to a subset of the singular regions, namely, collinear (and soft) emissions closer to beam $a$ will only affect $\Tau_a$, whilst emissions closer to beam $b$ will only affect $\Tau_b$.

Following the logic of section \ref{subsec:TauNsubtract}, we first write down the appropriate formula for the corresponding double-differential phase-space slicing:
\begin{align} \label{eq:diffslice}
\sigma(X)
&= \int_0^{\Tau_\delta}\!\!\df\Tau_a \int_0^{\Tau_\delta}\!\!\df\Tau_b\, \frac{\df \sigma^\sing(X)}{\df \Tau_a\, \df \Tau_b}
+  \int_{\Tau_\delta}\!\!\df\Tau_a \int_0^{\Tau_\delta}\!\!\df\Tau_b\, \frac{\df\sigma(X)}{\df \Tau_a\, \df \Tau_b}
\nn \\ & \quad
+ \int_0^{\Tau_\delta}\!\!\df\Tau_a \int_{\Tau_\delta}\!\!\df\Tau_b\, \frac{\df\sigma(X)}{\df \Tau_a\, \df \Tau_b}
+ \int_{\Tau_\delta}\!\!\df\Tau_a \int_{\Tau_\delta}\!\!\df\Tau_b\, \frac{\df\sigma(X)}{\df \Tau_a\, \df \Tau_b}
+ \ord{\delIR}
\,.\end{align}
Here, we substitute in the double-differential singular cross section when both $\Tau_a$ and $\Tau_b$ are below the IR cutoff $\Tau_\delta$, which is correct up to $\ord{\delIR}$. Having either $\Tau_a$ or $\Tau_b$ nonzero requires at least one additional emission, so the remaining three regions only require an NLO calculation. Of course, there are singularities in the second term as $\Tau_b \to 0$ with nonzero $\Tau_a$ (and similar singularities in the third term when $\Tau_a \to 0$ with nonzero $\Tau_b$), but these are handled as part of the NLO calculation.

\begin{figure}
\centering
 \includegraphics[scale=0.6]{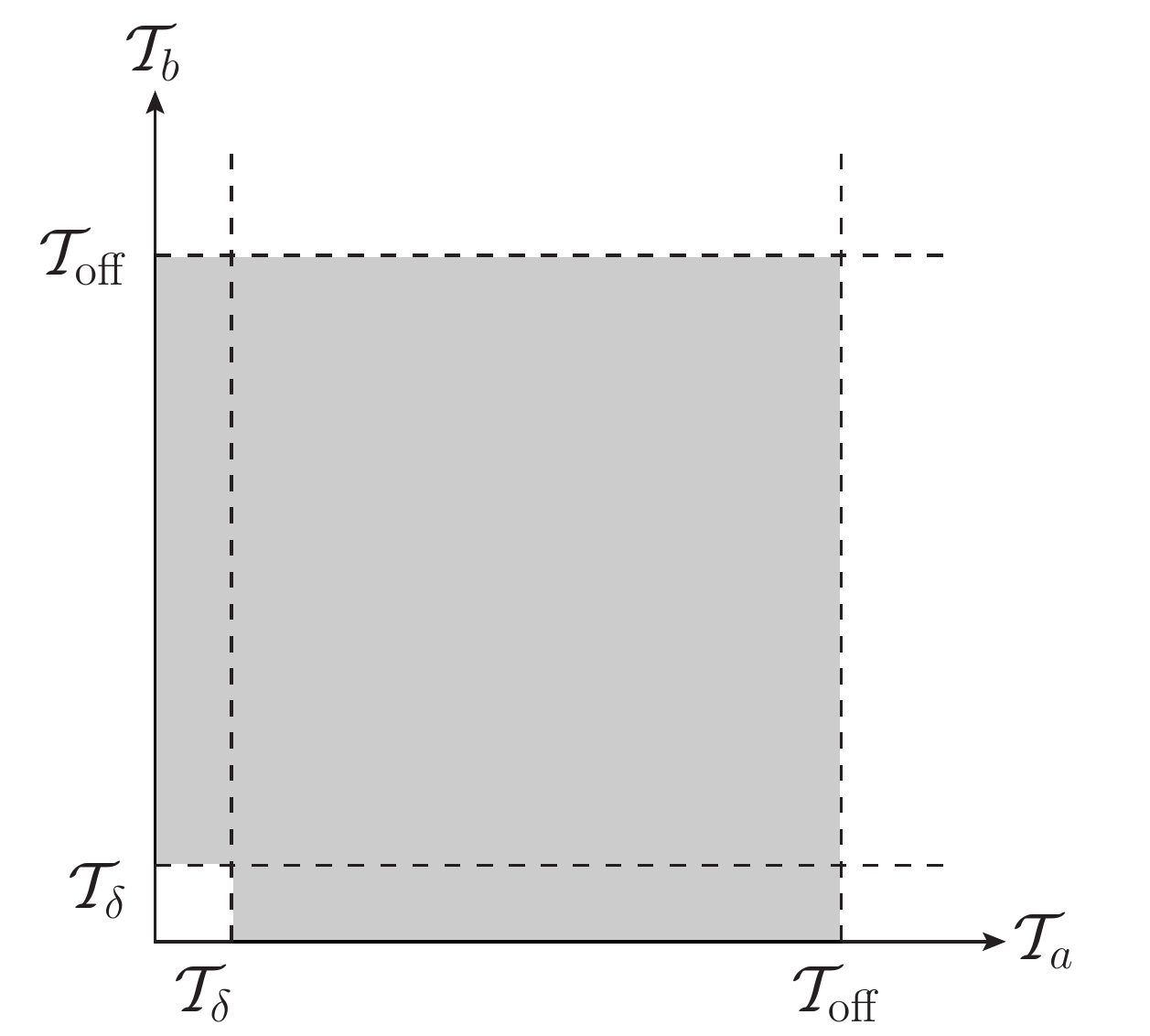}
\caption{Division of the $\Tau_a, \Tau_b$ phase space for double-differential $N$-jettiness subtractions.}
\label{fig:multidiffphase}
\end{figure}

Performing the slicing method using both $\Tau_a$ and $\Tau_b$ has no clear advantage over the slicing method using $\Tau_0$ alone, as in both methods one basically removes a small region of size $\Tau_\delta$ around $\Tau_0 = 0$, and handles it using the singular cross section. However, let us rewrite \eq{diffslice} as a subtraction by adding and subtracting the singular cross section for the shaded region in \fig{multidiffphase}, arranged in the following way:
\begin{align} \label{eq:diffsubtract}
\sigma(X)
&= \sigma^\sing(X,\Tau_a < \Tau_\off,\Tau_b < \Tau_\off)
\nn \\ & \quad
+ \int_{\Tau_\delta}^\Tauoff\!\df \Tau_b\,
\biggl[ \frac{\df\sigma(X, \Tau_a < \Tau_\delta)}{\df \Tau_b}
   - \frac{\df\sigma^\sing(X, \Tau_a < \Tau_\delta)}{\df \Tau_b} \biggr]
\nn \\ & \quad
+ \int_{\Tau_\delta}^\Tauoff\! \df \Tau_a\, \biggl[ \frac{\df\sigma(X, \Tau_b < \Tau_\delta)}{\df \Tau_a}
   - \frac{\df\sigma^\sing(X, \Tau_b < \Tau_\delta)}{\df \Tau_a} \biggr]
\nn \\ & \quad
+ \int_{\Tau_\delta}^\Tauoff\! \df\Tau_a \int_{\Tau_\delta}^\Tauoff\!\df\Tau_b\,
\biggl[ \frac{\df\sigma(X)}{\df \Tau_a\, \df \Tau_b} - \frac{\df \sigma^\sing(X)}{\df \Tau_a\, \df \Tau_b} \biggr]
\nn \\ & \quad
+ \int\! \df\Tau_a \int\!\df\Tau_b\,
\frac{\df\sigma(X)}{\df \Tau_a\, \df \Tau_b}\, \bigl[1 - \theta(\Tau_a < \Tauoff)\, \theta(\Tau_b < \Tauoff) \bigr]
+ \ord{\delIR}
\,.\end{align}
This equation is the two-variable analogue of \eq{TauNsubtraction}. The parameter $\Tau_\off$ controls again where we turn off the subtraction, and the dependence on it precisely cancels between all contributions.
The total cumulant in the first term contains the two-loop virtual corrections together with the corresponding integrated subtraction terms. The cross sections in the second and third terms are differential in one of the variables and integrated in the other. Since one of the variables is nonzero, while the other is integrated, they require an NLO calculation with one additional resolved emission. These terms contain all the real-virtual contributions and the singular cross section acts as the corresponding real-virtual subtraction.
The fourth term involves the double-differential cross sections and since both variables are nonzero only requires a LO calculation with two resolved emissions, one in each hemisphere. The double-differential singular cross section then acts as the corresponding double-real subtraction, which is point-by-point in both $\Tau_a$ and $\Tau_b$. (Contributions with two real emissions in the same hemisphere are part of the NLO calculations in the second and third terms.) Hence, by considering separately $\Tau_a$ and $\Tau_b$, one is able to disentangle different real-virtual and double-real contributions and also make the subtractions more local. The price one has to pay is that the double-differential singular cross section in the last term requires the double-differential NNLO soft function, which is more complicated. (For the beam and jet functions this requires no additional effort.)

A further important point to make is that $\Tau_a$ and $\Tau_b$ are defined such that requiring $\Tau_a > \TauIR$ or $\Tau_b > \TauIR$ forces the corresponding emission to be in hemisphere $a$ or $b$.
At NLO, there is only one real emission, so only one out of $\Tau_a$ and $\Tau_b$ can be nonzero. Then, the double-differential subtraction essentially splits the $\Tau_0$-subtraction into two pieces, acting in the two hemispheres. At NNLO, this splits the real-virtual contributions into the two pieces in the second and third lines of \eq{diffsubtract}. If this is undesired, one can instead consider the two variables $\Tau_\mathrm{min} = \min\{\Tau_a, \Tau_b \}$ and $\Tau_\mathrm{max} = \max\{\Tau_a, \Tau_b\}$. This effectively folds the phase space in \fig{multidiffphase} in half along the diagonal where $\Tau_a = \Tau_b$, and combines the second and third terms in \eq{diffsubtract} into one.

Now let us return to the general case with $N$ partons in the Born process. Then there are $N+2$ contributions $\TauN^a, \TauN^b,\TauN^1,\ldots,\TauN^N$, and one can consider the subtraction separately in all of them. At NLO, only one of them can be nonzero, while at NNLO at most two of them can be nonzero. This means that there will be many different contributions, where in each contribution only one or two of the $\TauN^i$ are differential and nonzero, while all the others are integrated over. Each $\TauN^i$ can only be nonzero when the corresponding emission is in the $i$th $N$-jettiness region. Hence, if desired, using the individual $\TauN^i$ as the resolution variables automatically yields a division of phase space into different singular regions around each of the $N$ partons, very similar to the phase-space divisions encountered in traditional local subtraction methods. On the other hand, if the proliferation of phase-space regions is undesired, one can still have the same gain at NNLO as in \eq{diffsubtract} by considering two combinations of all $\TauN^i$, e.g. the minimum and maximum nonzero $\TauN^i$, or the sum of all $\TauN^i$ together with the sum of all but the largest $\TauN^i$.

Instead of or in addition to splitting $\TauN$ into its different components, one can also increase the locality of the subtraction by performing it differentially in both $\TauN$ and another independent $N$-jet resolution variable. For example, one could look into each $N$-jettiness region $i$ and compute the scalar sum of transverse momenta with respect to the corresponding $N$-jettiness axis $E_{Ti} = \sum_{k \in i} \abs{p_{Tk}}$, performing the subtraction also differential in the $E_{Ti}$. Doing so resolves part of the radiation phase space, which would otherwise be integrated over when considering only $\TauN$ by itself. For the $X+0j$ case one could for example consider $\Tau_0$ together with the transverse momentum $p_T$ of the color-singlet final state $X$. The relevant factorization formulae differential in $\Tau_0$ and $p_T$ have been discussed and written down in \mycites{Jain:2011iu, Procura:2014cba} (see also \mycites{Collins:2007ph,Rogers:2008jk}), and the corresponding double-differential two-loop quark beam functions have been computed in \mycite{Gaunt:2014xxa}.

We have discussed several options how to extend the single-differential $N$-jettiness subtractions, but of course this is not an exhaustive list. Constructing such more-differential subtractions requires the appropriate singular cross section differential in all of the chosen jet resolution variables, and in order to experience the maximum advantage in terms of convergence, these differential cross sections should reproduce the correct singular behaviour in all of the relevant singular kinematic regimes. The factorization of multi-differential cross sections in SCET accurate in all relevant kinematic regimes is a topic that has received much interest recently, see e.g.\ \mycites{Bauer:2011uc, Jouttenus:2011wh, Jouttenus:2013hs, Procura:2014cba, Larkoski:2014tva, Larkoski:2015zka}, and it would be interesting to apply this work to the issue of calculating NNLO QCD cross sections.

\section{Practical Considerations and Implementation}
\label{sec:implementation}

In this section, we discuss in more detail how the singular cross section in \eq{dsigmasingPhiN2} can be implemented in practice as a subtraction term following our general discussion in \subsec{TauNsubtract}. We first discuss the NLO case in \subsec{NLOimpl}, where we also highlight the similarities to FKS subtractions, and then the NNLO case in \subsec{NNLOimpl}. In \subsec{NNLOresults}, we discuss numerical aspects using the NNLO rapidity spectrum in Drell-Yan and gluon-fusion Higgs production as an example.

\subsection{NLO}
\label{subsec:NLOimpl}

\subsubsection{FKS subtractions}

In the notation of \eq{sigmaLO}, the cross section at NLO is given by
\begin{align} \label{eq:sigmaNLO}
\sigma^\NLO(X)
&= \int\! \df\Phi_N\, B_N(\Phi_N)\, \meas(\Phi_N)
\nn \\ & \quad
+ \biggl[ \int\! \df\Phi_N\, V_N(\Phi_N)\, \meas(\Phi_N)
+ \int\! \df\Phi_{N+1}\, B_{N+1}(\Phi_{N+1})\, \meas(\Phi_{N+1}) \biggr]_{\epsilon\to 0}
\,,\end{align}
where $V_N$ is the $N$-parton virtual one-loop contribution and $B_{N+1}$ is the $N+1$-parton real-emission contribution,
\begin{align}
V_N(\Phi_N)
&= f_a\,f_b\,
\frac{\alpha_s}{4\pi}\sum_\mathrm{color} \bigl[\cA_{ab\to N}^{\dagger\zero} \cA_{ab\to N}^\one + \cA_{ab\to N}^{\dagger\one} \cA_{ab\to N}^\zero \bigr](\Phi_N)
\,, \nn \\
B_{N+1}(\Phi_{N+1})
&= f_a\,f_b\, \sum_\mathrm{color} \Abs{\cA^\zero_{ab\to N+1}(\Phi_{N+1})}^2\,
\,.\end{align}
The additional $\alpha_s$ in $B_{N+1}$ compared to $B_N$ is contained in $\cA^\zero(\Phi_{N+1})$.
As indicated in \eq{sigmaNLO}, the limit $\epsilon\to 0$ can only be taken in the sum of $V_N$ and integral over $B_{N+1}$.

When implementing \eq{sigmaNLO} using FKS subtractions~\cite{Mangano:1991jk, Frixione:1995ms, Frixione:1997np, Frederix:2009yq, Alioli:2010xd}, the cross section is obtained as follows:
\begin{align} \label{eq:NLOFKS}
\sigma^\NLO(X)
&= \int\! \df\Phi_N\, \biggl\{ (B_N + V_N^S)(\Phi_N)\, \meas(\Phi_N)
\nn \\ & \quad
+ \sum_k \int_\delIR\! \df\Phi_\rad\, \Bigl[ (B^k_{N+1} \meas)(\Phi^k_{N+1}) - S^k_{N+1}(\Phi_N, \Phi_\rad)\, \meas(\Phi_N) \Bigr] \biggr\}
+ \ord{\delIR}
\,, \nn \\
V_N^S(\Phi_N)
&= \biggl[V_N(\Phi_N) + \sum_k \int\! \df\Phi_\rad\, S^k_{N+1}(\Phi_N, \Phi_\rad) \biggr]_{\epsilon\to 0}
\,.\end{align}
Here, the phase space is first sampled over $\Phi_N$. For a fixed $\Phi_N$ point, one then further samples over the radiation phase space $\Phi_\rad$, where the sum over $k$ runs over all the different IR-singular regions. The real-emission contribution and measurement $(B_{N+1}X)(\Phi_{N+1}) \equiv B_{N+1}(\Phi_{N+1}) X(\Phi_{N+1})$ are evaluated at a constructed point $\Phi^k_{N+1} = \hat\Phi^k_{N+1}(\Phi_N, \Phi_\rad)$. The superscript $k$ on $B_{N+1}^k$ indicates that $B_{N+1}$ is divided up between the regions in such a way that it is precisely reproduced in the sum over all regions. The phase-space map $\hat\Phi^k_{N+1}$ and the subtraction terms $S_{N+1}^k$ are specific to each singular region. The $S_{N+1}^k$ are directly constructed in the singular limit, meaning they are functions of $\Phi_N$ and $\Phi_\rad$ only, and in particular do not depend on the actual map $\hat\Phi^k_{N+1}$. In practice, there is again a tiny IR cutoff $\delIR$ required on the $\Phi_\rad$ integral due to limited numerical precision and the fact that $B^k_{N+1}$ and $S^k_{N+1}$ each individually diverge. The subtracted virtual, $V_N^S$, contains the finite remainder after combining the virtual contributions with the integral of the subtractions and cancelling all $1/\epsilon$ IR poles.

\subsubsection{$\Tau_N$-subtractions}

As discussed in \subsec{TauNsubtract}, the full cross section for $X$ at $\TauN>0$ only requires a lower-order calculation. At NLO, we need its LO expression given by
\begin{equation} \label{eq:dsigmaNLOdTauN}
\frac{\df\sigma(X)}{\df\TauN} \bigg\vert_{\TauN>0}^\LO
= \int\!\df\Phi_{N+1}\, (B_{N+1} \meas)(\Phi_{N+1}) \,\delta[\TauN - \TauN(\Phi_{N+1})]
\,,\end{equation}
where it is obvious that this is a LO quantity.

Since the subtractions are used up to the upper cutoff $\TauN < \Tauoff$, as seen in \eq{TauNsubtraction}, it is most convenient to set $\xi = \Tauoff$ in the subtraction coefficients. For the singular spectrum at $\TauN>0$, we can simply drop $\C_{-1}$ and replace $\cL_n(\tau)\to \ln^n(\tau)/\tau$. The subtraction terms at NLO are then
\begin{align}
\sigma^\sing(\Phi_N, \Tauoff) &= \int_0^\Tauoff \!\df\TauN\,\frac{\df\sigma^\sing(\Phi_N)}{\df\TauN}
= \C_{-1}(\Phi_N, \Tauoff)
\,, \nn \\
\frac{\df\sigma^\sing(\Phi_N)}{\df\TauN}\bigg\vert_{\TauN>0}
&= \frac{1}{\TauN}\,\biggl[\C_0(\Phi_N, \Tauoff) + \C_1(\Phi_N, \Tauoff)\, \ln\Bigl(\frac{\TauN}{\Tauoff}\Bigr) \biggr] \theta(\TauN<\Tauoff)
\,,\end{align}
where for convenience we included the $\theta(\TauN<\Tauoff)$ in the singular spectrum.

Using the above with \eq{slicing}, the $\TauN$-slicing at NLO becomes
\begin{align}
\sigma^\NLO(X)
&= \int\!\df\Phi_N\, \sigma^\sing(\Phi_N, \TauIR) \, X(\Phi_N)
\nn \\ & \quad
+ \int\! \df\Phi_{N+1}\, (B_{N+1} \meas)(\Phi_{N+1})\, \theta[\Tau_N(\Phi_{N+1}) > \TauIR]
+ \ord{\delIR}
\,.\end{align}
This calculation is very easy from an implementation point of view, since it boils down to performing two LO phase-space integrals. As already eluded to, the main practical limitation are the large numerical cancellations between both terms, requiring the phase-space integrals to be evaluated to very high precision.

Using \eq{dsigmaNLOdTauN}, the differential $\TauN$-subtraction in \eq{TauNsubtraction} takes the form
\begin{align} \label{eq:sigmaNLOfixedTauN}
\sigma^\NLO(X)
&= \int\!\df\Phi_N\, \sigma^\sing(\Phi_N, \Tauoff) \, X(\Phi_N)
\nn \\ & \quad
+ \int_\TauIR\!\df\TauN\, \biggl\{\int\!\df\Phi_{N+1}\, (B_{N+1} \meas)(\Phi_{N+1}) \,\delta[\TauN - \TauN(\Phi_{N+1})]
\nn \\ & \qquad\qquad\qquad
- \int\!\df\Phi_N\, \frac{\df\sigma^\sing(\Phi_N)}{\df\TauN}\, X(\Phi_N) \biggr\}
+ \ord{\delIR}
\,.\end{align}
For a numerical implementation, one must be able to solve the $\delta$ function in the $\df\Phi_{N+1}$ integral, which amounts to being able to sample over all of $\Phi_{N+1}$ that gives a fixed $\TauN(\Phi_{N+1})$. One option to do so is to decompose $\Phi_{N+1}$ as
\begin{equation}
\Phi_{N+1} = \Phi_N \otimes \TauN \otimes \Omega_\rad
\,, \qquad
\df\Phi_{N+1} = \df\Phi_N\, \df\TauN\, \df\Omega_\rad
\,,\end{equation}
where $\Phi_N = \hat\Phi_N(\Phi_{N+1})$ is precisely the Born projection used to define $\TauN(\Phi_{N+1})$, see \subsec{TauNdef}. The $\Omega_\rad\equiv\Omega_\rad(\Phi_{N+1})$ contains the remaining information needed to fully specify $\Phi_{N+1}$, which includes the continuous angular radiation variables as well as the discrete information about flavor, spin, and in which $N$-jettiness region the additional emission goes. We can then rewrite \eq{sigmaNLOfixedTauN} as
\begin{align} \label{eq:sigmaNLOfixedTauNsolved}
\sigma^\NLO(X)
&= \int\!\df\Phi_N\, \biggl\{ \sigma^\sing(\Phi_N, \Tauoff) \, X(\Phi_N)
\\\nn & \quad
+ \!\int_\TauIR\!\!\!\df\TauN \biggl[ \int\!\!\df\Omega_\rad\, (B_{N+1} X)(\Phi_N\!\otimes\!\TauN\!\otimes\Omega_\rad)
- \frac{\df\sigma^\sing(\Phi_N)}{\df\TauN}\, X(\Phi_N) \biggr] \biggr\}
\!+ \ord{\delIR}
.\end{align}
One now samples first over $\Phi_N$ and then $\TauN$. For fixed $\Phi_N$ and $\TauN$, one further samples over $\Omega_\rad$ and evaluates the real-emission contribution at the $\Phi_{N+1}$ point reconstructed from all of these. Being able to reconstruct $\Phi_{N+1}(\Phi_N, \TauN, \Omega_\rad)$ is equivalent to inverting the Born projection. Recall however, that the singular contributions $\df\sigma^\sing(\Phi_N)/\df\TauN$ are independent of the Born projection. Therefore, one has the freedom to specifically choose the Born projection to facilitate this inversion, making it easily possible.

Note that the $\Omega_\rad$ integral contains a discrete sum over all $N$-jettiness axis/regions. If one were to separate $\TauN$ into its individual components $\TauN^i$ as discussed in \subsec{morediff}, this sum would become explicit and the single subtraction term $\df\sigma^\sing(\Phi_N)/\df\TauN$ would effectively separate into different subtraction terms for each region.

We also note the close similarity of \eq{sigmaNLOfixedTauNsolved} with the FKS subtraction in \eq{NLOFKS}. Basically, $\TauN\otimes\Omega_\rad$ now acts as $\Phi_\rad$, while the split up into singular regions is now determined by the definition of $\TauN$. The LO piece $\C_{-1}^\zero$ of $\sigma^\sing$ supplies the Born contribution $B_N$, and the NLO piece $\C_{-1}^\one$ corresponds to $V_N^S$.

\subsection{NNLO}
\label{subsec:NNLOimpl}

At NNLO, the subtraction terms are
\begin{align}
\sigma^{\sing}(\Phi_N, \Tauoff) &= \C_{-1}(\Phi_N, \Tauoff)
\,, \nn \\
\frac{\df\sigma^{\sing}(\Phi_N)}{\df\TauN}\bigg\vert_{\TauN>0}
&= \frac{1}{\TauN} \sum_{n=0}^3 \C_n(\Phi_N, \Tauoff)\, \ln^n\Bigl(\frac{\TauN}{\Tauoff}\Bigr)\, \theta(\TauN<\Tauoff)
\,.\end{align}
As at NLO, we have chosen $\xi = \Tauoff$ and included the $\theta(\TauN<\Tauoff)$ in the singular spectrum.

The full $\TauN$-differential cross section at $\TauN>0$ is now needed at NLO, where it is given by
\begin{align} \label{eq:dsigmaNNLOdTauN}
\frac{\df\sigma(X)}{\df\TauN} \bigg\vert_{\TauN>0}^\NLO
&= \biggl\{\int\!\df\Phi_{N+1} \, (B_{N+1}\meas + V_{N+1}\meas)(\Phi_{N+1})\, \delta[\TauN-\TauN(\Phi_{N+1})]
\nn \\ & \quad
+ \int\!\df\Phi_{N+2} (B_{N+2} \meas)(\Phi_{N+2})\, \delta[\TauN-\TauN(\Phi_{N+2})] \biggl\}_{\epsilon\to 0}
\nn \\
&= \int\! \df\Phi_{N+1}\, \biggl\{ (B_{N+1}\meas + V_{N+1}^S\meas)(\Phi_{N+1})\,\delta[\TauN-\TauN(\Phi_{N+1})]
\nn \\ & \quad
+ \sum_k\int\!\df\Phi_\rad\, \Bigl[ (B^k_{N+2} \meas)(\Phi^k_{N+2}) \,\delta[\TauN-\TauN(\Phi^k_{N+2})]
\nn \\ & \qquad
- S^k_{N+2}(\Phi_{N+1}, \Phi_\rad)\, \meas(\Phi_{N+1})\, \delta[\TauN - \TauN(\Phi_{N+1})] \Bigr] \biggr\}
\,.\end{align}
In the second equation we wrote it in the form of an NLO calculation with FKS-like subtractions, analogous to \eq{NLOFKS}, where now $\Phi_{N+2}^k = \hat\Phi_{N+2}^k(\Phi_{N+1}, \Phi_\rad)$.

In general, the $\hat\Phi^k_{N+2}$ map used in the $N+1$-jet NLO calculation will not preserve $\TauN$, that is, $\Tau_{N+1}[\hat\Phi_{N+2}^k(\Phi_{N+1},\Phi_\rad)] \neq \TauN(\Phi_{N+1})$. This means we have to be careful in implementing the $\TauIR$ cutoff, because in order for the neglected pieces to be nonsingular, the cutoff must be applied on the true $\TauN(\Phi_{N+2})$. We can do this by treating the cutoff analogous to the measurement $\meas$. That is, we define the NLO calculation
\begin{align}
\frac{\df\sigma^\NLO(X, \TauIR)}{\df\Phi_{N+1}}
&= (B_{N+1} \meas + V_{N+1}^S \meas)(\Phi_{N+1})\, \theta[\TauN(\Phi_{N+1}) - \TauIR]
\nn \\ & \quad
+ \sum_k \int\!\df\Phi_\rad\, \Bigl\{ (B_{N+1} \meas)(\Phi^k_{N+2})\, \theta[\TauN(\Phi^k_{N+2}) - \TauIR]
\nn \\ & \qquad
- S^k_{N+2}(\Phi_{N+1}, \Phi_\rad)\, \meas(\Phi_{N+1})\, \theta[\TauN(\Phi_{N+1}) - \TauIR] \Bigr\}
\,,\end{align}
which is fully-differential in $\Phi_{N+1}$ and satisfies
\begin{equation}
\int_\TauIR\!\df\TauN\,\frac{\df\sigma(X)}{\df\TauN} \bigg\vert^\NLO_{\TauN>0}
= \int\!\df\Phi_{N+1}\,\frac{\df\sigma^\NLO(X, \TauIR)}{\df\Phi_{N+1}}
\,.\end{equation}

Using the above together with \eq{slicing}, the $\TauN$-slicing at NNLO is given by
\begin{align} \label{eq:sigmaNNLOslicing}
\sigma^\NNLO(X)
&= \int\!\df\Phi_N\, \sigma^\sing(\Phi_N, \TauIR) \, X(\Phi_N)
+ \int\! \df\Phi_{N+1}\, \frac{\df\sigma^\NLO(X, \TauIR)}{\df\Phi_{N+1}}
+ \ord{\delIR}
\,.\end{align}
This is again quite easy to implement, only requiring a LO phase-space integral for the first term and a NLO calculation for the second term. The practical limitation is the achievable numerical precision in the NLO calculation and the $\Phi_{N+1}$ integral, which strongly limits how low $\TauIR$ can be pushed.

From \eq{TauNsubtraction}, the differential $\TauN$-subtraction at NNLO takes the form
\begin{align}
\sigma^\NNLO(X)
&= \int\!\df\Phi_N\, \sigma^\sing(\Phi_N, \Tauoff) \, X(\Phi_N)
\nn \\ & \quad
+ \int_\TauIR\!\df\TauN\, \biggl[ \frac{\df\sigma(X)}{\df\TauN}\bigg\vert^\NLO_{\TauN>0}
- \int\!\df\Phi_N\, \frac{\df\sigma^\sing(\Phi_N)}{\df\TauN}\, X(\Phi_N) \biggr]
+ \ord{\delIR}
\,.\end{align}
To implement this numerically, one must be able to compute the $\TauN$ spectrum $\df\sigma(X)/\df\TauN$ to NLO for a given $\TauN$, which requires to solve the $\delta$ functions in \eq{dsigmaNNLOdTauN}. For $\Phi_{N+1}$ we can use the same procedure as at NLO together with the $N+1$-jet NLO cross section $\df\sigma^\NLO(X,\TauIR)/\df\Phi_{N+1}$, giving
\begin{align} \label{eq:sigmaNNLOalmost}
\sigma^\NNLO(X)
&= \int\!\df\Phi_N\, \biggl\{ \sigma^\sing(\Phi_N, \Tauoff) \, X(\Phi_N)
+ \int_0\!\df\TauN\, \biggl[ \int\!\df\Omega_\rad\, \frac{\df\sigma^\NLO(X, \TauIR)}{\df\Phi_{N+1}}\bigg\vert_{\Phi_N \otimes \TauN\otimes\Omega_\rad}
\nn \\ & \quad
- \frac{\df\sigma^\sing(\Phi_N)}{\df\TauN}\, X(\Phi_N)\,\theta(\TauN - \TauIR) \biggr] \biggr\}
+ \ord{\delIR}
\,.\end{align}
This can be implemented like the NLO case in \eq{sigmaNLOfixedTauNsolved}, with the LO $B_{N+1}(\Phi_{N+1})$ replaced by $\df\sigma^\NLO(X, \TauIR)/\df\Phi_{N+1}$. The subtraction in \eq{sigmaNNLOalmost} is not completely local in $\TauN$, since the $\Phi_{N+2}$ points being integrated over in $\df\sigma^\NLO(X, \TauIR)/\df\Phi_{N+1}$ will generically not have the correct $\TauN$ value. However, a simple phase-space map $\hat\Phi^k_{N+2}$ that approximately preserves $\TauN$ might be sufficient in practice.

To achieve an exact point-by-point cancellation in $\TauN$, one also has to solve the $\delta(\TauN - \TauN(\Phi_{N+2})$ constraint in \eq{dsigmaNNLOdTauN}. This requires constructing a map $\hat\Phi^k_{N+2}$ for the $N+1$-jet NLO calculation that preserves $\TauN$ so $\TauN[\hat\Phi_{N+2}^k(\Phi_{N+1}, \Phi_\rad)] = \TauN(\Phi_{N+1})$ and is equivalent to inverting the Born projection $\hat\Phi_N(\Phi_{N+2})$ underlying the definition of $\TauN(\Phi_{N+2})$. This is quite a bit more challenging than at NLO. It has been achieved in \mycite{Alioli:2012fc} for a slightly modified version of $\TauN$. Assuming, we have a $\hat\Phi^k_{N+2}$ map like this, we can pull the $\TauIR$ cut out of the NLO calculation, such that
\begin{align}
\frac{\df\sigma^\NLO(X, \TauIR)}{\df\Phi_{N+1}}
&= \frac{\df\sigma^\NLO(X)}{\df\Phi_{N+1}}\,\theta[\TauN(\Phi_{N+1}) - \TauIR]
\\\nn
\frac{\df\sigma^\NLO(X)}{\df\Phi_{N+1}}&= (B_{N+1} \meas + V_{N+1}^S \meas)(\Phi_{N+1})
\\\nn & \quad
+ \sum_k \int\!\df\Phi_\rad\, \Bigl[ (B_{N+1} \meas)(\Phi^k_{N+2}) - S^k_{N+2}(\Phi_{N+1}, \Phi_\rad)\, \meas(\Phi_{N+1}) \Bigr]
\,.\end{align}
The differential $\TauN$-subtraction then becomes
\begin{align} \label{eq:sigmaNNLOsolved}
\sigma^\NNLO(X)
&= \int\!\df\Phi_N\, \biggl\{ \sigma^\sing(\Phi_N, \Tauoff) \, X(\Phi_N)
\nn \\ & \quad
+ \int_\TauIR\!\df\TauN\, \biggl[ \int\! \df\Omega_\rad\, \frac{\df\sigma^\NLO(X)}{\df\Phi_{N+1}}
\bigg\vert_{\Phi_N\otimes\TauN\otimes\Omega_\rad}
\!\!- \frac{\df\sigma^\sing(\Phi_N)}{\df\TauN}\, X(\Phi_N) \biggr] \biggr\}
+ \ord{\delIR}
\,.\end{align}
The subtraction is now fully localized in $\TauN$, and the only nonlocality is in the $\df\Omega_\rad$ variables.

In a practical implementation of \eq{sigmaNNLOalmost} or \eq{sigmaNNLOsolved}, the subtraction terms are easily evaluated and the most nontrivial ingredient is in fact the NLO calculation of $\df\sigma^\NLO(X)/\df\Phi_{N+1}$, for which one can use any existing FKS-like NLO calculation or one can iterate the $N$-jettiness subtractions and perform it using $\Tau_{N+1}$-subtractions. Note that in all cases above the $\meas$ measurement is performed inside $\df\sigma^\NLO(X)/\df\Phi_{N+1}$. If the $\hat\Phi_{N+2}^k$ map preserves $\meas$, so $X(\Phi_{N+2}^k) = X(\Phi_{N+1})$, then it can be pulled out of the $N+1$-jet NLO calculation.

\subsection{Example: NNLO rapidity spectrum for Drell-Yan and Higgs}
\label{subsec:NNLOresults}

\begin{figure}[t!]
\centering
\includegraphics[scale=0.35]{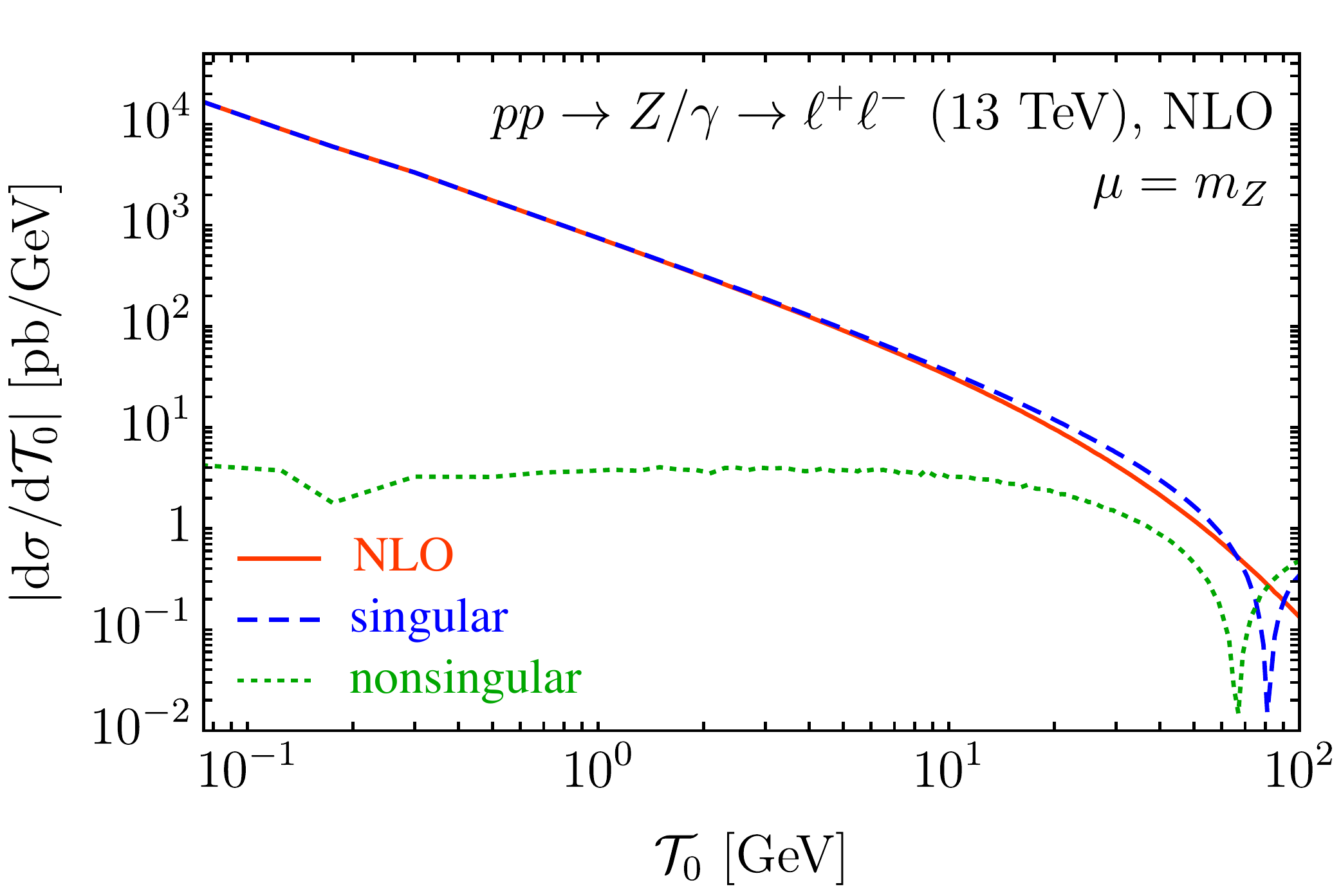}\hfill%
\includegraphics[scale=0.35]{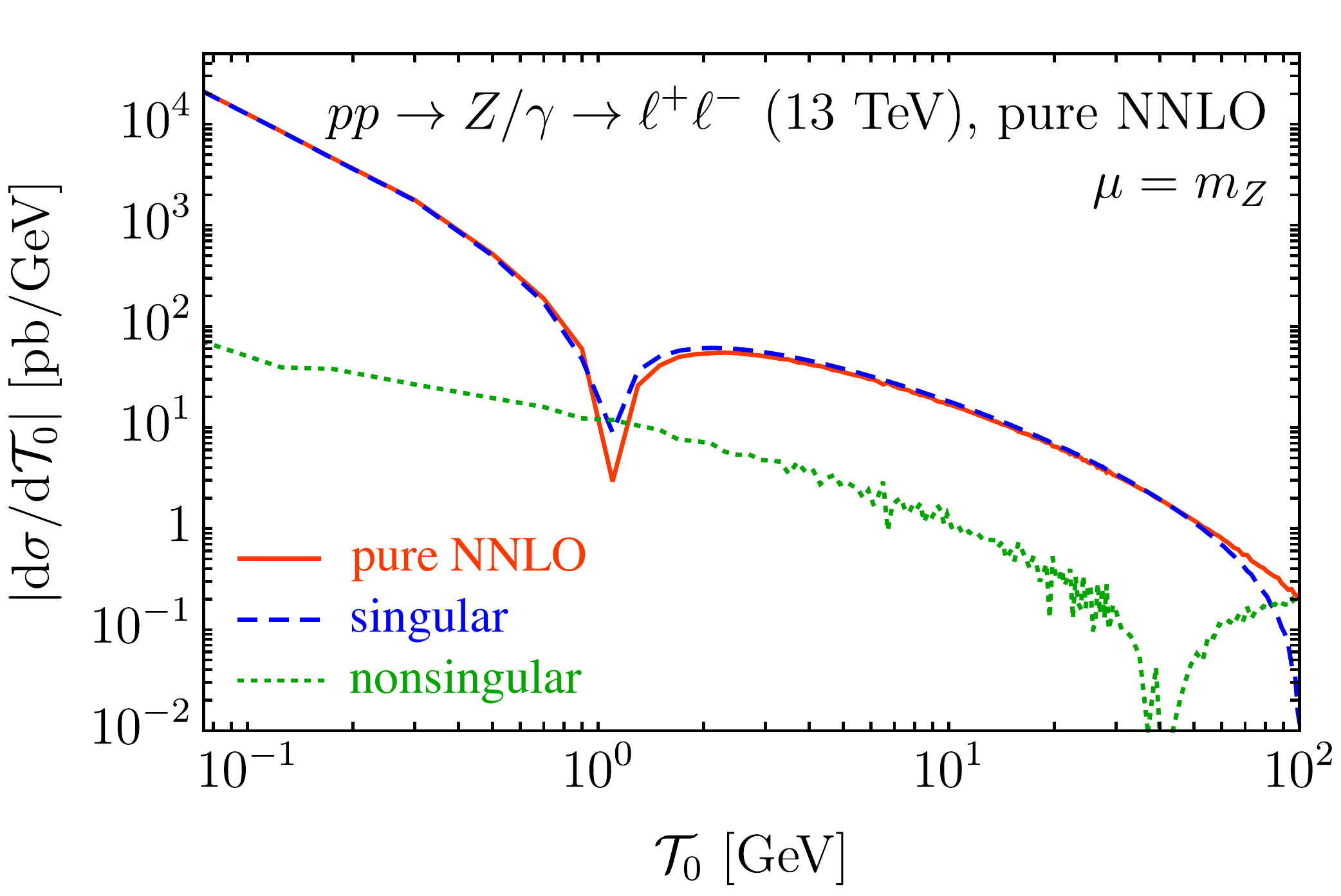}
\caption{The absolute value of the full, singular, and nonsingular contributions to the $\Tau_0$ spectrum for Drell-Yan production. The NLO $\ord{\alpha_s}$ corrections are shown on the left, and the pure NNLO $\ord{\alpha_s^2}$ corrections are on the right.}
\label{fig:Zpieces}
\end{figure}

\begin{figure}[t!]
\centering
\includegraphics[scale=0.35]{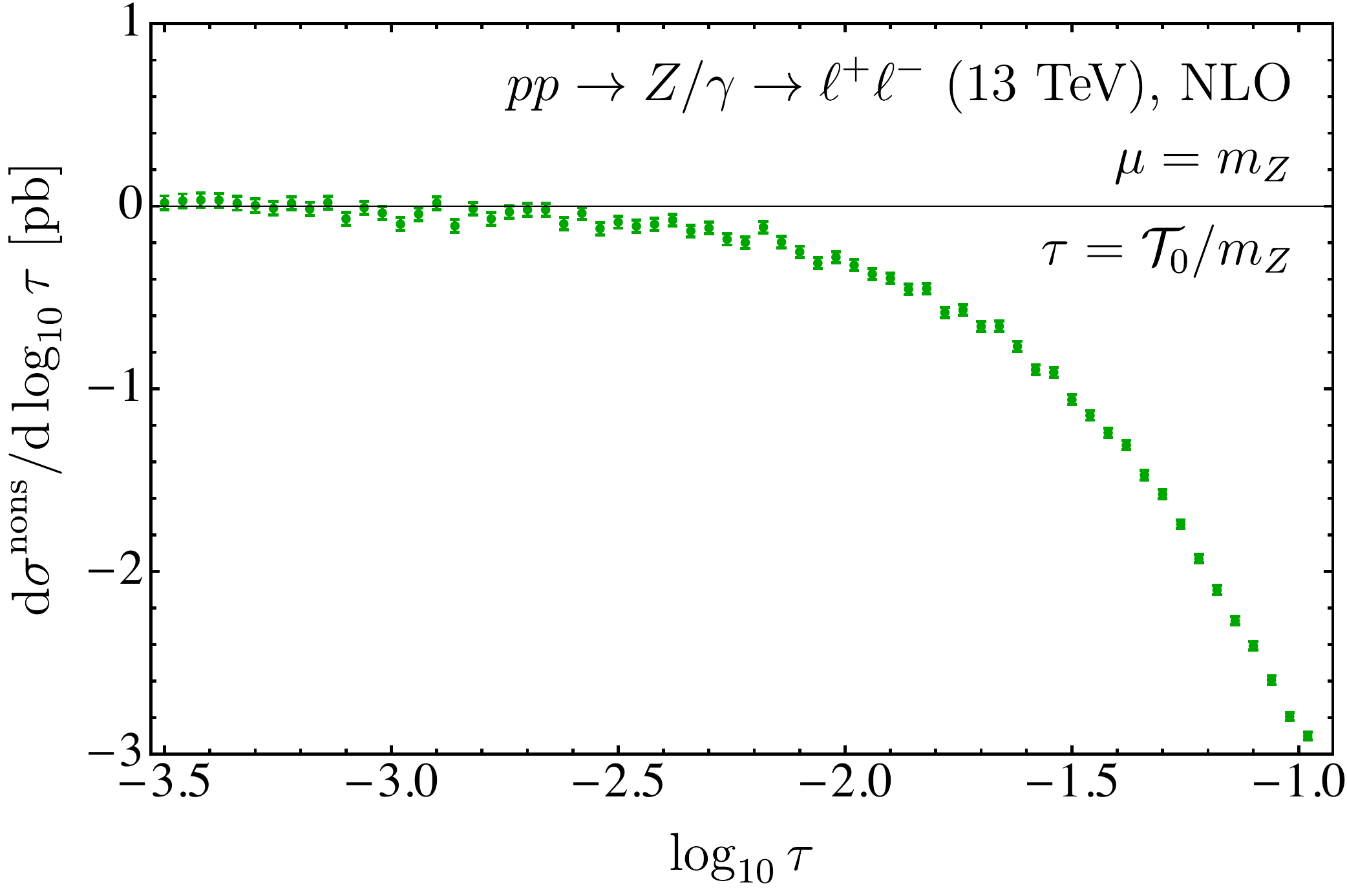}\hfill%
\includegraphics[scale=0.35]{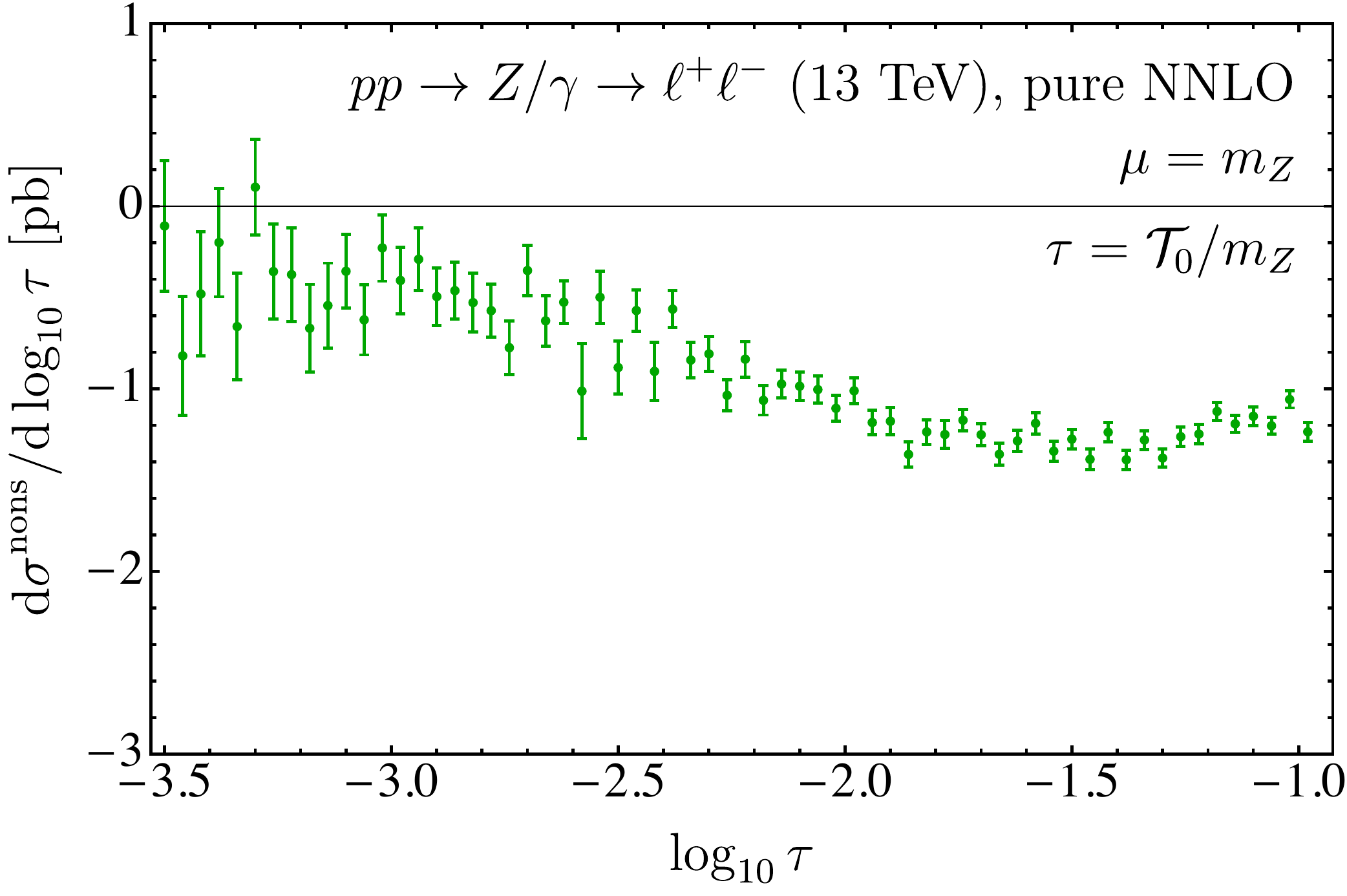}
\caption{The nonsingular $\Tau_0$ spectrum for Drell-Yan as a function of $\tau = \Tau_0/m_Z$. The NLO $\ord{\alpha_s}$ corrections are shown on the left, and the pure NNLO $\ord{\alpha_s^2}$ corrections are on the right.}
\label{fig:Znons}
\end{figure}

To illustrate our method with a nontrivial example, we consider the rapidity distribution of the vector boson in Drell-Yan production, $pp \to Z/\gamma \to \ell^+\ell^-$, and of the Higgs boson in gluon fusion, $gg \to H$, which are known to NNLO~\cite{Anastasiou:2003yy, Anastasiou:2003ds, Anastasiou:2004xq, Anastasiou:2005qj, Catani:2007vq, Grazzini:2008tf, Catani:2009sm}. Since the size of the perturbative corrections in the two cases are very different, they provide very useful and complementary test cases.

In both cases, 0-jettiness $\Tau_0$ is the resolution variable and all of the ingredients necessary to implement the $\Tau_0$-subtractions through NNLO are known. (We use the geometric measure with $\rho_i = 1$, see \eq{geometric}, which makes $\Tau_0$ identical to beam thrust.) The results are obtained for the LHC with a center-of-mass energy of $13\TeV$. We always use CT10 NNLO PDFs~\cite{Gao:2013xoa}. We choose common renormalization and factorization scales, $\mu_R = \mu_F = \mu$, with $\mu = m_Z$ for Drell-Yan production and $\mu = m_H$ for Higgs production. For the latter we use $m_H = 125\GeV$ and work in the top EFT limit. For the $Z+1$-jet and $H+1$-jet NLO calculations we use \mcfm~\cite{Campbell:2002tg, Campbell:2010ff}.

\begin{figure}[t!]
\centering
\includegraphics[scale=0.35]{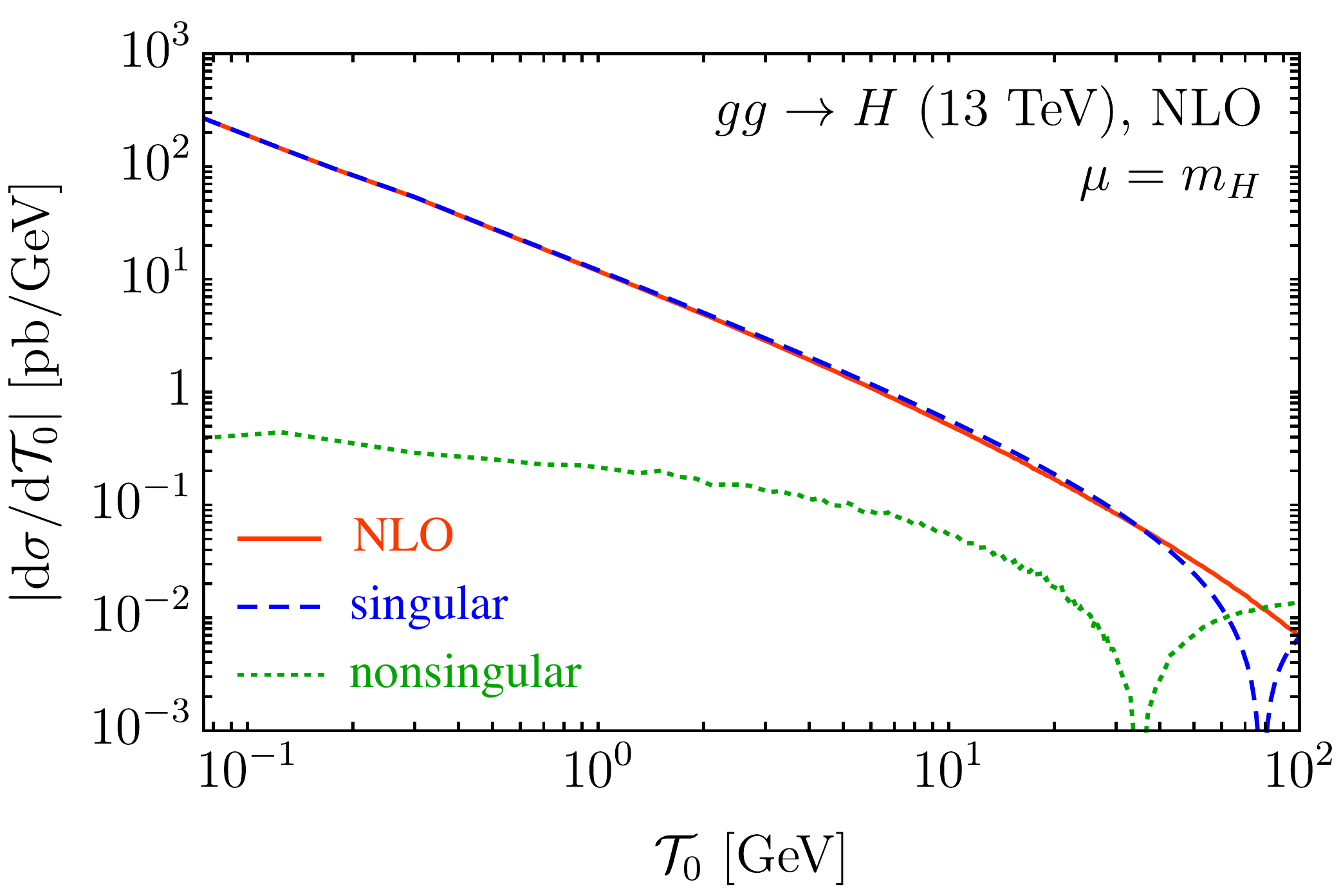}\hfill%
\includegraphics[scale=0.35]{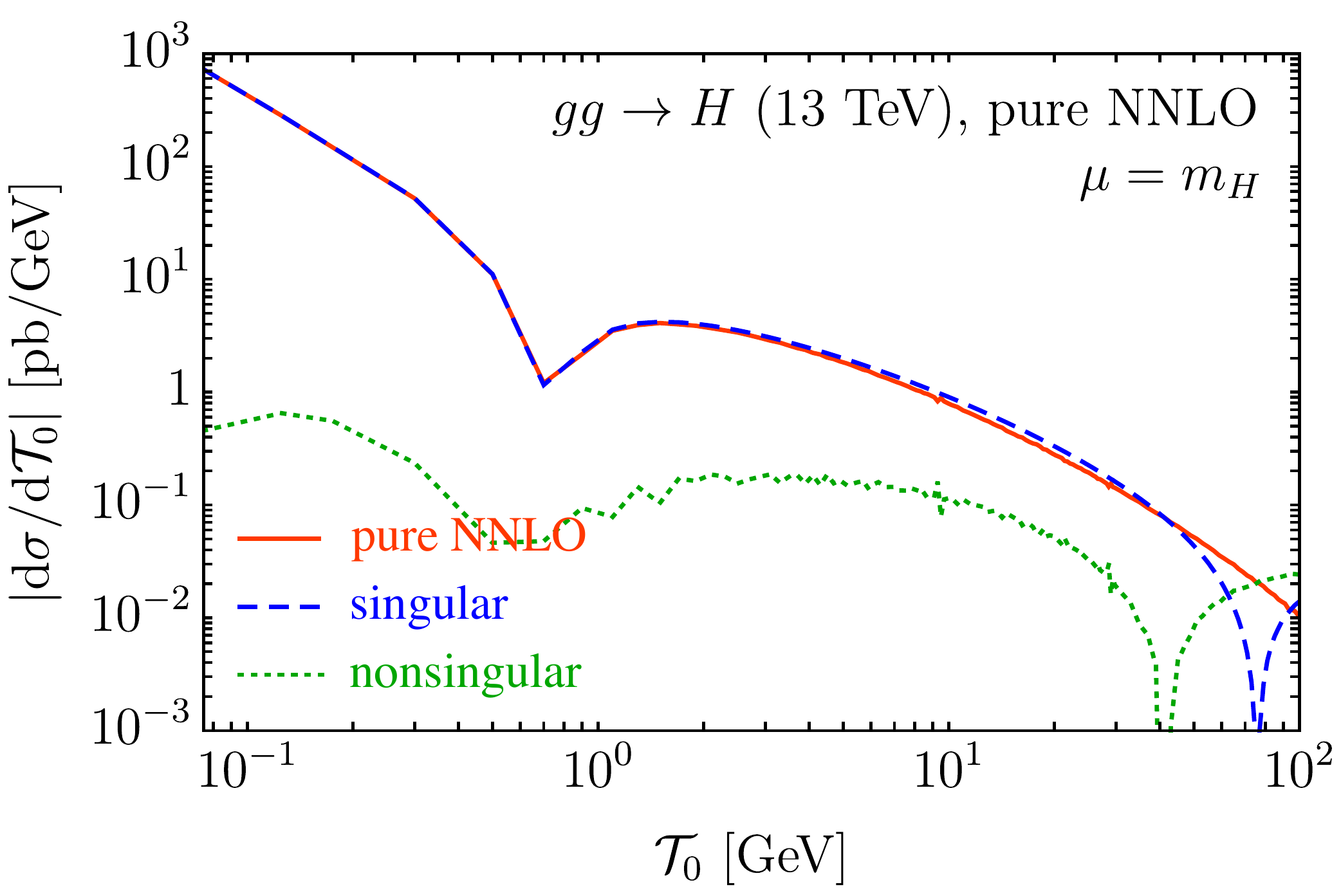}
\caption{The absolute value of the full, singular, and nonsingular contributions to the $\Tau_0$ spectrum for gluon-fusion Higgs production. The NLO $\ord{\alpha_s}$ corrections are shown on the left, and the pure NNLO $\ord{\alpha_s^2}$ corrections are on the right.}
\label{fig:Hpieces}
\end{figure}

\begin{figure}[t!]
\centering
\includegraphics[scale=0.35]{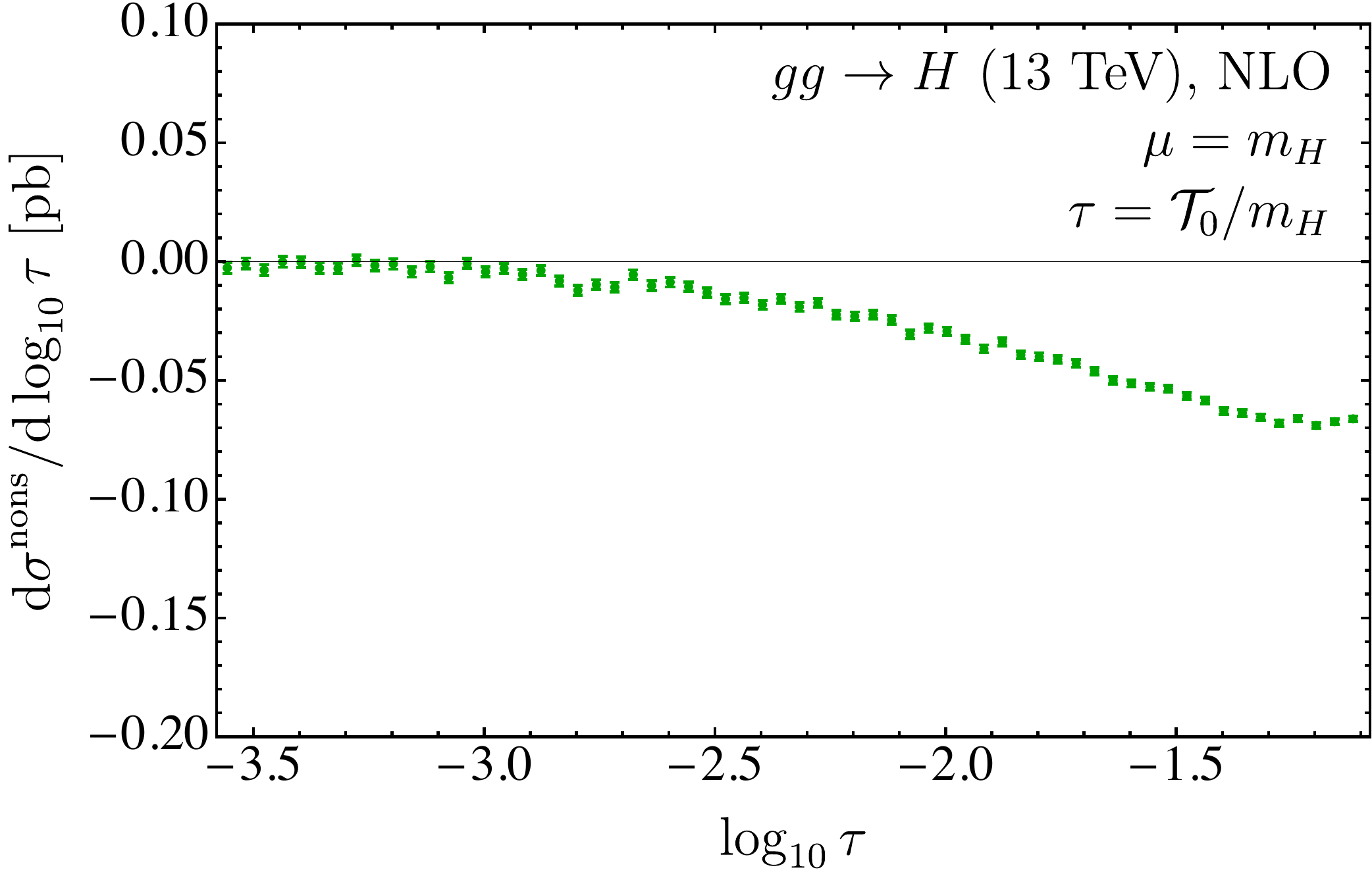}\hfill%
\includegraphics[scale=0.35]{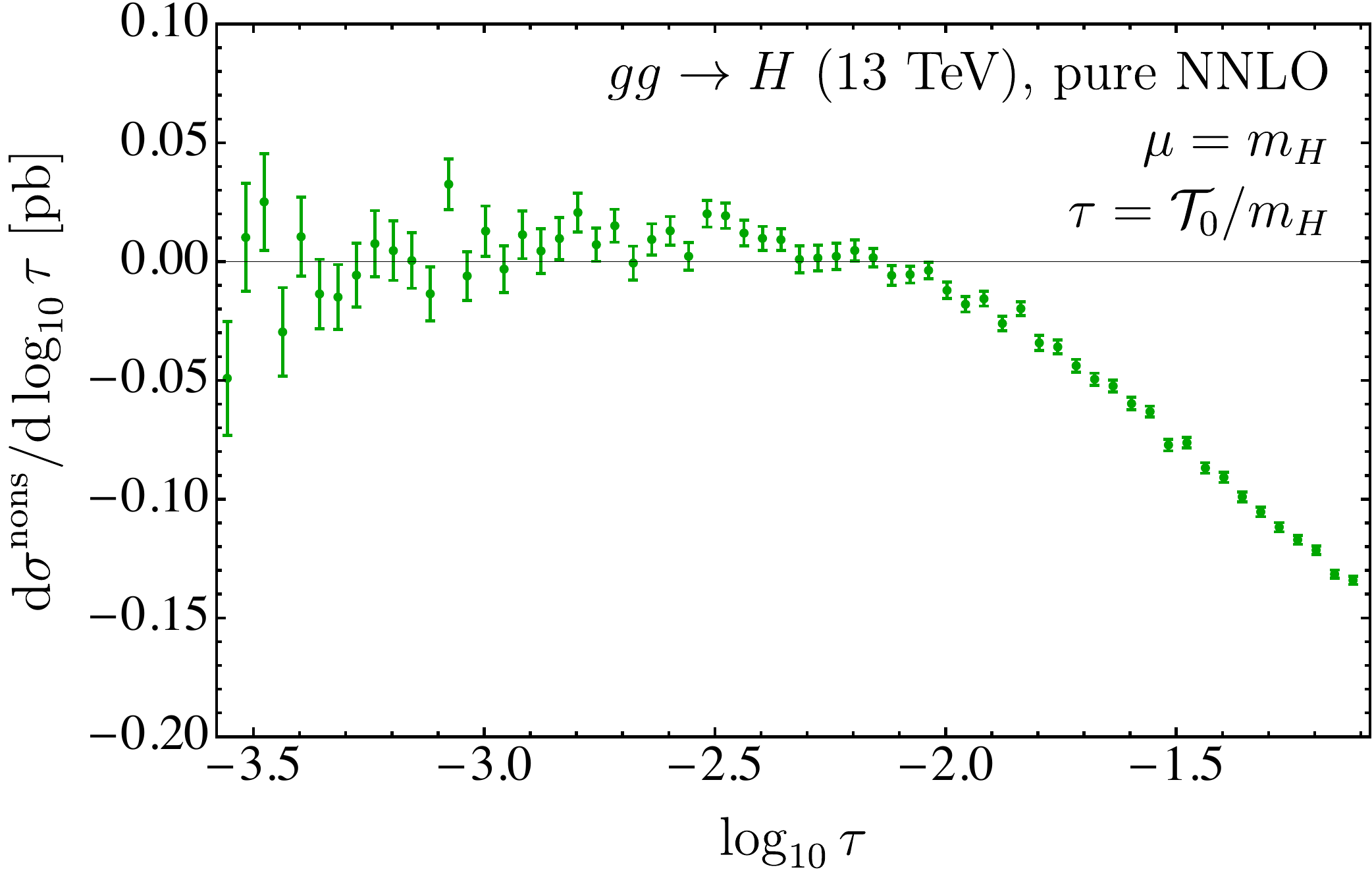}
\caption{The nonsingular $\Tau_0$ spectrum for gluon-fusion Higgs production as a function of $\tau = \Tau_0/m_H$. The NLO $\ord{\alpha_s}$ corrections are shown on the left, and the pure NNLO $\ord{\alpha_s^2}$ corrections are on the right.}
\label{fig:Hnons}
\end{figure}

An important validation of the $N$-jettiness subtractions is to confirm that the singular $\TauN$ spectrum is correctly describing the $\TauN\to 0$ singularities of the full QCD result.  This is done by calculating the nonsingular $\TauN$ spectrum as in \eq{nonsingular} as the difference of the full QCD and singular $\TauN$ spectra.  The decomposition of the $\Tau_0$ spectrum into singular and nonsingular components is shown in \figs{Zpieces}{Hpieces} for Drell-Yan and Higgs production, respectively, where we separately show the $\ord{\alpha_s}$ (NLO) and $\ord{\alpha_s^2}$ (pure NNLO) corrections, counted relative to the LO Born cross section.
(We plot the magnitudes of the contributions on a logarithmic scale, and the dips at large $\Tau_0$ and around $\Tau_0=1\GeV$ are due to the spectra going through 0. The small jitters in the pure NNLO nonsingular are due to numerical inaccuracies.) One can clearly see the large numerical cancellations between the full and singular results for small $\Tau_0$, where the nonsingular spectrum is several orders of magnitude smaller than the full and singular spectra.

As shown in \eq{TauNsubtraction}, the nonsingular spectrum is precisely the quantity that one integrates numerically when using the differential $\TauN$-subtractions. The fact that the nonsingular only contains integrable singularities is seen in \figs{Zpieces}{Hpieces} by its smaller slope toward $\Tau_0\to 0$. To check explicitly that the subtractions work and the nonsingular does not contain any $1/\TauN$ singularities, we consider the distribution $\df\sigma^\nons / \df \ln\tau = \tau \df\sigma^\nons / \df\tau$, which must go to 0 in the $\TauN\to 0$ limit. We plot it in \fig{Znons} for Drell-Yan and in \fig{Hnons} for Higgs production, again separately for the NLO and pure NNLO corrections, and using $\tau = \Tau_0 / m_Z$ and $\tau = \Tau_0 / m_H$, respectively. One can see that $\df\sigma^\nons / \df\ln \tau \to 0$ for $\tau \to 0$, as it must. The error bars come from the statistical integration uncertainties in the full result obtained from \mcfm. The numerical uncertainties in the singular result are negligible in comparison.

To obtain results for the NNLO rapidity spectrum, we use the simple $\Tau_0$-slicing in \eq{sigmaNNLOslicing}. As explained earlier, the missing $\ord{\delIR}$ contributions due to the $\TauIR$ cutoff are the same irrespective of how the subtractions are implemented. At NLO, we use $\TauIR = 0.03 \GeV$ ($\delIR \approx 3.2 \times 10^{-4}$ for Drell-Yan and $\delIR = 2.4 \times 10^{-4}$ for Higgs) and at NNLO we use $\TauIR = 0.1 \GeV$ ($\delIR \approx 1.1 \times 10^{-3}$ for Drell-Yan and $\delIR = 8 \times  10^{-4}$ for Higgs). These values are at the lower end of $\tau$ values plotted in \figs{Znons}{Hnons}, and are mainly limited by the \mcfm statistics.

Specifically, we use \mcfm to compute the NLO cross section for $\Tau_0 > \TauIR$ in the second term in \eq{sigmaNNLOslicing},
\begin{equation}
\int\!\df\Phi_{1}\, \frac{\df\sigma^\NLO(Y, \TauIR)}{\df\Phi_{1}}
\end{equation}
in bins of $Y$ for the processes $pp \to Z/\gamma \to \ell^+ \ell^- + \text{jet}$ and $pp \to H + \text{jet}$. Since there are no cuts on the final-state jets other than the requirement $\Tau_0 > \TauIR$, for small $\TauIR$ the calculation probes deep into the singular region and care must be taken to obtain reliable and numerically stable results.  This is combined with our own implementation of the NNLO singular cross section for $\Tau_0 < \TauIR$, $\sigma^{\sing}(Y, \TauIR)$, in the first term of \eq{sigmaNNLOslicing}.

\begin{figure}
\centering
 \includegraphics[scale=0.6]{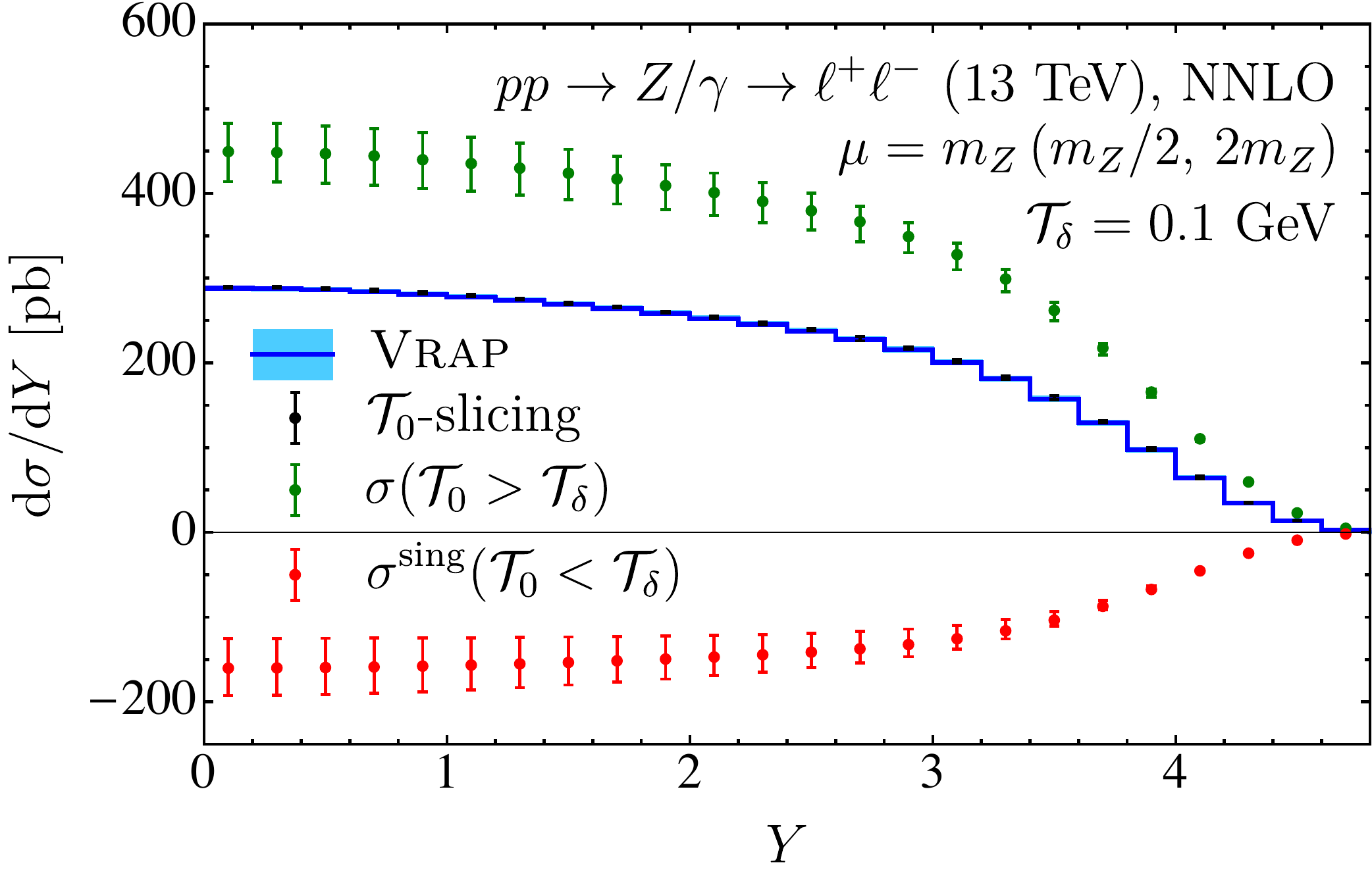}
\caption{The NNLO rapidity distribution in Drell-Yan production.  We plot the various ingredients in the $\Tau_0$-slicing method for $\TauIR = 0.1\GeV$, where in all cases the error bars correspond to the up and down scale variation. The blue histogram shows for comparison the NNLO result from \vrap.}
\label{fig:ZYdist}
\end{figure}

The results for the rapidity spectra are shown in \figs{ZYdist}{HYdist} for Drell-Yan and Higgs production, respectively. The two contributions from $\Tau_0 < \TauIR$ and $\Tau_0 > \TauIR$ are shown in red and green and the total result given by their sum in black. The error bars here show the scale variations up and down by a factor of two. Note that the relative size of the two contributions and the degree of cancellation between them can change significantly as the scale (or $\TauIR$ value) is changed. To validate the results from $\Tau_0$-slicing method, we compare to results from \vrap~\cite{Anastasiou:2003yy, Anastasiou:2003ds} for Drell-Yan and from \hnnlo~\cite{Catani:2007vq, Grazzini:2008tf} for Higgs, which are shown by the blue line and band. For both processes we find excellent agreement.

\begin{figure}
\includegraphics[scale=0.35]{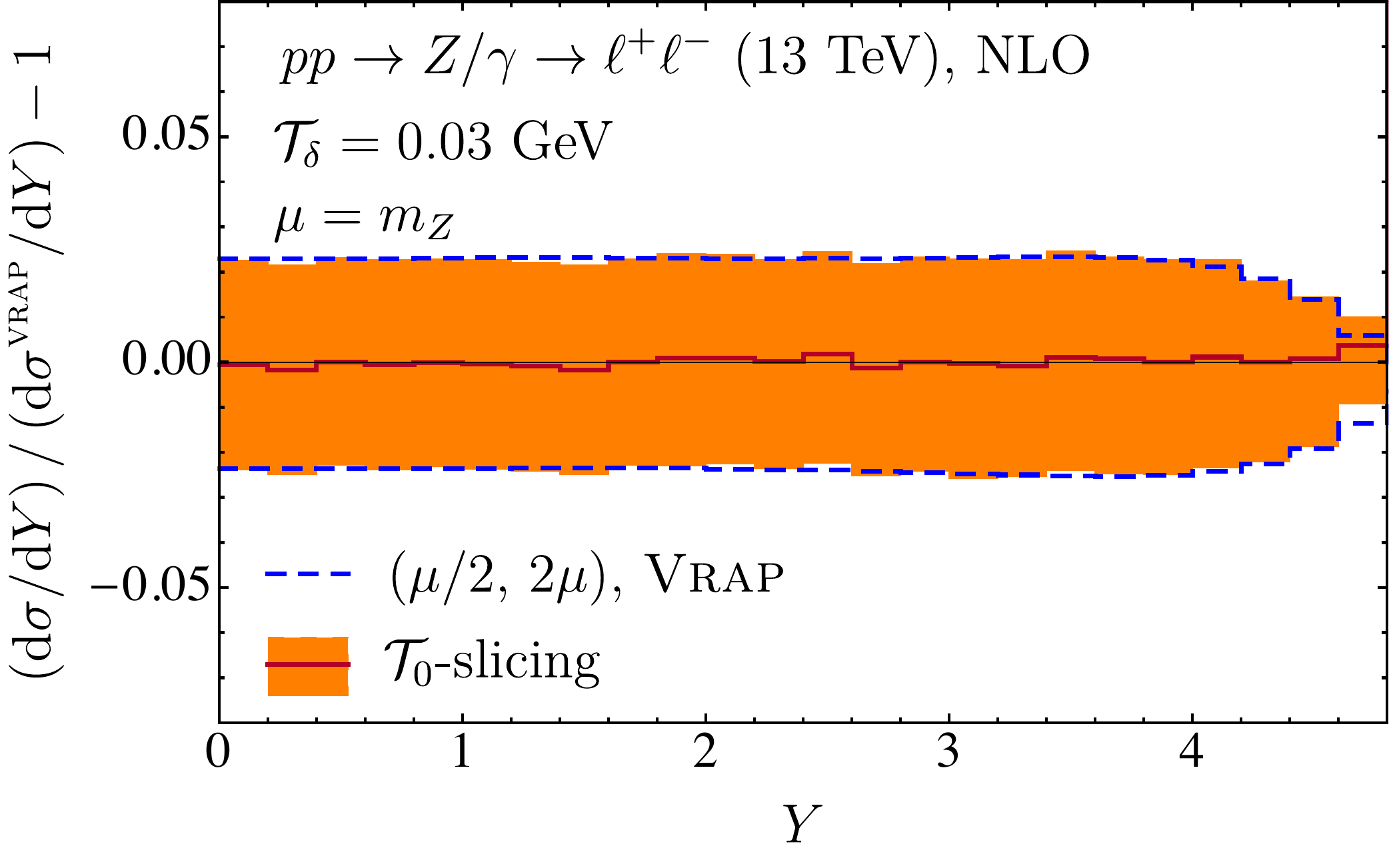}\hfill%
\includegraphics[scale=0.35]{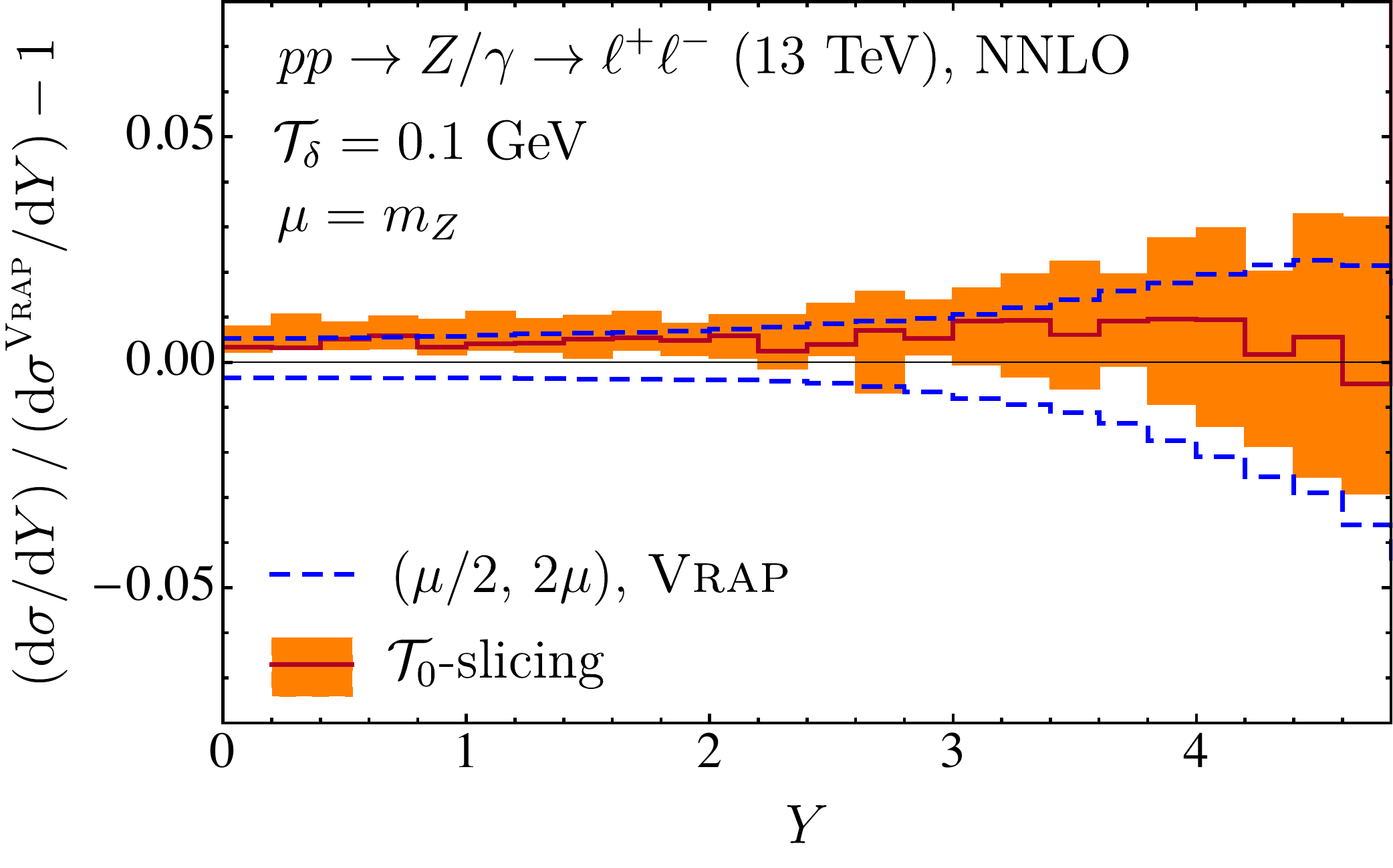}
\caption{The scale uncertainty band in the Drell-Yan rapidity distribution for both \vrap and $\Tau_0$-slicing, relative to the central scale from \vrap at NLO (right) and NNLO (left).}
\label{fig:Zfracerr}
\end{figure}

\begin{figure}
\centering
\includegraphics[scale=0.6]{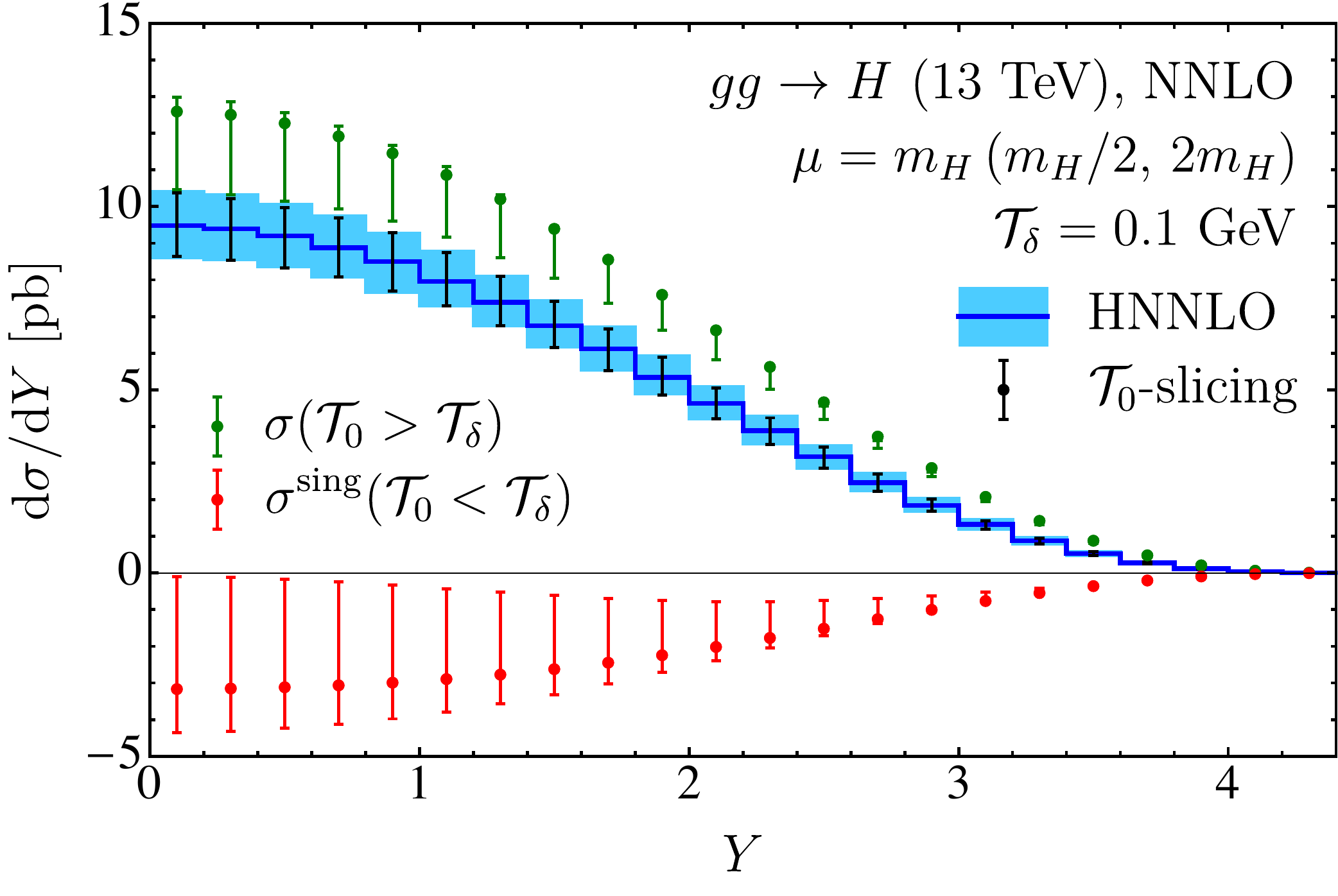}
\caption{The NNLO rapidity distribution in $gg\to H$ production. We plot the various ingredients in the $\Tau_0$-slicing method for $\TauIR = 0.1\GeV$, where in all cases the error bars correspond to the up and down scale variation. The blue histogram shows for comparison the NNLO result from \hnnlo.}\label{fig:HYdist}
\end{figure}

\begin{figure}
\includegraphics[scale=0.35]{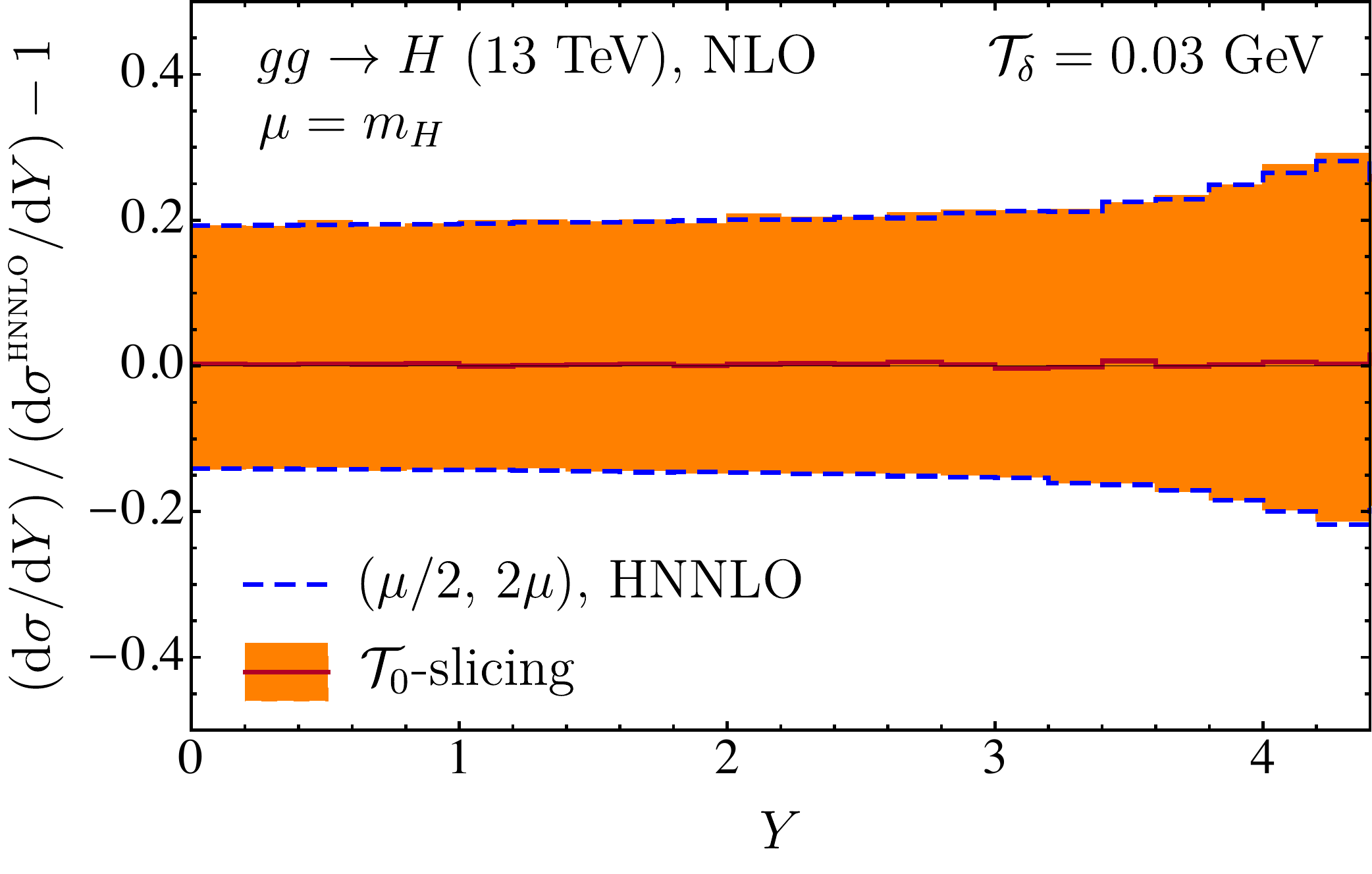}\hfill%
\includegraphics[scale=0.35]{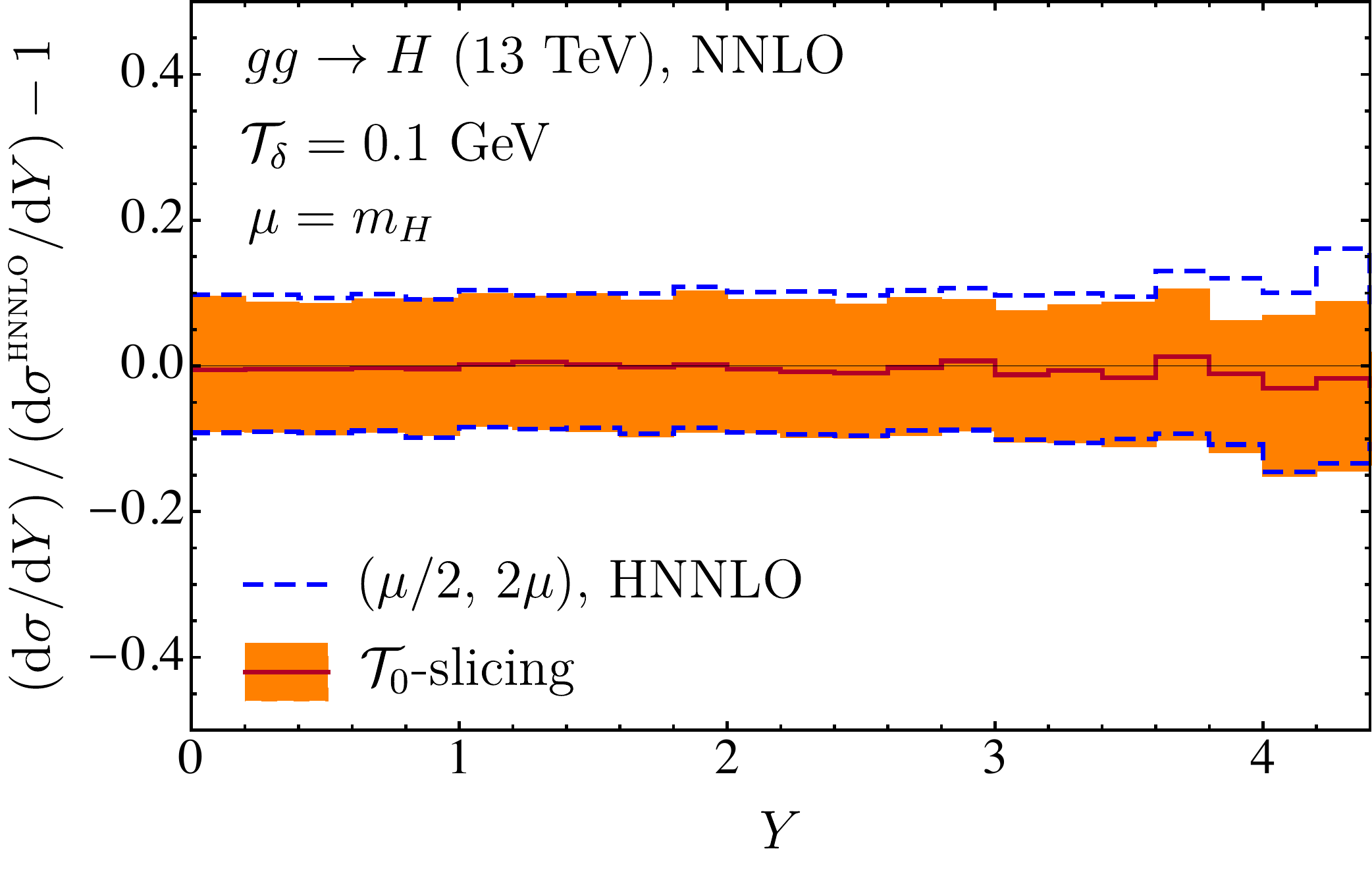}
\caption{The scale uncertainty band in the Higgs rapidity distribution for both \hnnlo and $\Tau_0$-slicing, relative to the central scale from \hnnlo at NLO (right) and NNLO (left).}
\label{fig:Hfracerr}
\end{figure}

In \figs{Zfracerr}{Hfracerr} we show the fractional difference of the $\Tau_0$-slicing results relative to \vrap and \hnnlo, respectively. At NLO with $\TauIR = 0.03\GeV$, the agreement is excellent. For Drell-Yan at NNLO, there is a small offset between the two results visible in \fig{Zfracerr}, representing $0.4\%$ of the total cross section. A similar offset of  $-0.2\%$ is also present in the Higgs case, but hardly visible because the scale variations are much larger. It is due to the missing $\ord{\delIR}$ nonsingular terms for $\Tau_0 < \TauIR = 0.1 \GeV$. A smaller value of $\TauIR$ would be needed to reduce this effect.  The size of the missing nonsingular terms we observe here is consistent with their expected size from our estimates in \subsubsec{accuracy}. Nevertheless, it is actually encouraging to see that even with the simple $\Tau_0$-slicing we are able to obtain this level of agreement. We would expect that an implementation of the differential $\Tau_0$-subtractions will allow one to use $\delIR$ values well below $10^{-4}$.

We conclude this discussion by noting that it is important, particularly for more complex processes, to carefully quantify the size of the neglected $\ord{\delIR}$ nonsingular contributions. In particular, as already seen in \fig{deltasigmaIR}, one cannot draw any conclusions for their possible size at NNLO from knowing their size at NLO. Also, the difference in the result when varying the $\TauIR$ value is not necessarily a good estimate of the absolute size of the missing nonsingular terms, because as discussed in \subsubsec{accuracy}, their scaling with $\delIR$ for $\delIR\to 0$ is much weaker than linear. A crucial check one should perform is to plot the nonsingular distribution as in \figs{Znons}{Hnons} and check its convergence toward zero.

\section{Conclusions}
\label{sec:conclusions}

Higher-order computations in QCD require the use of some subtraction technique that allows one to extract the collinear and soft phase-space divergences from the real-emission diagrams, and cancel these against the explicit divergences from the virtual loop diagrams. We explained how a subtraction scheme can be constructed using an IR safe $N$-jet resolution variable to control the approach to the IR-singular limit. The $N$-jettiness observable $\TauN$ is ideally suited for this task due to its known and simple factorization properties. Our resulting $N$-jettiness subtraction method is similar in spirit to the $q_T$ subtraction method introduced by Catani and Grazzini for color-singlet production, but may be applied to processes with arbitrarily many colored final-state partons (plus any color-singlet final state).

In our method, the subtraction term corresponds to the appropriate fixed-order expansion of the singular $N$-jettiness cross section, which can be efficiently computed using SCET. In this context, SCET allows the subtraction term to be broken down into various pieces (beam, jet, and soft functions) that are easier to compute, with the beam and jet functions being reusable for processes with any number of jets. The extension to N$^3$LO is possible and requires the calculation of the beam, jet, and soft functions at three-loop order.

We discussed in depth the details of the subtraction procedure, giving explicitly the equations and ingredients needed to construct the $\Tau_N$-subtraction terms at NLO and NNLO. The only ingredient which is not explicitly known is the $\mu$-independent constant term of the NNLO $N$-jet soft function for three or more $N$-jettiness axes. It can however be obtained relatively straightforward with existing technology. We also discussed how the $N$-jettiness subtractions can be implemented in practice. To demonstrate the method and study some of its numerical aspects, we presented NNLO results for the Drell-Yan and Higgs rapidity spectra computed using $0$-jettiness subtractions in its simplest form as a slicing method. The slicing method has been previously shown to be successful for NNLO
computations in \mycite{Gao:2012ja} and very recently in \mycites{Boughezal:2015dva, Boughezal:2015aha}. Given the viability of the $\TauN$-slicing, it will be very interesting to extend the implementations to the differential $N$-jettiness subtractions.

We have also suggested and discussed several different ways in which the numerical convergence of the $N$-jettiness subtraction method can be systematically improved. One option would be to include the leading nonsingular terms in the subtraction. These corrections are described by subleading factorization theorems for $N$-jettiness and SCET offers a systematic framework to compute them. Another way to improve the numerical convergence would be to make the subtraction more local, by performing the subtraction differentially in additional observables (such as $p_T$) and/or splitting the total $N$-jettiness observable into its components in the jet and beam regions. Much of the recent work in SCET on deriving factorization formulae for multi-differential cross sections can be very useful in this direction.

We only explicitly discussed the case of massless partons here. The construction of analogous $\TauN$-subtractions for processes involving massive quarks is possible with the same techniques. For $m_q \ll Q$, one would consider a massive quark jet with its own $N$-jettiness axis making use of the tools in SCET developed for the treatment of massive collinear quarks~\cite{Leibovich:2003jd, Fleming:2007qr, Fleming:2007xt, Jain:2008gb, Gritschacher:2013tza, Pietrulewicz:2014qza}. For $Q\sim m_q$, e.g. $t\bar{t}$ pair production and similar processes, an analogous approach to \mycites{Zhu:2012ts, Catani:2014qha} can be used. This amounts to treating the heavy quarks as part of the hard interaction (without its own $N$-jettiness axis) together with a more complicated soft function to account for soft gluon emissions from the heavy quarks. We leave further development in this direction to future work.

\begin{acknowledgments}
We thank Kirill Melnikov, Iain Stewart, and Fabrizio Caola for discussions and comments on the manuscript.
JRG and MS thank the theory group at LBL for hospitality during part of this work.
This work was supported by the DFG Emmy-Noether Grant No. TA 867/1-1
and by the Office of Science, Office of High Energy Physics,
of the U.S. Department of Energy (DOE) under Contract No. DE-AC02-05CH11231.
This research used resources of the National Energy Research Scientific Computing Center, which
is supported by the Office of Science of the DOE under Contract No. DE-AC02-05CH11231.

\end{acknowledgments}

\appendix

\section{Subtraction Ingredients}
\label{app:Ingredients}

We write the $\alpha_s$ expansion of the QCD beta function and the cusp and noncusp anomalous dimensions as
\begin{equation}
\mu\frac{\df}{\df\mu}\alpha_s(\mu) = \beta[\alpha_s(\mu)]
\,,\qquad
\beta(\alpha_s) =
- 2 \alpha_s \sum_{n=0}^\infty \beta_n\Bigl(\frac{\alpha_s}{4\pi}\Bigr)^{n+1}
\,,\end{equation}
and
\begin{equation} \label{eq:Gacusp}
\Gamma_\cusp(\alpha_s) = \sum_{n=0}^\infty \Gamma_n \Bigl(\frac{\alpha_s}{4\pi}\Bigr)^{n+1}
\,,\qquad
\gamma^i_F(\alpha_s) = \sum_{n=0}^\infty \gamma_{F\,n}^i \Bigl(\frac{\alpha_s}{4\pi}\Bigr)^{n+1}
\,.\end{equation}
The coefficients of the $\overline{\mathrm{MS}}$ beta function and cusp anomalous dimensions we need are
\begin{align} \label{eq:betaexp}
\beta_0 &= \frac{11}{3}\,C_A -\frac{4}{3}\,T_F\,n_f
\,,\nn\\
\beta_1 &= \frac{34}{3}\,C_A^2  - \Bigl(\frac{20}{3}\,C_A\, + 4 C_F\Bigr) T_F\,n_f
\,,\end{align}
and
\begin{align} \label{eq:Gacuspexp}
\Gamma^q_n &= C_F \Gamma_n
\,,\qquad
\Gamma^g_n = C_A \Gamma_n
\,,\nn\\
\Gamma_0 &= 4
\,,\nn\\
\Gamma_1
&= 4 \Bigl[ C_A \Bigl( \frac{67}{9} - \frac{\pi^2}{3} \Bigr)  - \frac{20}{9}\,T_F\, n_f \Bigr]
= \frac{4}{3} \bigl[ C_A (4 - \pi^2) + 5 \beta_0 \bigr]
\,.\end{align}

For the quark jet and beam functions in $\overline{\mathrm{MS}}$ we have~\cite{Becher:2006mr, Stewart:2010qs}
\begin{align} \label{eq:gammaBqexp}
\gamma_{J\,0}^q = \gamma_{B\,0}^q &= 6 C_F
\,, \nn \\
\gamma_{J\,1}^q = \gamma_{B\,1}^q
&= C_F \Bigl[
   C_A \Bigl(\frac{146}{9} - 80 \zeta_3\Bigr)
   + C_F (3 - 4 \pi^2 + 48 \zeta_3)
   + \beta_0 \Bigl(\frac{121}{9} + \frac{2\pi^2}{3} \Bigr) \Bigr]
\,.\end{align}
For the gluon jet and beam functions in $\overline{\mathrm{MS}}$ we have~\cite{Fleming:2003gt, Becher:2009th, Berger:2010xi}
\begin{align} \label{eq:gammaBgexp}
\gamma_{J\,0}^g = \gamma_{B\,0}^g &= 2 \beta_0
\,,\nn\\
\gamma_{J\,1}^g = \gamma_{B\,1}^g
&= C_A \Bigr[
   C_A \Bigl(\frac{182}{9} - 32\zeta_3\Bigr)
   + \beta_0 \Bigl(\frac{94}{9}-\frac{2\pi^2}{3}\Bigr) \Bigr]
   + 2\beta_1
\,. \end{align}

\subsection{Jet function}

We write the $\alpha_s$ expansion of the quark ($i = q$) and gluon ($i = g$) jet functions as
\begin{equation}
\label{eq:Jiexp}
J_i(s, \mu) = \sum_{n=0}^\infty \Bigl(\frac{\alpha_s(\mu)}{4\pi} \Bigr)^n\, J_i^{(n)}(s, \mu)
\,.\end{equation}
The coefficients have the form
\begin{align}
\label{eq:jetcoeffs}
J_i^{(m)}(s, \mu)
= J_{i,-1}^{(m)}\, \delta(s) + \sum_{n = 0}^{2m-1} J_{i,n}^{(m)}\, \frac{1}{\mu^2} \cL_{n}\Bigl(\frac{s}{\mu^2}\Bigr)
\,,\end{align}
where the $\cL_n(x)$ are plus distributions as defined in \eq{cLn}. The jet function is naturally a distribution in $s/\mu^2$, and this is the only $\mu$ dependence of the coefficients. Rescaling the arguments of the distributions using \eq{cLn_rescale}, we have
\begin{align} \label{eq:Jrescale}
J_i^{(m)}(Q_i k_i, \mu)
&= \frac{1}{Q_i}\, J_{i,-1}^{(m)}\Bigl(\frac{Q\xi}{\mu^2}\Bigr)\, \delta(k_i)
+ \frac{1}{Q_i} \sum_{n=0}^{2m-1} J_{i,n}^{(m)}\Bigl(\frac{Q_i \xi}{\mu^2}\Bigr)\, \frac{1}{\xi} \cL_n\Bigl(\frac{k_i}{\xi}\Bigr)
\,, \nn \\
J_{i,-1}^{(m)}(\la) &=  J_{i,-1}^{(m)} + \sum_{n = 0}^{2m-1} J_{i,n}^{(m)}\,\frac{\ln^{n+1}\!\la}{n+1}
\,,\nn\\
J_{i,n}^{(m)}(\la) &= J_{i,n}^{(m)} + \sum_{k = 1}^{2m-1-n} \frac{(n+k)!}{n!\,k!}\, J_{i,n+k}^{(m)}\,\ln^k\! \la
\,,\end{align}
where $\xi$ is an arbitrary dimension-one parameter, which exactly cancels between the different rescaled coefficients and that we can choose at our convenience. The $J_{i,n}^{(m)}(\la)$ are the coefficients appearing in the explicit expressions for the subtraction terms in \subsubsecs{NLOsub}{NNLOsub}.

The jet function coefficients in \eq{jetcoeffs} read up to two loops
\begin{align} \label{eq:Jn}
J_{i,1}^\one &= \Gamma_0^i
\,,\nn\\
J_{i, 0}^\one &= - \frac{\gamma^i_{J\,0}}{2}
\,,\nn\\
J_{i,3}^\two &= \frac{(\Gamma_0^i)^2}{2}
\,,\nn\\
J_{i,2}^\two &=
-\frac{\Gamma_0^i}{2} \Bigl(\frac{3\gamma^i_{J\,0}}{2} + \beta_0 \Bigr)
\,,\nn\\
J_{i,1}^\two &=
\Gamma_1^i - (\Gamma_0^i)^2 \frac{\pi^2}{6}
+ \frac{\gamma^i_{J\,0}}{2} \Bigl( \frac{\gamma^i_{J\,0}}{2} + \beta_0\Bigr)
+ \Gamma_0^i\, J_{i,-1}^\one
\,,\nn\\
J_{i,0}^\two
&= (\Gamma_0^i)^2 \zeta_3 + \Gamma_0^i \gamma^i_{J\,0}\, \frac{\pi^2}{12} - \frac{\gamma^i_{J\,1}}{2}
- \Bigl(\frac{\gamma^i_{J\,0}}{2} + \beta_0\Bigr) J_{i,-1}^\one 
\,.\end{align}
The $\delta(s)$ pieces for the quark jet function are~\cite{Bauer:2003pi, Becher:2006qw}
\begin{align}
J_{q,-1}^\zero &= 1
\,, \nn \\
J_{q,-1}^\one &= C_F (7 - \pi^2)
\,, \nn \\
J_{q,-1}^\two
&= C_F \biggl[
   C_F \Bigl(\frac{205}{8} - \frac{67\pi^2}{6} + \frac{14\pi^4}{15} - 18\zeta_3 \Bigr)
   + C_A \Bigl(\frac{1417}{108} - \frac{7\pi^2}{9} - \frac{17\pi^4}{180} - 18\zeta_3 \Bigr)
\nn\\ & \quad
   + \beta_0 \Bigl(\frac{4057}{216} - \frac{17\pi^2}{9} - \frac{4\zeta_3}{3}\Bigr) \biggr]
\,,\end{align}
and for the gluon jet function they are~\cite{Fleming:2003gt, Becher:2009th, Becher:2010pd}
\begin{align}
J_{g,-1}^\zero &= 1
\,, \nn \\
J_{g,-1}^\one &= C_A \Bigl(\frac{4}{3} - \pi^2\Bigr) + \frac{5}{3}\, \beta_0
\,, \nn \\
J_{g,-1}^\two
&= C_A^2 \Bigl(\frac{4255}{108} - \frac{26 \pi^2}{9} + \frac{151 \pi^4}{180} - 72 \zeta_3\Bigr)
   + C_A\beta_0 \Bigl(-\frac{115}{108} - \frac{65 \pi^2}{18} + \frac{56 \zeta_3}{3} \Bigr) 
\\ \nn & \quad
   + \beta_0^2 \Bigl(\frac{25}{9} - \frac{\pi^2}{3}\Bigr)
   + \beta_1 \Bigl(\frac{55}{12} - 4 \zeta_3 \Bigr)
\,.\end{align}

\subsection{Beam function}

The beam function is given by~\cite{Stewart:2009yx, Stewart:2010qs}
\begin{equation}
B_i(t, x, \mu) = \sum_j \int\!\frac{\df z}{z}\,\cI_{ij}(t, z, \mu, \mu_F)\, f_j\Bigl(\frac{x}{z}, \mu_F\Bigr)
\,,\end{equation}
where $f_j(x, \mu_F)$ are the standard PDFs and $\cI_{ij}(t, z, \mu, \mu_F)$ are perturbative matching coefficients. Here, we have explicitly separated the $\mu_F$ dependence, which cancels between the matching coefficients and the PDFs, such that the beam function is $\mu_F$ independent up to higher orders in $\alpha_s(\mu)$. (Usually, one takes $\mu_F = \mu$ in the fixed-order beam function, since these are not really formally distinct scales.) For our purposes, the $\mu_F$ dependence in the beam function determines the complete $\mu_F$ factorization scale dependence in the singular fixed-order cross section, while the $\mu$ dependence contributes to the usual renormalization scale dependence.

We expand the beam function matching coefficients as
\begin{equation} \label{eq:Iijexp}
 \cI_{ij}(t,z,\mu, \mu_F)
 = \sum_{n=0}^{\infty}  \cI_{ij}^{(n)}(t,z,\mu, \mu_F)\, \Bigl(\frac{\alpha_s(\mu)}{4\pi}\Bigr)^n
\,.\end{equation}
The perturbative coefficients have the structure
\begin{align}
\label{eq:beamcoeffs}
\cI_{ij}^{(m)}(t, z, \mu, \mu_F)
= \cI_{ij,-1}^{(m)}\Bigl(z, \frac{\mu^2}{\mu_F^2}\Bigr)\, \delta(t)
   + \sum_{n = 0}^{2m-1} \cI_{ij,n}^{(m)}\Bigl(z, \frac{\mu^2}{\mu_F^2}\Bigr)\, \frac{1}{\mu^2} \cL_{n}\Bigl(\frac{t}{\mu^2}\Bigr)
\,,\end{align}
where the $\cL_n(x)$ are the plus distributions defined in \eq{cLn}. The beam function is naturally a distribution in $t/\mu^2$. Rescaling the arguments of the distributions using \eq{cLn_rescale}, we have
\begin{align} \label{eq:Iijrescale}
\cI_{ij}^{(m)}(Q k, z, \mu, \mu_F)
&= \frac{1}{Q}\, \cI_{ij,-1}^{(m)}\Bigl(z, \frac{\mu^2}{\mu_F^2}, \frac{Q\xi}{\mu^2}\Bigr)\, \delta(k)
+ \frac{1}{Q} \sum_{n=0}^{2m-1} \cI_{ij,n}^{(m)}\Bigl(z, \frac{\mu^2}{\mu_F^2}, \frac{Q \xi}{\mu^2}\Bigr)\, \frac{1}{\xi} \cL_n\Bigl(\frac{k}{\xi}\Bigr)
\,, \nn \\
\cI_{ij,-1}^{(m)}(z, \la_F, \la) &=  \cI_{ij,-1}^{(m)}(z, \la_F) + \sum_{n = 0}^{2m-1} \cI_{ij,n}^{(m)}(z, \la_F)\,\frac{\ln^{n+1}\!\la}{n+1}
\,,\nn\\
\cI_{ij,n}^{(m)}(z, \la_F, \la) &= \cI_{ij,n}^{(m)}(z, \la_F) + \sum_{k = 1}^{2m-1-n} \frac{(n+k)!}{n!\,k!}\, \cI_{ij,n+k}^{(m)}(z, \la_F)\,\ln^k\!\la
\,,\end{align}
where $\xi$ is an arbitrary dimension-one parameter, which exactly cancels between the different rescaled coefficients and that we can choose at our convenience. From these coefficients we also define the corresponding beam function coefficients as
\begin{equation}
B_{i,n}^{(m)}(x, \mu,\mu_F, \la) = \sum_j\int\!\frac{\df z}{z}\, \cI_{ij,n}^{(m)}\Bigl(z, \frac{\mu^2}{\mu_F^2}, \la \Bigr)\, f_j\Bigl(\frac{x}{z},\mu_F\Bigr)
\,,\end{equation}
which are the coefficients appearing in the explicit expressions for the subtraction terms in \subsubsecs{NLOsub}{NNLOsub}.

The results for the coefficients in \eq{beamcoeffs} are as follows. At LO, we simply have
\begin{equation}
\cI_{ij, -1}^\zero(z,\la_F)
= \delta_{ij}\delta(1 - z)
\,.\end{equation}
The NLO coefficients have been computed in \mycites{Stewart:2009yx, Stewart:2010qs, Berger:2010xi}, and are given by
\begin{align} \label{eq:Iijone}
\cI_{ij,1}^\one(z,\la_F) &= \Gamma_0^i \, \delta_{ij}\delta(1 - z)
\,, \nn \\
\cI_{ij, 0}^\one(z,\la_F) &= - \frac{\gamma^i_{B\,0}}{2}\, \delta_{ij}\delta(1-z) + 2P_{ij}^\zero(z)
\,, \nn \\
\cI_{ij, -1}^\one(z, \la_F) &= 2 I^\one_{ij}(z) + \ln \la_F\,2P_{ij}^\zero(z)
\,.\end{align}
The NNLO coefficients have been computed in \mycites{Gaunt:2014xga, Gaunt:2014cfa}, and read
\begin{align} \label{eq:Iijtwo}
\cI_{ij,3}^\two(z,\la_F) &= \frac{1}{2}\, (\Gamma_0^i)^2\, \delta_{ij}\delta(1-z)
\,,\nn\\
\cI_{ij,2}^\two(z,\la_F) &=
\Gamma_0^i \Bigl[- \Bigl(\frac{3}{4} \gamma_{B\,0}^i + \frac{\beta_0}{2} \Bigr) \delta_{ij}\delta(1-z) + 3P^\zero_{ij}(z) \Bigr]
\,,\nn\\
\cI_{ij,1}^\two(z, \la_F) &=
\Bigl[\Gamma_1^i - (\Gamma_0^i)^2 \frac{\pi^2}{6}
   + \frac{\gamma^i_{B\,0}}{2} \Bigl( \frac{\gamma^i_{B\,0}}{2} + \beta_0\Bigr) \Bigr] \delta_{ij}\delta(1-z) 
   + 2\Gamma_0^i\, I^\one_{ij}(z)
\nn \\ & \quad
   - 2 (\gamma_{B\,0}^i + \beta_0) P^\zero_{ij}(z)
   + 4 \sum_k P^\zero_{ik}(z)\conv_z P^\zero_{kj}(z)
   + \ln \la_F\, 2 \Gamma_0^i\, P^\zero_{ij}(z)
\,,\nn\\
\cI_{ij,0}^\two(z, \la_F)
&= \Bigl[(\Gamma_0^i)^2 \zeta_3 + \Gamma_0^i \gamma_{B\,0}^i \frac{\pi^2}{12} - \frac{\gamma_{B\,1}^i}{2} \Bigr]
   \delta_{ij} \delta(1-z)
   - \Gamma_0^i \frac{\pi^2}{3} P^\zero_{ij}(z)
   - (\gamma_{B\,0}^i + 2\beta_0) I^\one_{ij}(z)
\nn \\ & \quad
   + 4 \sum_k I^\one_{ik}(z)\conv_z P^\zero_{kj}(z)
   + 4 P^\one_{ij}(z)
\nn \\ & \quad
   + \ln \la_F \Bigl[ -\gamma_{B\,0}^i \, P^\zero_{ij}(z)
   + 4 \sum_k P^\zero_{ik}(z)\conv_z P^\zero_{kj}(z) \Bigr]
\,,\nn\\
\cI_{ij,-1}^\two(z,\la_F)
&= 4 I^\two_{ij}(z)
   + \ln \la_F \Bigl[4 \sum_k I^\one_{ik}(z)\conv_z P^\zero_{kj}(z) + 4 P^\one_{ij}(z) \Bigr]
\nn \\ & \quad
   + \ln^2\! \la_F\, \Bigl[ \beta_0\, P^\zero_{ij}(z) + 2 \sum_k P^\zero_{ik}(z)\conv_z P^\zero_{kj}(z) \Bigr]
\,.\end{align}
Explicit results for the matching functions $I_{ij}^\one(z)$ and $I_{ij}^\two(z)$ as well as the splitting functions $P_{ij}^\zero(z)$ and $P_{ij}^\one(z)$ and all required convolutions between them can be found in \mycites{Gaunt:2014xga, Gaunt:2014cfa} in the same notation that we use here.

\subsection{Single-differential soft function}
\label{app:singsoft}

The single-differential $N$-jettiness soft function is related to the one of \eq{fact}, which is multi-differential in the soft contributions to the $\Tau_N^i$,  by
\begin{align} \label{eq:softsinglediff}
\hS_\kappa(k,\{\hat q_i\}, \mu)
= \int\! \Bigl[\prod_i \df k_i\Bigr]\, \delta \Bigl(k-\sum_i k_i \Bigr)\, \hS_\kappa(\{k_i\},\{\hat q_i\},\mu)
\,.\end{align}
Recall that the subscript $\kappa$ encodes the information on the Born partonic channel. For the soft function, it specifies the color space of the external partons in which it acts.

We expand the soft function in $\alpha_s(\mu)$ as
\begin{equation} \label{eq:softexp}
 \hS_\kappa(k,\{\hat q_i\},\mu) = \sum_n \Bigl(\frac{\alpha_s(\mu)}{4 \pi}\Bigr)^{\!n} \, \hS_\kappa^{(n)}(k,\{\hat q_i\},\mu)
\,,\end{equation}
where the perturbative coefficients can be written as
\begin{equation} \label{eq:softLexp}
\hS^{(m)}_\kappa(k,\{\hat q_i\}, \mu)
= \hS_{\kappa,-1}^{(m)}(\{\hat q_i\})\, \delta(k) + \sum_{n = 0}^{2m-1} \hS_{\kappa, n}^{(m)}(\{\hat q_i\})\, \frac{1}{\mu} \cL_{n}\Bigl(\frac{k}{\mu}\Bigr)
\,.\end{equation}
The soft function is naturally a distribution in $k/\mu$ and this is the only $\mu$ dependence of the coefficients. Rescaling the arguments of the plus distributions using \eq{cLn_rescale}, we have
\begin{align} \label{eq:Srescale}
\hS^{(m)}_\kappa(k,\{\hat q_i\}, \mu)
&= \hS_{\kappa,-1}^{(m)}\Bigl(\!\{\hat q_i\},\frac{\xi}{\mu}\Bigr)\, \delta(k)
+ \sum_{n=0}^{2m-1} \hS_{\kappa, n}^{(m)}\Bigl(\!\{\hat q_i\},\frac{\xi}{\mu}\Bigr)\, \frac{1}{\xi} \cL_n\Bigl(\frac{k}{\xi}\Bigr)
\,, \nn \\
\hS_{\kappa,-1}^{(m)}(\{\hat q_i\},\la)
&= \hS_{\kappa,-1}^{(m)}(\{\hat q_i\}) + \sum_{n = 0}^{2m-1} \hS_{\kappa, n}^{(m)}(\{\hat q_i\})\,\frac{\ln^{n+1}\!\la}{n+1}
\,,\nn\\
\hS_{\kappa, n}^{(m)}(\{\hat q_i\},\la)
&= \hS_{\kappa, n}^{(m)}(\{\hat q_i\}) + \sum_{k = 1}^{2m-1-n} \frac{(n+k)!}{n!\,k!}\, \hS_{\kappa, n+k}^{(m)}(\{\hat q_i\})\,\ln^k\! \la
\,.\end{align}
The dimension-one parameter $\xi$ is again arbitrary and exactly cancels between the coefficients. The coefficients $\hS_{\kappa, n}^{(m)}(\{\hat q_i\},\la)$ are those appearing in the explicit expressions for the subtraction terms in \subsubsecs{NLOsub}{NNLOsub}.
In the rest of this subsection the dependence on the jet axes $\hat q_i$ of the soft function and its anomalous dimension is always understood and we often suppress the explicit $\{\hq_i\}$ argument.

The renormalization scale dependence of the soft function is subject to the renormalization group equation derived in \mycite{Jouttenus:2011wh},
\begin{align} \label{eq:softRGE}
\mu \frac{\df}{\df \mu} \hS_\kappa (k,\mu)
&= \frac12 \int \!\df k^\prime\, \Bigl[ \hga_S(k-k^\prime) \, \hS_\kappa(k^\prime) + \hS_\kappa(k-k^\prime)\, \hga_S^\dagger(k^\prime) \Bigr]
\,,\end{align}
with the soft anomalous dimension
\begin{align}
\label{eq:softanomdim}
\hga_S(k,\mu)
&= 2 \Gamma_\cusp[\alpha_s(\mu)] \biggl\{ \frac{1}{\mu} \cL_0\Bigl(\frac{k}{\mu}\Bigr) \sum_i {\bf T}_i^2
+ \delta(k) \frac{1}{2}\sum_{i \ne j} \Tij \ln\bigl[(-1)^{\Delta_{ij}} \hat s_{ij} + \img 0\bigr] \biggr\}
\nn \\ & \quad
+ \hga_S[\alpha_s(\mu)]\,\delta(k)
\nn \\
&= \Gamma_\cusp[\alpha_s(\mu)] \biggl\{ 2\bC \frac{1}{\mu} \cL_0\Bigl(\frac{k}{\mu}\Bigr)
+ \delta(k) \bigl[\bL(\{\hs_{ij}\}) + \bI \bigr] \biggr\}
+ \hga_S[\alpha_s(\mu)]\,\delta(k)
\,.\end{align}
Here, $\Delta_{ij} = 1$ if the partons $i$ and $j$ are both incoming or both outgoing and $\Delta_{ij} = 0$ if one of them is incoming and the other one outgoing. The invariant
\begin{align}
\label{eq:shat}
\hs_{ij} \equiv \frac{2q_i\cdot q_j}{Q_i\, Q_j} = 2 \hq_i \cdot \hq_j
\end{align}
is always positive with our conventions and corresponds to an angular measure between any two partons (depending on the precise choice of the $Q_i$). Note that $\Gamma_\cusp(\alpha_s)$ here has the overall color factor removed, see \eqs{Gacusp}{Gacuspexp}. To write the last line in \eq{softanomdim}, we defined the abbreviations
\begin{align}
\bC &= \sum_i {\bf T}_i^2 = \id_\kappa \sum_i C_i
\qquad (\text{with } C_q = C_{\bar q} = C_F,\, C_g = C_A)
\,, \nn \\
\bL(\{\hs_{ij}\}) &\equiv
\sum_{i\ne j} \Tij\, \ln \hat s_{ij}
\,, \nn \\
\bI &\equiv
 \img\pi \sum_{i\ne j} \Tij\, \Delta_{ij}
= \img\pi \bigl[2(\bT_a + \bT_b)^2 -\bC \bigr]
\,.\end{align}
Note that for $ee$ and $ep$ collisions, $\bI$ is always proportional to $\id_\kappa$ and can be ignored, as it drops out of \eq{softRGE}. Similarly, for $pp$ collisions it can be ignored for $0$-jet and $1$-jet processes where the color space is still trivial.
Up to two loops the noncusp soft anomalous dimension is given by
\begin{align}
\hga^\kappa_S(\alpha_s) &= 0 + \bC\, \gamma_{S\,1} \Bigl(\frac{\alpha_s}{4\pi}\Bigr)^2 + \ord{\alpha_s^3}
\,, \nn \\
\gamma_{S\,1} &= C_A \Bigl(-\frac{64}{9} + 28 \zeta_3\Bigr) + \beta_0 \Bigl(-\frac{56}{9} + \frac{\pi^2}{3}\Bigr)
\,.\end{align}

The fixed-order coefficients in \eq{softLexp} are as follows. At leading order, we have
\begin{equation}
 \hS_{\kappa,-1}^\zero(\{\hat q_i\}) = \id_\kappa
\,.\end{equation}
The one-loop coefficients are given by~\cite{Jouttenus:2011wh}
\begin{align}
\hS_{\kappa,1}^\one(\{\hat q_i\}) &= -2 \Gamma_0 \, {\bf C}
\,, \nn\\
\hS_{\kappa,0}^\one(\{\hat q_i\}) &=  - \Gamma_0 \, \bL(\{\hs_{ij}\})
\,, \nn\\
\hS_{\kappa,-1}^\one(\{\hat q_i\}) &= \sum_{i\ne j} \Tij \Bigl[ \ln^2 \hat s_{ij} -\frac{\pi^2}{6} + 4 \sum_{m \ne i,j} \! I_{ij,m}(\{\hq_i\}) \Bigr]
\,,\end{align}
where
\begin{align}
I_{ij,m}(\{\hq_i\})
&= I_0 \Bigl(\frac{\hat s_{jm}}{\hat s_{ij}},\frac{\hat s_{im}}{\hat s_{ij}},\Big\{ \frac{\hat s_{jl}}{\hat s_{jm}},\frac{\hat s_{il}}{\hat s_{im}}, \phi_{lm} \Big\}_{l\ne i,j,m} \Bigr) \ln \frac{\hat s_{jm}}{\hat s_{ij}}
\nn \\ & \quad
+  I_1 \Bigl(\frac{\hat s_{jm}}{\hat s_{ij}},\frac{\hat s_{im}}{\hat s_{ij}},\Big\{ \frac{\hat s_{jl}}{\hat s_{jm}},\frac{\hat s_{il}}{\hat s_{im}}, \phi_{lm} \Big\}_{l\ne i,j,m} \Bigr)
\,.\end{align}
The $I_0$ and $I_1$ are finite phase-space integrals, which are required for three or more $N$-jettiness axes. They are not known fully analytically, but can be evaluated numerically for a given set  $\{\hq_i\}$. Their explicit expressions and an algorithm to reduce them to simple one-dimensional numerical integrals for arbitrary $N$ is provided in \mycite{Jouttenus:2011wh}. With only three $N$-jettiness axes, the integrals are still planar. Starting from four axes, the angles $\phi_{lm}$ also enter, which are the azimuthal angles between the $\hq_m$ and $\hq_l$ axes in the plane transverse to the $\hq_i$ and $\hq_j$ axes.

Iteratively solving the RGE in \eq{softRGE}, we obtain the two-loop coefficients
\begin{align}
\hS_{\kappa,3}^\two(\{\hat q_i\}) &= 2 \Gamma_0^2\,\bC^2
\,,\nn\\
\hS_{\kappa,2}^\two(\{\hat q_i\}) &= \Gamma_0\, \bC\, \bigl[3 \Gamma_0\, \bL + 2 \beta_0\bigr]
\,,\nn\\
\hS_{\kappa,1}^\two(\{\hat q_i\})
&= \Gamma_0^2 \Bigl( \bL^2 + \frac{1}{2} [\bI, \bL] - \frac{2 \pi^2}{3} \bC^2 \Bigr) + 2 \Gamma_0 \bigl( \beta_0 \bL - \bC\, \hS_{\kappa,-1}^\one(\{\hat q_i\}) \bigr) - 2 \Gamma_1{\bf C}
\,,\nn\\
\hS_{\kappa,0}^\two(\{\hat q_i\})
&= \Gamma_0^2\,\bC \Bigl( 4\bC\, \zeta_3 - \frac{\pi^2}{3} \bL \Bigr)
  - \Gamma_1 \bL
  - \bC \gamma_{S\,1}
\nn\\ &\quad
  - \frac{\Gamma_0}{2} \Bigl( \bigl\{ \bL,\, \hS_{\kappa,-1}^\one(\{\hat q_i\}) \bigr\}
  +  \bigl[\bI,\, \hS_{\kappa,-1}^\one(\{\hat q_i\}) \bigr] \Bigr)
  - 2 \beta_0\, \hS_{\kappa,-1}^\one(\{\hat q_i\})
\,.\end{align}

For two external partons, $\kappa = q\bar q$ and $\kappa = gg$, the result for the two-loop constant is known analytically~\cite{Kelley:2011ng, Monni:2011gb, Hornig:2011iu} and does not depend on the whether the partons are incoming or outgoing, i.e., it is the same for $0 \to q\bar q$, $q\to q$, and $q\bar q\to 0$, and similarly for two gluons~\cite{Kang:2015moa},
\begin{align}
\label{eq:softtwoloopconstqbarq} 
\hS_{q\bar q,-1}^\two
&= C_F \biggl[
   C_A \Bigl(-\frac{640}{27} + \frac{4\pi^2}{3} + \frac{22 \pi^4}{45} \Bigr)
   - C_F \frac{3 \pi^4}{10}
   + \beta_0 \Bigl(-\frac{20}{27} - \frac{37 \pi^2}{18} + \frac{58 \zeta_3}{3}\Bigr)
\biggr]
\,, \\
\label{eq:softtwoloopconstgg}
\hS_{gg,-1}^\two
&= C_A \biggl[
   C_A \Bigl(-\frac{640}{27} + \frac{4\pi^2}{3} + \frac{17 \pi^4}{90} \Bigr)
   + \beta_0 \Bigl(-\frac{20}{27} - \frac{37 \pi^2}{18} + \frac{58 \zeta_3}{3}\Bigr)
\biggr]
\,.\end{align}

The two-loop constants $\hS_\kappa^\two$ ($\kappa= ggg, q\bar q g$) required e.g. for $1$-jettiness in $pp$ collisions, have recently been computed numerically in \mycite{Boughezal:2015eha}. The two-loop constant for arbitrary $N$-jet processes can in principle be obtained numerically from known results for two-loop soft amplitudes as outlined in \mycite{Boughezal:2015eha}.

\bibliographystyle{../jhep}
\bibliography{../NNLO}

\end{document}